\documentclass[a4paper,11pt]{book}

\newcommand{\titolo}{Description of bound and unbound \\ many--body systems at the drip--lines}

\title{\titolo}
\author{Laura Moschini}
\usepackage{amsmath}
\usepackage{amsfonts}
\usepackage{amssymb}
\usepackage[english]{babel}
\usepackage{booktabs}
\usepackage[bf]{caption}
\usepackage{color}
\usepackage{fancyhdr}
\usepackage[T1]{fontenc}
\usepackage{lmodern}
\usepackage[top=35mm, bottom=35mm, left=29mm, right=29mm]{geometry}
\usepackage{graphicx, subfloat}
\usepackage[utf8x]{inputenc}
\usepackage{subfig}
\usepackage{pstricks} 
\usepackage{pst-circ}
\usepackage{textcomp}
\usepackage{multirow}
\usepackage{epstopdf}
\usepackage[babel]{csquotes}
\usepackage{amsmath}
\usepackage{graphicx}
\usepackage{caption}
\usepackage{float}
\usepackage[italian]{varioref} 
\usepackage{textcomp} 
\usepackage{booktabs} 
\usepackage[strict]{changepage} 
\usepackage[titletoc]{appendix}
\begin{document}
\begin{titlepage}

	\begin{center} 
		\begin{figure}[hc!]
		\begin{center}
			\includegraphics[scale=0.12]{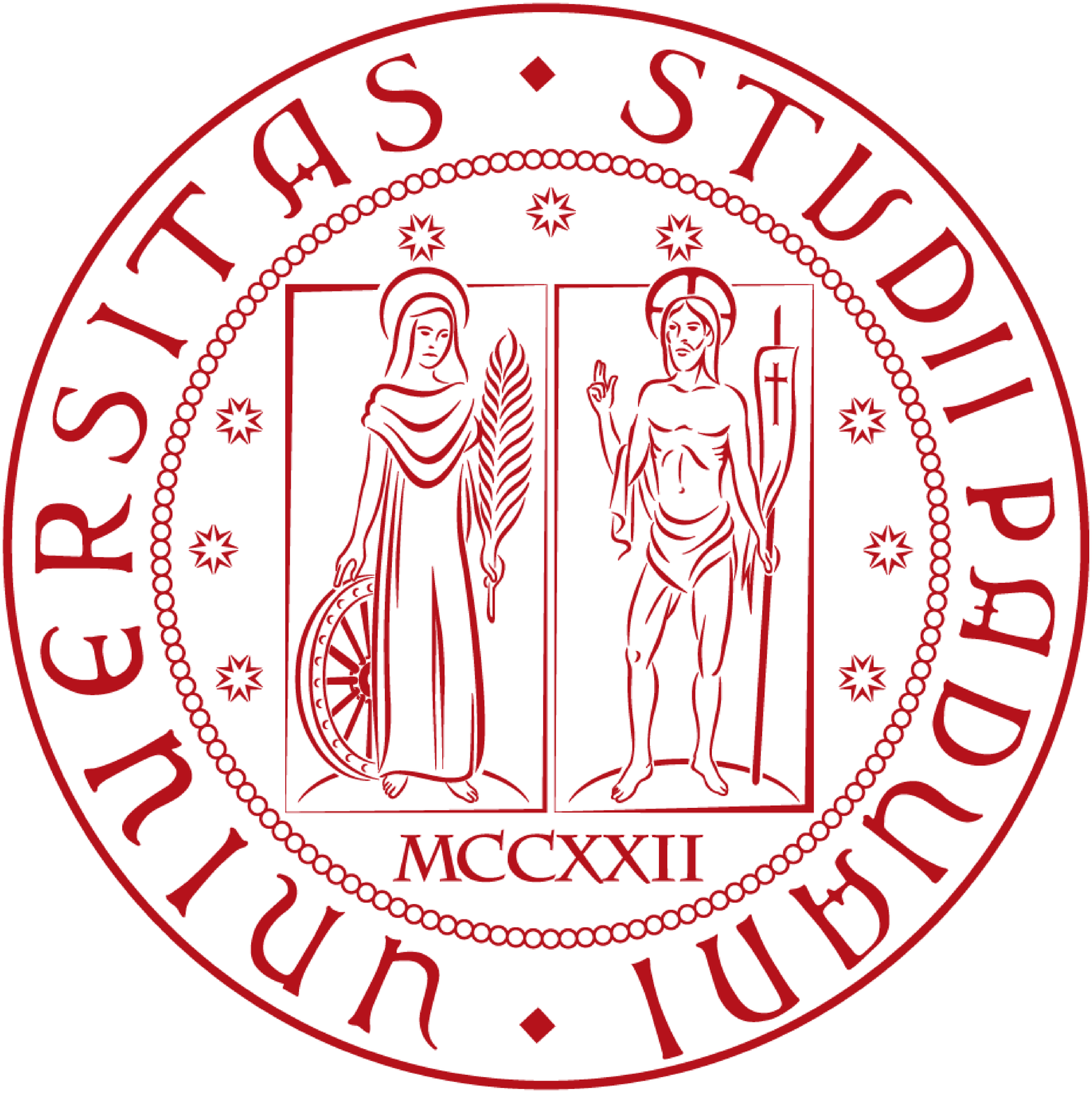}
		\end{center}
		\end{figure}
		
		\vspace{0.3cm}
		\textbf{\Large{UNIVERSITÀ DEGLI STUDI DI PADOVA}}\\
		\vspace{0.3cm}
		\Large{Dipartimento di Fisica e Astronomia ``Galileo Galilei''}\\
		\vspace{1cm}
		\large{\textsc{Tesi di Laurea Magistrale in Fisica}}
	\end{center}
	

\begin{center}
	\vspace{1.5cm}
	\huge{\textbf{Description of Bound and Unbound \\
						\vspace*{0.15cm}
						 Many--Body Systems at the Drip--Lines}}
	\vspace{4.4cm}
	\large{
		\begin{flushleft}
			\textbf{\textsc{Relatore}}: Prof.\ Andrea Vitturi\\
			\vspace*{0.3cm}
			\textbf{\textsc{Correlatore}}: Prof.\ Francisco B.\ Per\'ez Bernal\\
		\vspace{1cm}
		\end{flushleft}
		\begin{flushright}
			\textbf{\textsc{Laureanda}}: Laura Moschini\\
		\end{flushright}
	}
	\vspace{2.5cm}
	\rule{\textwidth}{0.5pt}
		\textsc{Anno Accademico 2012/2013}\\
\end{center}

\end{titlepage}
\tableofcontents
\chapter{Introduction}
        \newpage

        \null 

        \thispagestyle{empty} 

        \newpage
One of the hottest research lines in Nuclear Structure nowadays is the investigation and measurement of nuclei under extreme conditions and, in particular, nuclei far from stability. An example of such systems is a nucleus with many neutrons, with the barely bound outermost ones creating what is called a halo.
Nuclei that do not accept more neutrons mark on the isotope chart the neutron dripline, and along this line truly enticing and striking novel nuclear structure phenomena are being observed \cite{Tanihata}. Nowadays measuring the properties of such nuclei is the goal of the main experimental nuclear facilities around the world.\\

\begin{figure}[!htc]
\begin{center}
\scalebox{0.32}{\includegraphics{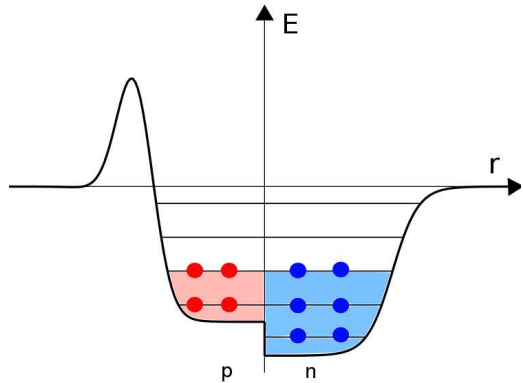}}
\caption{Schematic configuration of a stable nuclear system.\label{fig_intro_1}}
\end{center}
\end{figure}

\begin{figure}[!htc]
\begin{center}
\scalebox{0.32}{\includegraphics{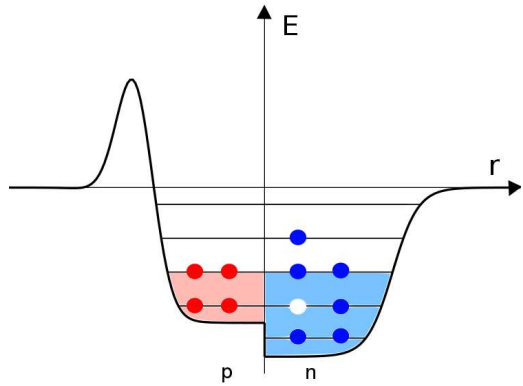}}
\caption{Schematic hole--particle excitation in a stable nuclear system.\label{fig_intro_2}}
\end{center}
\end{figure}

\begin{figure}[!htc]
\begin{center}
\scalebox{0.32}{\includegraphics{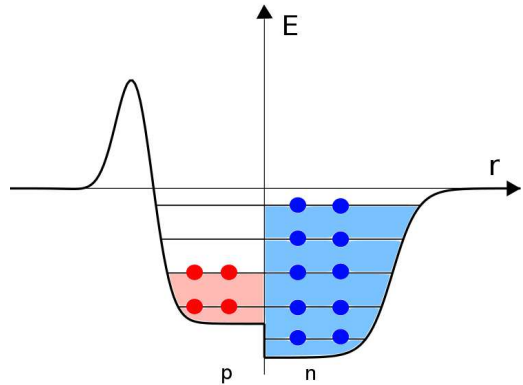}}
\caption{Schematic weackly--bound nuclear system close to the dripline.\label{fig_intro_3}}
\end{center}
\end{figure}

The theoretical description of halo nuclei is hindered by its weakly--bound nature. Bound nuclei in the vicinities of the stability valley can be modeled with a mean field potential partially filled with nucleons (protons and neutrons). In this simplified model a stable configuration is schematically depicted in Fig.\ \ref{fig_intro_1}, where all particles occupy the lowest energy state available. Nuclear excitations in this picture can be seen as the occupation by one or more nucleons of a higher energy state, leaving a hole behind, as shown in Fig.\ \ref{fig_intro_2}.
How is this simplified picture affected as nuclei get close to the neutron dripline? In this case the Fermi energy approaches to zero, as shown in Fig.\ \ref{fig_intro_3} and last nucleons are so weakly--bound that including the role of the continuum in the system description is absolutely necessary.\\

\vspace{0.5cm}
\vspace{0.5cm}
\vspace{0.5cm}

In accordance with this picture we present in this memory results for a one--dimensional toy model of a particular kind of halo nuclei: Borromean two--neutron halo nuclei. Borromean systems are three--body systems (core plus two particles) that are bound systems (with only one bound state in general) and they have no bound state in the possible binary subsystems. The best known examples of Borromean nuclei are $^6He$ and $^{11}Li$. Other Borromean nuclear systems are $^{14}Be$ and $^{22}C$.\\

The proposed two--body system main ingredients are a mean field 1D Woods--Saxon potential that happens to be totally filled plus two extra neutrons added to this core. The resulting system would be unbound, however the action of a point contact matter density--dependent residual interaction produces a final bound state.\\

It is clear that the description of such a system requires an adequate consideration of the role of the continuum. In order to do so we first deal  with one--body mean field Woods--Saxon potential in Chap.\ \ref{1b_probl}. We select a set of Woods--Saxon potential parameters that qualitatively model a nuclear structure problem with three bound states, and discretize the system continuum, using a finite set of normalizable (square--integrable) pseudostates; three different approaches have been followed. The first one consists on diagonalizing the Woods--Saxon potential Hamiltonian matrix in a truncated Harmonic Oscillator basis, the second makes use of a local scale transition to construct a truncated Transformed Harmonic Oscillator basis, while the third one uses a rigid wall box to achieve continuum discretization.
We present results obtained for eigenvalues, eigenfunctions and other quantities of interest (Total Strength and Energy Weighted Sum Rules, Electric Dipole and Quadrupole Transition Intensities) in Chap.\ \ref{1b_probl}. We also compare in this chapter the obtained results with the output of other approaches.\\

Once the pseudostate description of the model Woods--Saxon potential has been set up, we proceed in Chap.\ \ref{2b_probl} to state the two--body problem (two neutrons plus Woods--Saxon core) in 1D, to define an appropriate basis and to construct and diagonalize the system Hamiltonian with the different continuum discretization obtained in Chap.\ \ref{1b_probl}. Apart from a discussion on the convergence of eigenvalues and eigenfunctions we compute other quantities of interest as the Anomalous Density and Electric Dipole and Quadrupole Transition Transition Intensities.\\

Though the simplification achieved devising a 1D model precludes us from comparing directly to experimental data, we expect that the present toy model, despite its simplicity, contains the basic physical ingredients for a correct description of the problem under study. A solution of a simplified toy model frequently cast light upon a physical problem whose full solution is hindered by mathematical complexities or is plainly not possible. Reduced dimensionality models  of application in Nuclear Physics can be found e.\ g.\ in Refs.\ \cite{Bert1,Bert2,Dasso}. In particular, some results for the model investigated in the present memory can be found in \cite{Hanigo}.\\

Chapter \ref{HOvsTHO_cfr} is devoted to a short discussion of the comparison between the methods developed.\\

The fifth and last chapter of the memory contains some some concluding remarks and suggestions of future investigations along the present line of research.\\

As a last remark we would like to mention that the calculations presented in this memory have been carried out using \textsc{fortran90}, \textsc{gnu}-Octave, and \textsc{perl} codes\footnote{We have also made use of \textsc{nag}, \textsc{atlas}, \textsc{lapack95}, and \textsc{lapack} libraries.} developed for this purpose and that the codes are available under request\footnote{In these programs the following values of the
		relevant physical constants are used
		\begin{subequations}
		\begin{align}
		\hbar c &= 197.32858\,{\mbox MeV fm}~,\\
		amu &= 938.92635\, {\mbox MeV}/c^2~,\\
		\hbar^2/amu &= 41.4713768\, {\mbox MeV fm}^2~.
		\end{align}
		\label{physconst}
		\end{subequations}
	}.

\chapter{One--body problem\label{1b_probl}}
        \newpage

        \null 

        \thispagestyle{empty} 

        \newpage
In this chapter we present different pseudostate methods to model weakly--bound one--body quantum systems. In such systems, e.\ g.\ halo nuclei, it is mandatory to include continuum effects in the calculations. In the pseudostate methods the continuum wave functions are obtained as the eigenstates of the system Hamiltonian matrix in a truncated basis of square-integrable wave functions. We present three different continuum discretization methods, and test their results with a 1D Woods--Saxon potential with parameters chosen in order to mimic the nuclear interaction in a 1D system. We compare also the obtained results with a further approach to to continuum discretization, the average method, based on the definition of bins and building normalizable wave functions superposing the continuum wave functions.

\section{Continuum Discretization with Pseudostates\label{discr_methods}}
The problem of particles moving in a one dimensional Woods--Saxon mean field can be solved diagonalizing the Hamiltonian in different bases.
In particular we have made calculations using Harmonic Oscillator basis (HO), Transformed Harmonic Oscillator basis (THO) and Woods--Saxon in a Box.
Solutions of the BOX problem has been performed in two different ways. With the same spirit than the previous two cases, we have used a  Infinite Square Well basis (ISQW) to define an Hamiltonian matrix and diagonalize it.
The obtained results are satisfactorily compared with the numerical solution of the 1D time independent Schroedinger equation (TISE) using a Numerov approach.

We solve the 1D TISE 
\begin{equation}
\hat H_{1b} \psi(x) = E \psi(x) \rightarrow \left[ -\frac{\hbar^2}{2\mu} \frac{d^2}{dx^2} + V_{WS}(x) \right] \psi(x) = E \psi(x),
\label{1Dham}
\end{equation}
building and diagonalizing an Hamiltonian matrix with each of the bases considered to obtain the wave functions and the corresponding energies of the problem.\\
As our aim is to model weakly--bound systems, we obtain not only the existing bound
states, but also pseudostates in the continuum (positive energy states). 
The 1D Woods--Saxon potential is: \\
\begin{equation}
V_{WS}(x) = \dfrac{V_{0}}{1+e^{\frac{|x|-R}{\alpha}}}
\label{wsaxon}
\end{equation}
and the potential parameters we have chosen are the following: \\
\begin{subequations}
\begin{align}
V_0 &= -50.00\, {\mbox MeV}\\
R &= 2.00\,{\mbox fm}\\
a &= 0.40\, {\mbox fm^{-1}}\\
\mu &= 0.975\, {\mbox amu}
\end{align}
\label{wspotpar}
\end{subequations}

This particular set of parameters has been chosen to set up a toy model of a weakly--bound 1D nucleus.
With the parameters in Eq.\ (\ref{wspotpar}) we are not aiming to describe quantitatively any particular nucleus. The simple 1D nature of the model hinders any direct comparison with experimental data. However, as often occurs with toy models we expect to encompass the main physical ingredients of the weakly--bound system problem leaving aside a substantial part of the mathematical complexity that precludes a more detailed description.

\begin{figure}[!htc]
\begin{center}
\scalebox{0.32}{\includegraphics{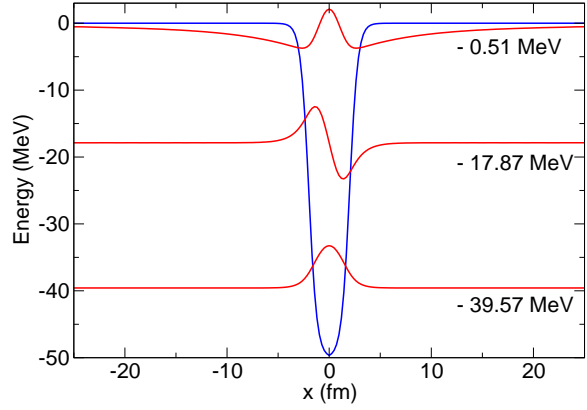}}
\caption{The Woods--Saxon 1D potential with the bound states. They have been computed with the methods that are presented in this section (in particular with a HO basis with N = 50).\label{pot_fig}}
\end{center}
\end{figure}

With this choice of parameters the system has three bound states whose energies are presented in Fig.\ \ref{pot_fig}.\\

\vspace{0.5cm}

In the present memory only the results obtained with a Woods--Saxon potential are reported, as they are requested for the description of the 1D many--body problem, but other potentials have also been tested in order to check the developed computational tools: Woods-Saxon with different parameters and with barriers, Morse, Hazi--Taylor, Ginocchio, and Poeschl--Teller potentials.\\

In the next subsections we present the different bases considered for continuum discretization. 
\subsection{\label{sec_HO}Harmonic Oscillator Basis}
The HO is a powerful tool in Quantum Mechanics which has been use as a basis to construct pseudostates \cite{hazi_taylor}.
To solve the TISE in a 1D Harmonic Oscillator (HO) basis, the starting point is a truncated $N$ dimensional basis set of 1D HO
wave functions 
\begin{equation}
\phi^{HO}_i(x) = \langle x| i\rangle = M_i \sqrt{a} H_i(a x)
e^{(-a^2x^2/2)}, ~i = 0, \ldots, N-1.
\label{hobasis}
\end{equation}
The parameter $a$ is the inverse
of the oscillator length, see Appendix \ref{inv_osc_len} for information about the optimization of this parameter.
 $H_i(a x)$ is the $i$-th Hermite polynomial,
and $M_i$ is a normalization constant
\begin{equation}
M_i = \frac{1}{\sqrt{2^i i!\sqrt{\pi} }}.
\end{equation}

The HO basis can be easily constructed using Hermite polynomials
recurrence relation. Also the necessary integral calculations are
simplified making use of the following Hermite polynomial properties
\cite{abram}
\begin{align}
\frac{d}{dx} H_n(x) &= 2 n H_{n-1}(x)~,\notag\\
H_{n+1}(x) &= 2 x H_{n}(x) - 2 n H_{n-1}(x)~,\label{abramrel}\\
x H_j(x) &= \frac{1}{2} H_{j+1}(x) + j H_{j-1}(x)~,\notag\\
x^2 H_j(x) &= \frac{1}{4} H_{j+2}(x) + \frac{2j+1}{2} H_{j}(x) +j(j-1) H_{j-2}(x)~.\notag
\end{align}

The resulting basis is
\begin{align}
\phi^{HO}_0(x) &= \sqrt{\frac{a}{\sqrt{\pi}}} \exp{(-a^2x^2/2)}~,\notag \\
\phi^{HO}_1(x) &= \sqrt{\frac{a}{2\sqrt{\pi}}} (2ax) \exp{(-a^2x^2/2)}~,\notag \\
&\ldots\notag\\
\phi^{HO}_n(x) &= \sqrt{\frac{a}{2^nn!\sqrt{\pi}}} \frac{1}{\sqrt{2}}
H_n(a x) \exp{(-a^2x^2/2)}~,\notag \\
\phi^{HO}_{n+1}(x) &= \sqrt{\frac{a}{2^{n+1}(n+1)!\sqrt{\pi}}} [2 a x
H_n(a x) - 2n H_{n-1}(a x)] \exp{(-a^2x^2/2)}~,\notag \\
&\ldots\notag\\
\phi^{HO}_k(x) &= \sqrt{\frac{2}{k}} \,a \,x\, \phi^{HO}_{k-1}(a x)-\sqrt{\frac{k-1}{k}}\phi^{HO}_{k-2}(a x) ~.
\end{align}

The 1D TISE for Harmonic Oscillator states is
\begin{equation}
\left[-\frac{\hbar^2}{2\mu} \frac{d}{dx^2} + V_{HO}(x)\right]\phi^{HO}_n(x) = \hslash \omega (n+\frac{1}{2}) \phi^{HO}_n(x)~,
\end{equation}
\noindent where
\begin{align}
V_{HO}(x) &= \frac{1}{2}K x^2~, \\
a &= \sqrt[4]{\frac{\mu  K}{\hbar^2}}~,\\
\hbar \omega &= \hbar \sqrt{\frac{K}{\mu}} = \frac{\hbar^2 a^2}{\mu}~.
\end{align}

In order to obtain solutions of Eq.\ (\ref{1Dham}) for $ V_{WS} $ (\ref{wsaxon}, \ref{wspotpar}) the first step is to compute the matrix elements of the kinetic energy in the HO basis. This part is common to any potential.
\begin{align*}
 \langle \phi^{HO}_{i} |\hat T| \phi^{HO}_{j} \rangle &= -\frac{\hbar^2}{2\mu} a N_i N_j
 \int_{-\infty}^{+\infty}dx\,\exp{(-a^2x^2/2)} H_i(a
 x)\frac{d^2}{dx^2}H_j(a x)\exp{(-a^2x^2/2)}  \\
 &=  -\frac{\hbar^2}{2\mu} a^2 N_i N_j
 \int_{-\infty}^{+\infty}dy\,\exp{(-y^2/2)} H_i(y)\frac{d^2}{dy^2}H_j(y)\exp{(-y^2/2)} ~,
\end{align*}
\noindent with $y = ax$.

Making use of the relations (\ref{abramrel})
\begin{displaymath}
\frac{d^2}{dy^2}H_j(y)\exp{(-y^2/2)} = \left[\frac{1}{4}H_{j+2}(y)+j(j-1)H_{j-2}(y)-\frac{2j+1}{2}H_{j}(y)\right]\exp{(-y^2/2)}~.
\end{displaymath}

The inclusion of this expression in the integral gives as a result,
taking into account the Hermite polynomials orthonormality
\begin{equation}
 \langle \phi^{HO}_{i} |\hat T| \phi^{HO}_{j} \rangle = -\frac{\hbar^2}{2\mu} a^2 \frac{\sqrt{(j+1)(j+2)}}{2}\,\delta_{i,j+2}-\frac{\hbar^2}{2\mu} a^2 \frac{\sqrt{j(j-1)}}{2}\,\delta_{i,j-2}+\frac{\hbar^2}{2\mu} a^2 \frac{2j+1}{2}\,\delta_{i,j} ~.\label{tij}
\end{equation}

Note that an identical result is obtained taking into account
  that $\hat T = \hat H - \hat V_{HO}$ and thus $\langle \phi^{HO}_{i} |\hat T| \phi^{HO}_{j} \rangle = \left(j+\frac{1}{2}\right)\frac{\hbar^2 a^2}{\mu}\,\delta_{i,j} - \frac{K}{2}
  \langle i| x^2|  j\rangle$. The values of the physical constant
  employed in the codes are given in Eq.\ (\ref{physconst}).\\

The second step is the calculation of the potential matrix elements in the HO basis
\begin{equation}
 \langle \phi^{HO}_{i} |\hat V| \phi^{HO}_{j} \rangle = 
\int _{-\infty} ^{+\infty} dx ~\phi ^{(HO) \ast} _{i} ~V_{WS}~\phi^{(HO)}_{j}.
\label{Vij}
\end{equation}
These and other integrals in the developed codes were computed using a third order finite differential formula with a subroutine from the NAG library.
Due to symmetry reasons $ \langle \phi^{HO}_{i} |\hat V| \phi^{HO}_{j} \rangle = 0 $ if the basis wave functions have different parity (symmetric or antisymmetric).

Once the matrix is diagonalized, we obtain a set of one--body eigenvalues $ E_{WS,i} ^{(HO)}$ and eigenfunctions $ \vert  \psi_{WS,i} ^{(HO)} \rangle $ that are linear combinations of the basis elements

\begin{equation}
\psi_{WS,i} ^{(HO)}(x) = \sum_{k=0}^{N-1}\alpha ^{(HO)} _{ik}\phi^{HO}_{k}(x),~~i~=~0,~1,~2,~...,~N-1.
\label{1bodywfHO}
\end{equation} 
\subsection{Transformed Harmonic Oscillator Basis}
A disadvantage of the HO basis when used as a basis in variational methods to model bound states is the Gaussian asymptotic behavior, compared to the exponential behavior of the bound states.
This is even more stressed in continuum pseudostates. This fact and the rigidity of the HO basis explain the success of the THO basis.
A Transformed HO (THO) basis consists of a HO basis to which a local scale transformation (LST) $ s(x) $ has been applied. The aim of this transformation is to alter the HO wave functions asymptotic behavior. See, e.\ g.\, Refs.\ \cite{tho_lst_2,tho_lst,tho_lst_3}.

To solve the problem in a 1D Transformed Harmonic
Oscillator (THO) basis the starting point is a truncated $N$ dimensional basis set $ \phi^{HO}_i(x) $ of 1D HO
wave functions (see section \ref{sec_HO}), which has to be scaled into the new basis
\begin{equation}
\phi^{THO}_i(x) = \sqrt{\frac{ds(x)}{dx}}~\phi^{HO}_i(s(x))
~i = 0, \ldots, N-1~,
\label{thobasis}
\end{equation}
according to the analytical LST function (see Ref.\ \cite{tho_lst_4})
\begin{equation}
s(x) = ( x^{-m} + (\gamma\sqrt{x})^{-m} )^{-\frac{1}{m}}
\label{scaling}
\end{equation}
that is valid for $ x > 0 $; for negative $ x $ values we impose that $ s(x) $ is an odd function: $ s(-x) = -s(x) $.
The quantity $\gamma $ is a parameter of the LST and is a variable of the calculation; $ b $ is the oscillator length which is determined by minimizing the ground state energy with respect to this parameter for a $N = 1$ basis. The power $m$, according to Ref.\ \cite{tho_lst} is set to $ m = 4 $; though the results depend very weakly on this parameter.\\

The matrix elements of the kinetic energy operator in this THO basis can be computed as
\begin{equation}
 \langle \phi^{THO}_{i} |\hat T| \phi^{THO}_{j} \rangle = - \frac{\hslash^2}{2\mu}~\int_{-\infty}^{+\infty} dx ~\sqrt{\frac{ds(x)}{dx}}~\phi^{HO}_i[s(x)]~\frac{d^2}{dx^2}\sqrt{\frac{ds(x)}{dx}}~\phi^{HO}_j[s(x)].
\label{iTj_THO}
\end{equation}
Making use of the relations
\begin{equation}
\frac{ds(x)}{dx} = \frac{s(x)}{2x}~ \left[ 1+ \left( \frac{s(x)}{x} \right) ^m \right] ,
\label{scaling_der}
\end{equation}
and
\begin{equation}
\frac{d^2 s(x)}{dx^2} = - \frac{s(x)}{4x}~\frac{x^{\frac{m}{2}-1}}{\gamma^m + x^{\frac{m}{2}}}~\left[ 1+ (m+1) \left( \frac{s(x)}{x} \right)^m \right],
\label{scaling_2_der}
\end{equation}
it can be finally evaluated as
\begin{align*}
 \langle \phi^{THO}_{i} |\hat T| \phi^{THO}_{j} \rangle  = & \frac{1}{4}\frac{\hslash^2}{2\mu}~\int_{-\infty}^{+\infty} dx ~ \left( \frac{d^2 s(x)}{dx^2} \right)^2~\left( \frac{ds(x)}{dx}\right) ^{-1}~\phi^{HO}_i(s(x))\phi^{HO}_j(s(x))\\
& + \frac{1}{2}\frac{\hslash^2}{2\mu} a~\int_{-\infty}^{+\infty} dx ~  \frac{d^2 s(x)}{dx^2}~ \frac{ds(x)}{dx} ~\phi^{HO}_i(s(x))~\left[ \sqrt{\frac{j}{2}}\phi^{HO}_{j-1}(s(x)) - \sqrt{\frac{j+1}{2}}\phi^{HO}_{j+1}(s(x))\right]\\
& + \frac{1}{2}\frac{\hslash^2}{2\mu} a~\int_{-\infty}^{+\infty} dx ~ \frac{d^2 s(x)}{dx^2}~\frac{ds(x)}{dx}~\phi^{HO}_j(s(x))~\left[ \sqrt{\frac{i}{2}}\phi^{HO}_{i-1}(s(x)) - \sqrt{\frac{i+1}{2}}\phi^{HO}_{i+1}(s(x))\right]\\
& + \frac{\hslash^2}{2\mu} a^2~\int_{-\infty}^{+\infty} dx ~\left( \frac{ds(x)}{dx}\right)^{3}~ \left[\sqrt{\frac{i}{2}}\phi^{HO}_{i-1}(s(x)) - \sqrt{\frac{i+1}{2}}\phi^{HO}_{i+1}(s(x))\right]\\
& + \frac{\hslash^2}{2\mu} a^2~\int_{-\infty}^{+\infty} dx ~\left( \frac{ds(x)}{dx}\right)^{3}~\left[\sqrt{\frac{j}{2}}\phi^{HO}_{j-1}(s(x)) - \sqrt{\frac{j+1}{2}}\phi^{HO}_{j+1}(s(x))\right],\\
\end{align*}
where following the notation of the previous subsection, $a = b^{-1}$ is the inverse oscillator length.\\

Then the second step is the calculation of the potential matrix elements $ \langle \phi^{THO}_{i} |\hat V| \phi^{THO}_{j} \rangle $.

Once the TISE for the 1D mean field Woods--Saxon
potential (\ref{wsaxon}) is solved using the basis (\ref{thobasis}), 
we obtain a set of discrete eigenvalues $ E_{WS,i} ^{(THO)}$ and eigenfunctions $ \vert  \psi_{WS,i} ^{(THO)} \rangle $ that can be written as a linear combinations of the THO basis states
\begin{equation}
\psi_{WS,i} ^{(THO)}(x) = \sqrt{\frac{ds(x)}{dx}}~\sum_{k=0}^{N-1}\alpha ^{HO} _{ik}\phi^{HO}_{k}(s(x)),~~i~=~0,~1,~2,~...,~N-1.
\label{1bodywfTHO}
\end{equation} 
\subsection{Woods--Saxon in a Box}
A third possible way to obtain a continuum discretization is making use of a box of radius $ x_b $, with $ x_b $ being large enough compared to the potential range

\begin{equation}
V_{b}(x)=\begin{cases}
V_{WS} & \text{if $|x|< x_b $} ,\\
\infty & \text{if  $|x| \geq x_b $}.
\end{cases}
\label{boxpot_b}
\end{equation}
As previously explained, this potential--in--a--box problem can be solved using standard numerical techniques for the solution of differential equations.
However, in order to define an approach similar to the one presented in the HO and THO cases, more algebraic, we proceed to build and diagonalize the system Hamiltonian matrix in a truncated Infinite Square Well basis (ISQW).
Results are compared with the standard approach in the next sections.

In this case the basis used to build the 1D Hamiltonian matrix is  a truncated N dimensional basis of eigenstates 
basis of an ISQW) at $\pm x_b$
\begin{equation}
V_{ISQW}(x)=\begin{cases}
0& \text{if $|x|<x_b $} ,\\
\infty& \text{if  $|x|>x_b $}.
\end{cases}
\label{boxpot}
\end{equation}

The basis set wave functions can be written as 
\begin{equation}
\phi^{B}_k(x) = \begin{cases}
\frac{1}{\sqrt{x_b}} \cos{\left(\frac{k \pi x}{2 x_b}\right)}& \text{if
    $k$ is odd},\\
\frac{1}{\sqrt{x_b}} \sin{\left(\frac{k \pi x}{2 x_b}\right)}& \text{if
    $k$ is even,}
\end{cases}
\label{boxbasisapp}
\end{equation}
\noindent with $k = 0,1,2 \ldots, N-1$. The parameter $x_b$ is the
box potential radius. 

The basis and Hamiltonian matrix element calculation is computationally simple.

The first step is to compute the matrix elements of the kinetic energy
term in the ISQW. As in the previous cases, this part is common to
any potential, and due to symmetry considerations, the matrix elements $\langle \phi^{B}_{j} |\hat T|
\phi^{B}_{k} \rangle$ are zero when $j$ and $k$ do not have the same parity. 
In case that both $j$ and $k$ have the same parity 
\begin{align*}
 \langle \phi^{B}_{j} |\hat T| \phi^{B}_{k} \rangle &= -\frac{\hbar^2}{2\mu} \frac{1}{x_b}
 \int_{-x_b}^{x_b}dx\,\sin{\left(\frac{j \pi x}{2x_b}\right)} 
 \frac{d^2}{dx^2}\sin{\left(\frac{k \pi x}{2
       x_b}\right)} \\
 &= \frac{\hbar^2}{2\mu} \frac{1}{x_b}\frac{k^2\pi}{2 x_b}
 \int_{-\pi/2}^{\pi/2}dy\,\sin{\left(j y\right)} \sin{\left(k y\right)} \\
 &=  \frac{\hbar^2}{2\mu} \frac{k^2\pi^2}{4 x_b^2}~,
\end{align*}
\noindent with $y = \frac{\pi x}{2 x_b}$.

Then the second step is the calculation of the potential matrix elements $ \langle \phi^{B}_{j} |\hat V| \phi^{B}_{k} \rangle $.
If the one-body mean field potential is symmetrical, in the
calculation of the potential elements $\langle \phi^{B}_{j} |\hat V| \phi^{B}_{k} \rangle = 0$
if $j$ and $k$ have different parities, as in the previous cases.\\

Once the 1D TISE for the mean field
Woods--Saxon potential (\ref{wsaxon}) is solved, we obtain the
eigenenergies $ E_{WS,i} ^{(B)}$ and the one-body eigenfunctions \\
\begin{equation}
\psi_{WS,i} ^{(B)} (x) = \sum_{k=0}^{N-1}\alpha ^{B} _{ik}\phi ^{B}_{k}(x),~~i~=~0,~1,~2,~...,~N-1.
\label{1bodywfBOX}
\end{equation} 
\section{Model Woods--Saxon Potential Energies and Wavefunctions}

As previously stated, the pseudostate (PS) approach that we follow implies the diagonalization of the model potential in different basis sets.
The obtained results depend on the basis dimension as well as on other parameters (e.\ g.\ oscillator length $b=a^{-1}$, LST parameter $\gamma$, or box radius $x_b$).
We proceed to study the convergence of eigenvalues and eigenfunctions when these parameters varied.

In particular, the chosen Woods--Saxon potential has a very weakly--bound state $\Psi_2(x)$ ($E_2 = -0.51 MeV$) shown in Fig.\ \ref{pot_fig}, whose energy and wave function correct calculation is an exigent test for the approach.

We check the energy convergence and the variation of the asymptotic wave functions values with the parameters.
A proper description of the wave functions tails is crucial to reasonably describe the structure of a nuclear system; in fact these results are most important for the reaction calculations.\\

\vspace{0.5cm}
\begin{figure}[!htc]
\begin{center}
\scalebox{0.42}{\includegraphics{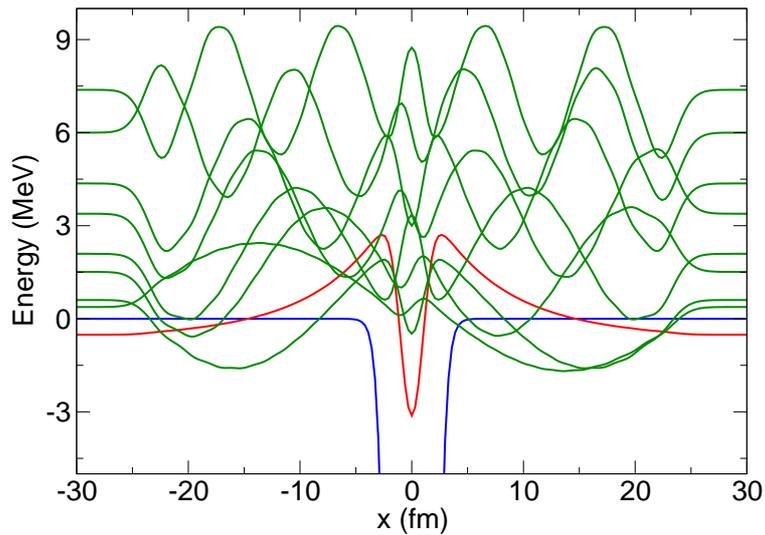}}
\caption{The Woods--Saxon model potential (blue), last bound wave function (red) and the first continuum PSs (green) obtained with the Hamiltonian matrix diagonalization (in particular HO with N = 50).\label{fig_pot_continuum}}
\end{center}
\end{figure}

\paragraph{Harmonic Oscillator Basis}
In the HO case the main parameter is the basis dimension N; so we check results convergence with the basis size.
Apart from N, a second parameter is the inverse oscillator length $a$.
In general its value is fixed by optimizing the system ground state energy with an $N = 1$ basis. However, for cases like the model Woods--Saxon potential with weakly--bound states an algorithm has been devised to estimate $a$ values that improve the convergence with N (see App.\ \ref{inv_osc_len}).

\begin{figure}[!htc]
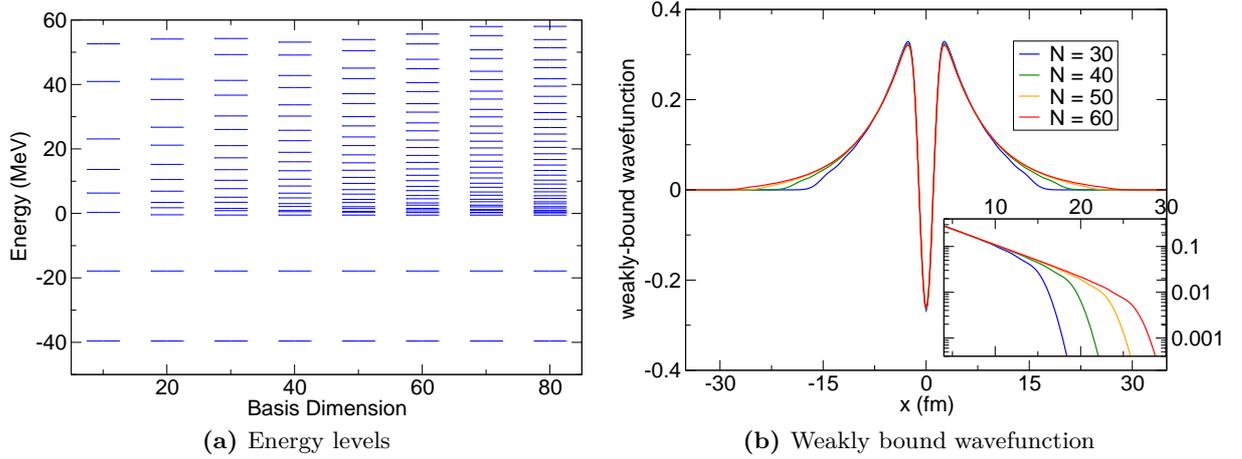

\begin{center}
\subfloat[Energy levels]{
\scalebox{0.32}{\includegraphics{spectra_ho_func_of_N}}
}
\hspace*{0.1cm}
\subfloat[Weakly bound wavefunction]{
\scalebox{0.32}{\includegraphics{weakly_wf_ho_func_of_N}}
}
\caption{Panel (a): Eigenvalues of the Woods--Saxon model potential as a function of the number of basis states. Panel(b): weakly--bound state wave function as a function of x for different N values.\label{fig_ho}}
\end{center}
\end{figure}

In Fig.\ \ref{fig_ho}a we depict the eigenvalues of the model Woods--Saxon potential as a function of the basis dimension N.
Negative energy levels converge to the bound state  energies quite fast. We consider that an energy level is converged when $\Delta E = 5 keV$ for a dimension increment $\Delta N = 10$.
Convergence in this case is obtained for $N = 50$. Of course the convergence is much faster for the ground state and first excited state than for the weakly--bound second state.

In Fig. \ref{fig_ho}b we show the weakly--bound state wave function for different N values.
As expected, the major differences are in the wave function tails, that requires large N values to extend towards large $x$ values (see inset panel).

\vspace{0.5cm}

\begin{figure}[!htc]
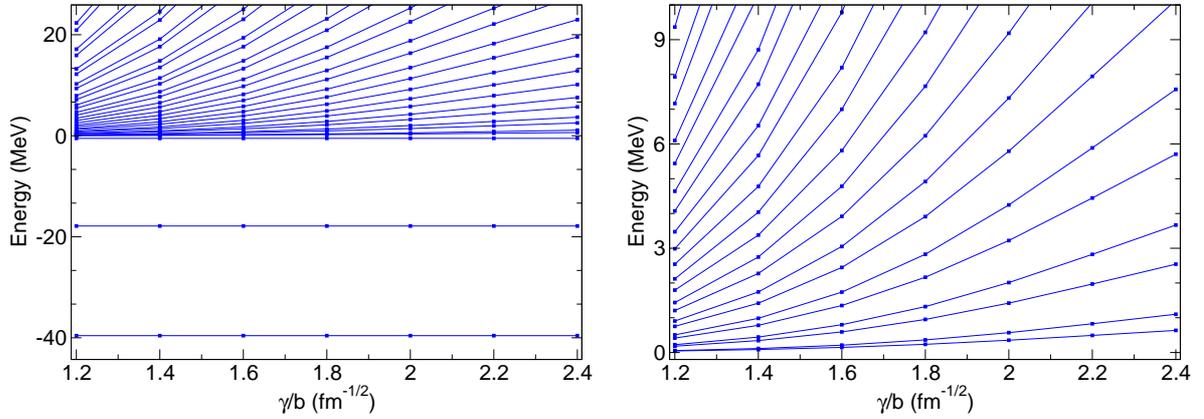

\begin{center}
\subfloat{
\scalebox{0.32}{\includegraphics{spectra_tho_func_of_ratio_N50}}
}
\hspace*{0.1cm}
\subfloat{
\scalebox{0.32}{\includegraphics{spectra_tho_func_of_ratio_N50_zoom}}
}
\caption{Eigenvalues of the model Woods--Saxon potential as a function of the $\gamma/b$ ratio for a THO basis with $N = 50$. The right panel is a zoom of the first continuum levels. \label{fig_tho_levels_with_Ratio}}
\end{center}
\end{figure}

\paragraph{Transformed Harmonic Oscillator Basis}
In the THO basis case results depend on the parameters $\gamma/b$ and N.
The second parameter of the LST, $m$, affects very lightly to the results and has been fixed to a constant $m = 4$ value. The value of $b$ is optimized 
minimizing the system ground state energy with an $N = 1$ basis.
The ratio $ \gamma/b = \left( \frac{8\mu\varepsilon}{\hbar^{2}} \right)^{1/4} $ and gives an extra degree of freedom to the approach compared to the HO case.
The value of $\frac{\gamma ^2}{2 b^2}$ can be seen as an effective momentum value $k_{eff}$ and the asymptotic value of the basis functions is $e^{- \frac{\gamma ^2}{2 b^2} |x|}$. 
As $\gamma/b$ increases (decreases) the basis spatial extension increases (decreases).
Therefore, for large $\gamma/b$ values the positive eigenvalues tend to concentrate at higher energies, and the $\gamma/b$ ratio controls the density of PSs as a function of the excitation energy. 
This is an useful property of the THO basis that makes this approach more flexible than the HO to deal with different physical problems \cite{tho_lst_3}.
In Fig.\ \ref{fig_tho_levels_with_Ratio} we depict the system energies as a function of $\gamma/b$.
It can be easily appreciated the varying PS density and the piling of PSs at low energies for small $\gamma/b$ values.
For large $\gamma/b$ values, as can also be seen in Fig.\ \ref{fig_tho_levels_with_Ratio}, the THO reaches the HO limit.

The improved asymptotic wavefunction behavior is ascertained computing the bound states energy convergence.
In order to achieve the $5 keV$ energy convergence goal, an $N = 50$ basis is needed for $\gamma/b = 2.4 fm^{-1/2}$ (as in the HO case) while it is enough with an $N = 20$ basis for $\gamma/b = 1.2 fm^{-1/2}$.

\begin{figure}[!htc]
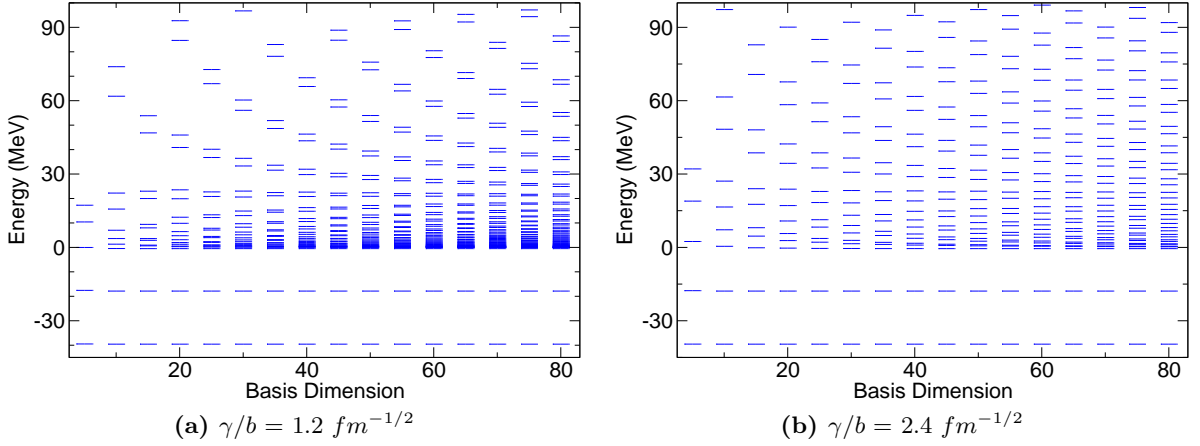

\begin{center}
\subfloat[$\gamma/b$ = 1.2 $fm^{-1/2} $]{
\scalebox{0.32}{\includegraphics{spectra_tho_func_of_N_ratio1c2}}
}
\hspace*{0.1cm}
\subfloat[$\gamma/b$ = 2.4 $fm^{-1/2} $]{
\scalebox{0.32}{\includegraphics{spectra_tho_func_of_N_ratio2c4}}
}
\caption{Model Woods--Saxon potential energy levels as a function of the THO basis dimensions N for $\gamma/b = 1.2 fm^{-1/2}$ (panel (a)) and $\gamma/b = 2.4 fm^{-1/2}$ (panel(b)).\label{fig_tho_levels_with_N}}
\end{center}
\end{figure}

In Fig.\ \ref{fig_tho_levels_with_N} the energy spectrum for the two previous values of $\gamma/b$ is depicted as a function of N.
In both cases the appearence of symmetry pairs is clear.
In Fig.\ \ref{fig_tho_levels_with_N}a $\gamma/b = 1.2 fm^{-1/2}$ and in Fig.\ \ref{fig_tho_levels_with_N}b $\gamma/b = 2.4 fm^{-1/2}$.
In the $\gamma/b = 1.2 fm^{-1/2}$ case the weakly--bound level converges faster to its true energy value and when both figures are compared it is remarkable the difference in PS densities, mainly in the low energy region.\\

\begin{figure}[!htc]
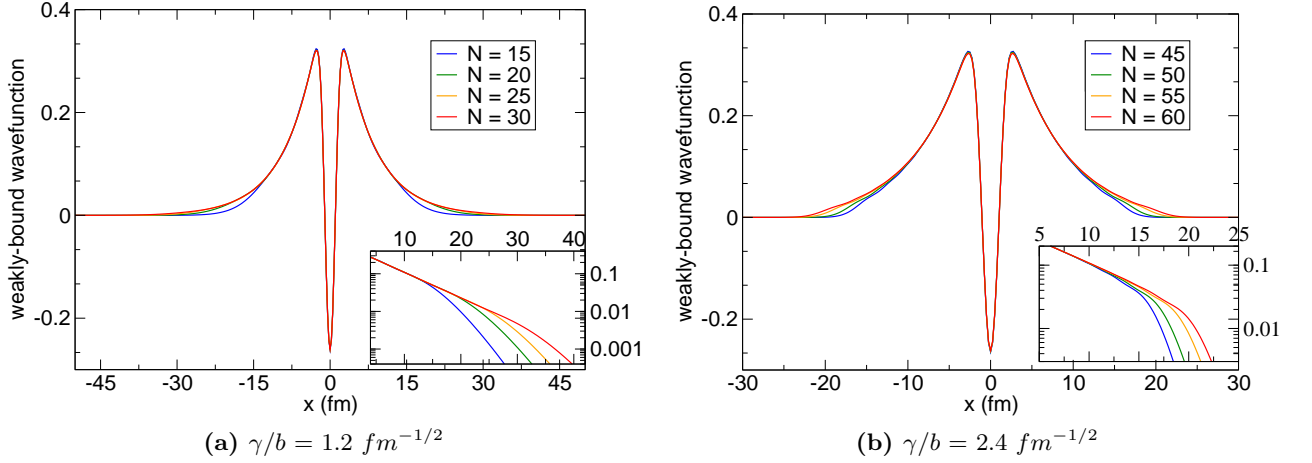

\begin{center}
\subfloat[$\gamma/b$ = 1.2 $fm^{-1/2} $]{
\scalebox{0.32}{\includegraphics{weakly_wf_tho_func_of_N_ratio1c2}}
}
\hspace*{0.1cm}
\subfloat[$\gamma/b$ = 2.4 $fm^{-1/2} $]{
\scalebox{0.32}{\includegraphics{weakly_wf_tho_func_of_N_ratio2c4}}
}
\caption{Wavefunction of the third bound state (weakly--bound state) of the model Woods--Saxon potential for different basis dimensions and $\gamma/b = 1.2 fm^{-1/2}$ (panel (a)) and $\gamma/b = 2.4 fm^{-1/2}$ (panel(b)).\label{fig_tho_weakly_wf_with_N}}
\end{center}
\end{figure}

As in the HO case, the wave function $\Psi_2(x)$ (weakly--bound state) extends to larger x values and is more difficult to reproduce than the other two bound states wave functions. 
In Fig.\ \ref{fig_tho_weakly_wf_with_N} we show the wave function of this weakly--bound state for different basis dimension N values and $\gamma/b = 1.2 fm^{-1/2}$ (Fig.\ \ref{fig_tho_weakly_wf_with_N}a) and $\gamma/b = 2.4 fm^{-1/2}$ (Fig.\ \ref{fig_tho_weakly_wf_with_N}b).
As expected the larger $\gamma/b$ case is very similar to the HO result depicted in Fig.\ \ref{fig_ho}, while the asymptotic behavior is greatly improved for $\gamma/b = 1.2 fm^{-1/2}$.\\

A comparison between the THO and HO results is presented in Chap.\ \ref{HOvsTHO_cfr}.

\vspace{4cm}

\vspace{1cm}

\newpage
\paragraph{Potential in a Box}

\begin{figure}[!htc]
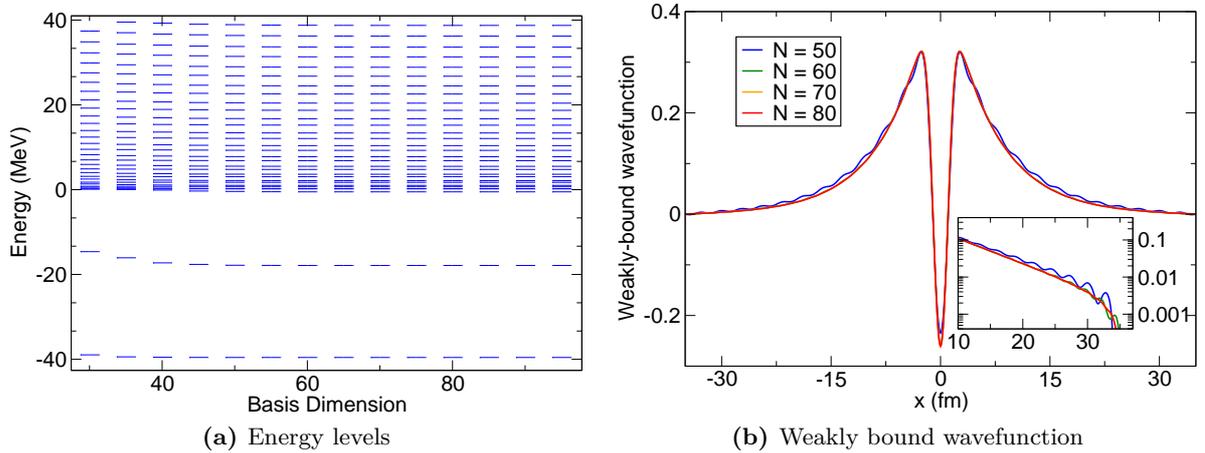

\begin{center}
\subfloat[Energy levels]{
\scalebox{0.32}{\includegraphics{spectra_isqw_func_of_N_Xb35}}
}
\hspace*{0.1cm}
\subfloat[Weakly bound wavefunction]{
\scalebox{0.32}{\includegraphics{weakly_wf_isqw_func_of_N}}
}
\caption{In the left part is reported the spectrum as a function of N, for $ x_{b} $ fixed at the convergence value of 35 fm. In the right part: weakly bound wave function again fixing $x_b = 35 fm$ and varying the basis dimension N.\label{fig_isqw_1}}
\end{center}
\end{figure}

The main parameter when solving the problem of a potential in a rigid wall box is $x_b$, the location of the box. Nevertheless, as discussed in the previous section, we build the system Hamiltonian matrix using an ISQW basis. Therefore we need to set a two--step procedure.
We first study the results convergence with N for a fixed $x_b$ value and, afterwards, we study the results dependence with $x_b$.
If we fix $x_b = 35 fm$, the bound state $\Delta E = 5 keV$ energy convergence is achieved for $N = 85$.
The eigenvalues dependence with N is shown in Fig.\ \ref{fig_isqw_1}a. 
It is important to emphasize the difference between this case and the HO and THO cases.
In the last two cases as N increases the bound state energies converge to their true values while the PS energy density increase, and positive energy values tend to fill the positive energy region, as can be seen in Figs.\ \ref{fig_ho}a and \ref{fig_tho_levels_with_N}.
The box transforms the problem and true continuum states disappear, being replaced by eigenstates of the Box plus Woods--Saxon potential (\ref{boxpot_b}).
This is apparent in Fig.\ \ref{fig_isqw_1}a where not only Woods--Saxon bound states tend to constant value but also positive energy states do so.
Of course the PS density, once N reaches a sufficiently large value, is constant.
Thus in this case convergence in eigenvalues can be forced both for negative and positive energies and the value of N depends on the number of PSs included.

\begin{figure}[!htc]
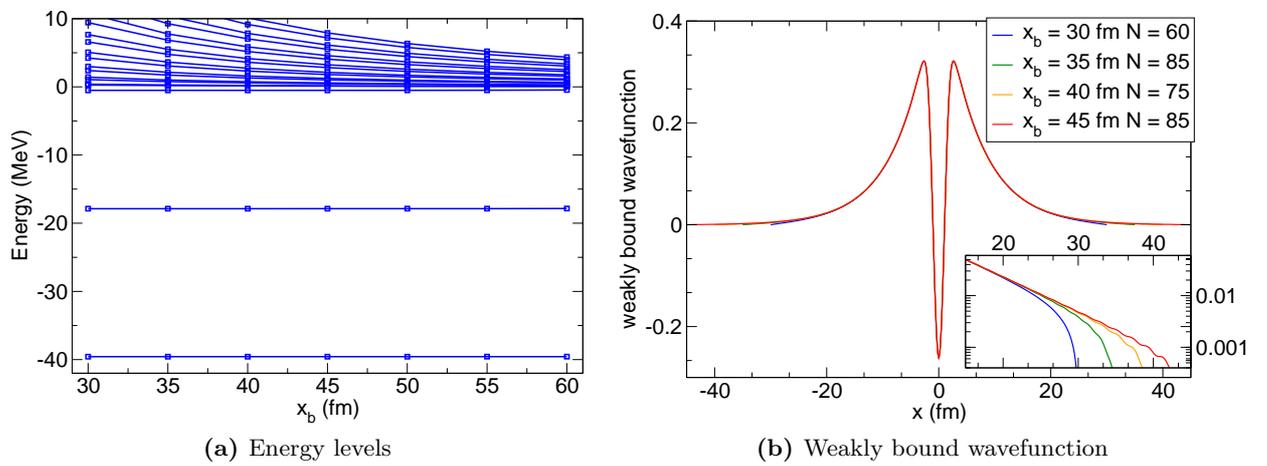

\begin{center}
\subfloat[Energy levels]{
\scalebox{0.32}{\includegraphics{spectra_isqw_func_of_Xb_N90}}
}
\hspace*{0.1cm}
\subfloat[Weakly bound wavefunction]{
\scalebox{0.32}{\includegraphics{weakly_wf_isqw_func_of_Xb}}
}
\caption{In the left part is reported the spectrum as a function of  $ x_{b} $ fixing N = 90. In the right part: weakly bound wave function varying $x_b$ (with the corresponding number of basis states at which convergence is reached).\label{fig_isqw_2}}
\end{center}
\end{figure}

In Fig.\ \ref{fig_isqw_1}b we plot the weakly--bound state $\Psi_2(x)$ wave function for $x_b = 35 fm$ and different ISQW basis dimension N.
The dependence on the box radius $x_b$ is shown in Fig.\ \ref{fig_isqw_2}.
In the left panel the Woods--Saxon potential in a box eigenvalues are plotted as a function of $x_b$.
In this case, as the box radius increases the value of N necessary to obtain convergence results also increases, Woods--Saxon bound states tend to their true values and the density of PS increases in the low energy region.

In the right panel of Fig.\ \ref{fig_isqw_2} we show the $\Psi_2(x)$ wave function for different box sizes.
In the plot legend is also included the basis dimensions N required to obtain converged values.
In this case it is very important that the $x_b$ value is large enough in order to truncate the bound state wave function.
In fact, the oscillations in the tail probably depend on the computational truncation of the basis, as N increases they reduce.

\vspace{0.5cm}
\vspace{0.5cm}
\section{Results for Sum Rules, Electric Transitions and Resonances\label{1b_results}}
In the present section we present results from calculations other than the eigenvalues and eigenfunctions of the one--body model Woods--Saxon potential.

We first compute the dependence with the basis size of the Total Strength (${\cal S}_T ^{(b)}$) and the Energy--Weighted Sum Rule ($ {\cal E}_W ^{(b)}$).
These two quantities should converge to their true values as the basis size increases and give an estimation of the goodness of our PS continuum discretization.

In second place, we compute Dipole and Quadrupole Electric Transition Moment integrals for the three bound states and its associated transition probabilities for the model Woods--Saxon potential, comparing the results obtained with the different approaches considered.

In third and final place, we briefly show how to deal with situations with resonances in a simple and computationally efficient way in the PS formalism.

\subsection{Sum Rules}
We define the Total Strength for the operator $ {\cal O}$ and the b-th bound state, $ {\cal S}_T ^{(b)} ({\cal O}, N) $, as follows
\begin{equation}
{\cal S}_T ^{(b)}({\cal O}, N) = 
\sum_{i=0} ^{N-1} \mid  \langle \Psi_{b} \vert  {\cal O}  \vert \Psi_{i} \rangle \mid ^2 
\label{completeness}
\end{equation}
where the $ \vert \Psi_{b} \rangle $ is the bound state wave function and $ { \vert \Psi_{i} \rangle }_{i=0} ^{N-1} $ is the set of bound states plus PSs\footnote{The set $ { \vert \Psi_{i} \rangle }_{i=0} ^{N-1} $ can be replaced by the corresponding set of basis states $ { \vert \Phi_{i} ^{BAS} \rangle }_{i=0} ^{N-1} $ }.

In the large N limit, due to the basis completeness
\begin{equation}
{\cal S}_T^{(b)} ({\cal O}) = \lim_{N\rightarrow\infty} {\cal S}_T ^{(b)}({\cal O}, N) = \langle \Psi_{b} \vert {\cal O}^{2} \vert \Psi_{b} \rangle .
\label{sum_rules_limit_1}
\end{equation}
The convergence of $ {\cal S}_T ^{(b)} ({\cal O}, N) $ as N increases to $ {\cal S}_T^{(b)} ({\cal O}) $ measures the completeness of the basis \cite{sum_rules}.

The Energy--Weighted Sum Rule is defined as
\begin{equation}
{\cal E}_W  ^{(b)} ({\cal O}, N) = \sum _{i=0} ^{N-1}  (E_i - E_b) \mid  \langle \Psi_{b} \vert {\cal O}  \vert \Psi_{i} \rangle \mid ^2 ,
\label{ew_sum_rule}
\end{equation}
where again $ \vert \Psi_{b} \rangle $ is a bound state and $ { \vert \Psi_{i} \rangle }_{i=0} ^{N-1} $ encompasses all eigenstates of the model Woods--Saxon potential for an N-dim basis.

In the case $ {\cal O} = x $
\begin{equation}
{\cal E}_W  ^{(b)} (x) = \lim_{N\rightarrow\infty} {\cal E} ^{(b)} (x, N) = \frac{1}{2} \frac{\hbar ^2}{\mu}
\label{sum_rules_limit_2}
\end{equation}
a demonstration of Eq.\ (\ref{sum_rules_limit_2}) can be found in App.\ \ref{app_ewsr}.

In Tab.\ \ref{tab_sum_rules} we present the basis dimension N required to obtain a ${\cal S}_T ^{(b)} (x, N)$ converged value.
In the BOX case the convergence of the total Strength $ {\cal S}_T ^{(b)} (x, N) $ is computed for two different box radii, $x_b = 45 fm$ and $ 55 fm$.
As expected, the convergence is slower for the least bound state, $ \vert \Psi_{2} \rangle $, and the use of the THO basis with $\gamma/b = 1.2 fm^{-1/2}$  improves the results. The $\gamma/b = 2.4 fm^{-1/2}$ basis results are close to the HO ones, improving them slightly.

Numerical results for HO and THO cases under consideration for $ {\cal E}_W  ^{(gs)}$ can be found in Tab.\ \ref{tab_ew_sum_rules}.
The convergence for  $ {\cal E} _W ^{(gs)} (x, N)$ is particularly fast .

\begin{table}[!htc]
\begin{center}
\begin{tabular}{c|c|c|c|c|c}
\hline
\hline 
 &  HO  & THO$^{(1)}$ & THO$^{(2)}$ & BOX ($45 fm$) & BOX ($55 fm$) \\
\hline
$ {\cal S}_T ^{(gs)} (x, N) = 0.978 fm^2 $ &  10 &  20  &   10 &  70 & 90 \\
$ {\cal S}_T ^{(1)} (x, N) = 3.19 fm^2 $ &  20 &  10  &   20 &  80 & 90 \\
$ {\cal S}_T ^{(2)} (x, N) = 34.4 fm^2 $ &  100 &  40  &  115 &  110 & 140 \\
\hline
\hline
\end{tabular} 
\caption{Basis dimension N at which ${\cal S}_T ^{(b)} (x, N)$ converges to constant value for the different methods. THO$^{(1)}$ indicates the case with $\gamma/b = 2.4 fm^{-1/2}$ and  THO$^{(2)}$ the case with $\gamma/b = 1.2 fm^{-1/2}$.\label{tab_sum_rules}}
\end{center}
\end{table}

\begin{table}[!htc]
\begin{center}
\begin{tabular}{c|c|c|c}
\hline
\hline 
N & HO & THO$^{(2)}$ & THO$^{(1)}$ \\
\hline
3 &   0.512 &   0.513  &    0.500  \\
4 &   0.494 &   0.514  &    -    \\
5 &   0.500 &   0.511  &    -  \\
6 &   - &   0.534  &    -  \\
7 &   - &   0.502  &    -   \\
\hline
\hline
\end{tabular} 
\caption{$ {\cal E}_W  ^{(gs)} (x, N)$ convergence for the HO and THO methods varying N. THO$^{(1)}$ indicates the case with $\gamma/b = 2.4 fm^{-1/2}$ and  THO$^{(2)}$ the case with $\gamma/b = 1.2 fm^{-1/2}$. $ {\cal E}_W  ^{(gs)} (x, N)$ result is reported in the $ \frac{\hbar ^2}{\mu} $ unities. \label{tab_ew_sum_rules}}
\end{center}
\end{table}

\newpage
\subsection{\label{1b_E1_E2}Dipole and Quadrupole transition intensities}
The low binding energy of weakly--bound systems affects their response to different probes.
In particular, its electromagnetic transition intensity patterns differ greatly from the intensities for systems far from the influence of the continuum.
We proceed to compute the electric dipole (E1) and quadrupole (E2) transition intensities for the bound states of the model Woods--Saxon potential.
The 1D multipole electric operator for an A--body system is defined as
\begin{equation}
{\cal M} (E \lambda) = e \sum _{i = 1} ^A Z_i x_i ^{\lambda}
\label{MElambda}
\end{equation}
where $e Z_i$ is the charge of the i-th component and $x_i$ the coordinate of the i-th body in the center of mass system.

The transition probability between bound states or between a bound state and a continuum pseudostate (see the discussion in \cite{thesis_lay}) can be written as
\begin{equation}
B(E \lambda)_{\Psi_i \rightarrow \Psi_f} = | \langle \Psi_{i} \vert {\cal M} (E \lambda) \vert \Psi_{f} \rangle | ^2.
\end{equation}
In our case, for E1 and E2 electric transitions, we need to compute the transition moment integrals
\begin{equation}
\langle \Psi_{b} \vert {\cal O} \vert \Psi_{i} \rangle = \int _{- \infty} ^{+\infty} dx 
\Psi_{b} ^{\ast} (x) {\cal O} \Psi_{i} (x),
\end{equation}
with ${\cal O} = x $ and $x^2$.
These integrals can be computed numerically.
Taking into consideration that the eigenstates are linear combination of the basis states
\begin{equation}
\Psi_j (x) = \sum _{k = 0} ^{N-1} \alpha_{jk} \Phi_k ^{BAS} (x), 
\end{equation}
with $BAS = HO,~THO,~BOX$, the matrix elements $ \langle \Phi_{k}^{BAS}  \vert {\cal O} \vert \Psi_{k'}^{BAS} \rangle $ can be computed analytically in the HO and BOX cases.
In the HO case the matrix elements can be easily computed using the relations
(\ref{abramrel}), though it is more direct to make use of the creation
and annihilation operator formalism. In any case
\begin{align}
 \langle i|x| j\rangle &= a^{-1}
 \left[\sqrt{\frac{j}{2}}\delta_{i,j-1} + \sqrt{\frac{j+1}{2}}\delta_{i,j+1} \right]~,\label{xhomatel}\\
 \langle i|x^2| j\rangle &= a^{-2}
 \left[ \frac{2j+1}{2}\delta_{i,j} +
   \frac{\sqrt{j(j-1)}}{2}\delta_{i,j-2} + 
   \frac{\sqrt{(j+1)(j+2)}}{2}\delta_{i,j+2}\right]~.  \label{x2homatel}
\end{align}

In the BOX case both integrals can be easily computed using the properties of
trigonometric functions and elementary integration techniques. In the
$x$ operator case the symmetry of the integrand only left as nonzero
cases those where the $j$ and $k$ state labels have different
parities
\begin{align}
 \langle j|x| k\rangle &= x_b^{-1} \int_{-x_b}^{x_b} dx\, x
 \cos{ \left( \frac{j \pi x}{2 x_b} \right) } \sin{ \left( \frac{k
       \pi x}{2 x_b} \right) } \\ 
 &= \frac{4x_b}{\pi^2}\left[\frac{(-1)^{\frac{k-j-1}{2}}}{(k-j)^2}+\frac{(-1)^{\frac{k+j-1}{2}}}{(k+j)^2}\right]\label{xBOXmatel}
\end{align}

Symmetry considerations in the calculation of matrix elements of the
$x^2$ operator imply that in the different parity case the result is
zero, while for the same parity case and $j\neq k$
\begin{equation}
 \langle j|x^2| k\rangle =
 \frac{8x_b^2}{\pi^2}\left[\frac{(-1)^{\frac{j-k}{2}}}{(j-k)^2} \pm \frac{(-1)^{\frac{k+j}{2}}}{(k+j)^2}\right]~,\label{x2BOXmatel}
\end{equation}
\noindent where the plus sign applies in the odd-odd case, the minus
sign in the even-even case. If $j=k$ the valid formula is
\begin{equation}
 \langle j|x^2| j\rangle =
 2x_b^2\left[\frac{1}{6} - \frac{1}{\pi^2j^2}\right]~.\label{x2BOXdiagmatel}
\end{equation}
As already mentioned,
the rest of possible cases are zero.\\
In the THO case results have been calculated integrating the wave functions, and just to have a comparison and a confirmation, also integrating the basis states.

\begin{figure}[!htc]
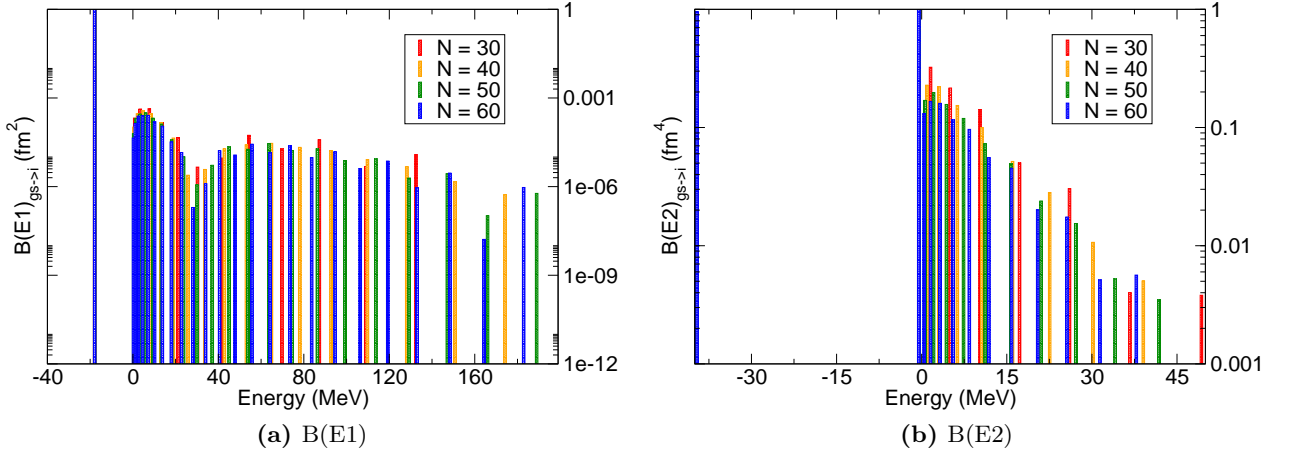

\begin{center}
\subfloat[B(E1)]{
\scalebox{0.32}{\includegraphics{1b_E1_HO_gs}}
}
\hspace*{0.1cm}
\subfloat[B(E2)]{
\scalebox{0.32}{\includegraphics{1b_E2_HO_gs}}
}
\caption{B(E1) and B(E2) results calculated starting from the ground state $\Psi_{gs}$  as a function of the levels energies for different basis dimension N in the HO case. 
\label{E1_E2_HO_gs}}
\end{center}
\end{figure}

\begin{figure}[!htc]
\begin{center}
\subfloat[B(E1)]{
\scalebox{0.32}{\includegraphics{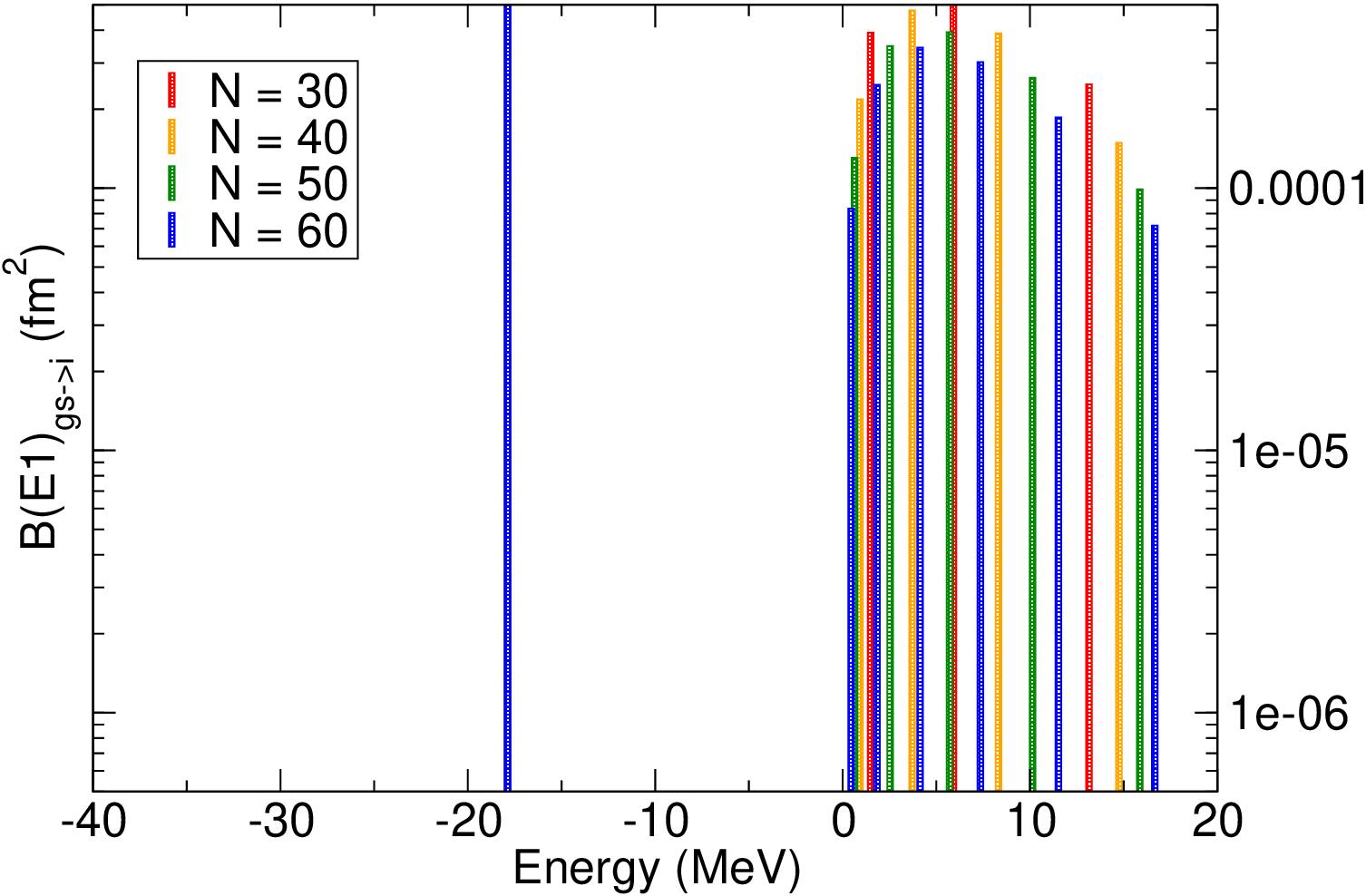}}
}
\hspace*{0.1cm}
\subfloat[B(E2)]{
\scalebox{0.32}{\includegraphics{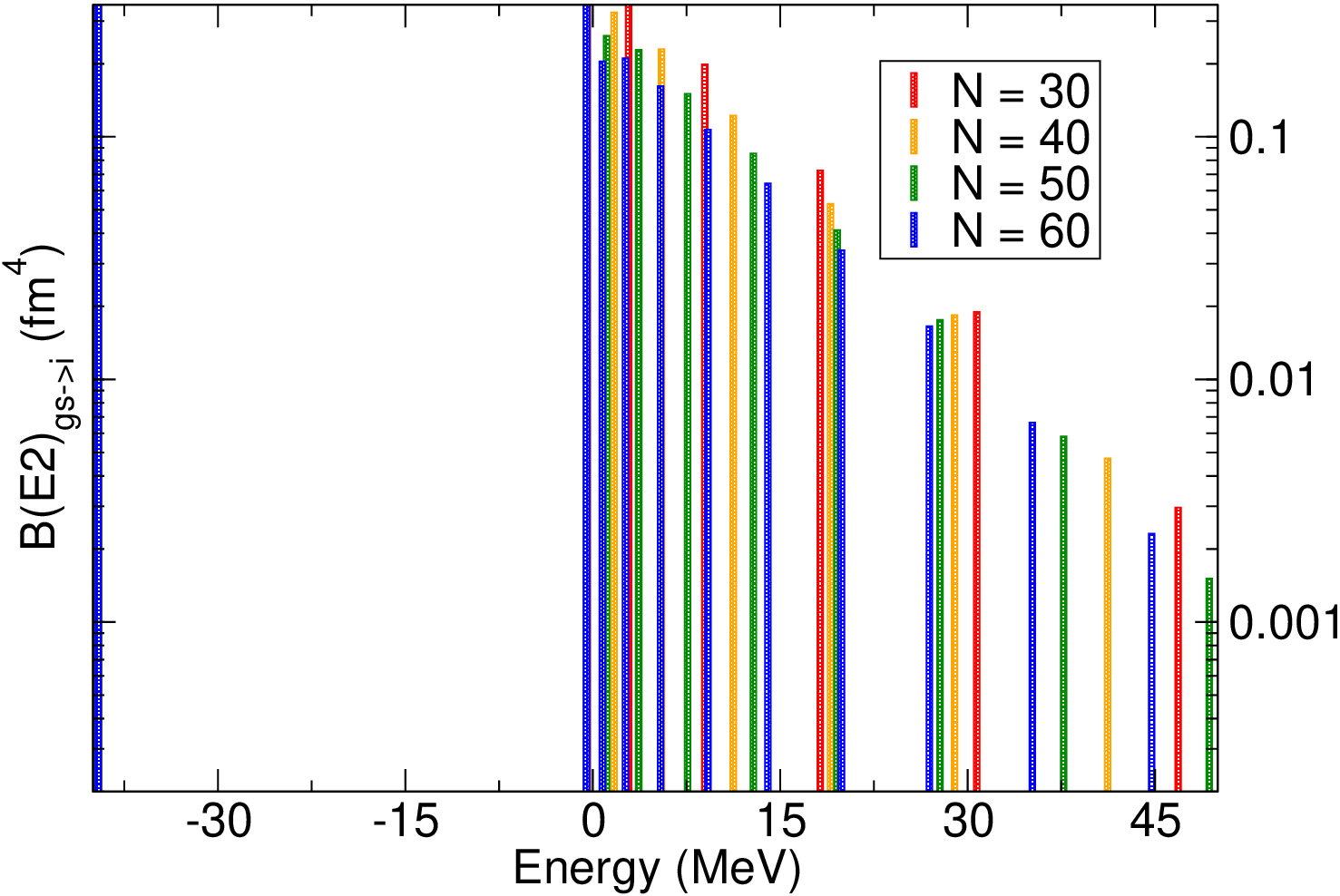}}
}
\caption{B(E1) and B(E2) results calculated starting from the ground state $\Psi_{gs}$  as a function of the levels energies for different basis dimension N in the THO case with $\gamma/b = 2.4 fm ^{-1/2}$. 
\label{E1_E2_THO_2c4_gs}}
\end{center}
\end{figure}

\begin{figure}[!htc]
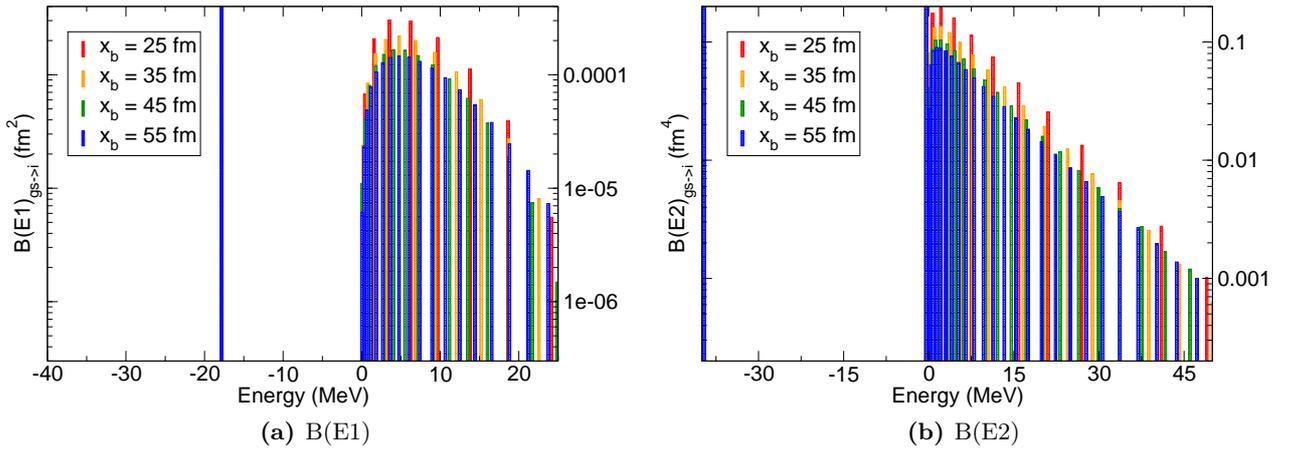

\begin{center}
\subfloat[B(E1)]{
\scalebox{0.32}{\includegraphics{1b_E1_ISQW_gs}}
}
\hspace*{0.1cm}
\subfloat[B(E2)]{
\scalebox{0.32}{\includegraphics{1b_E2_ISQW_gs}}
}
\caption{B(E1) and B(E2) results calculated starting from the ground state $\Psi_{gs}$  as a function of the levels energies for different box radii $x_b$ in the BOX case. 
\label{E1_E2_BOX_gs}}
\end{center}
\end{figure}

\begin{figure}[!htc]
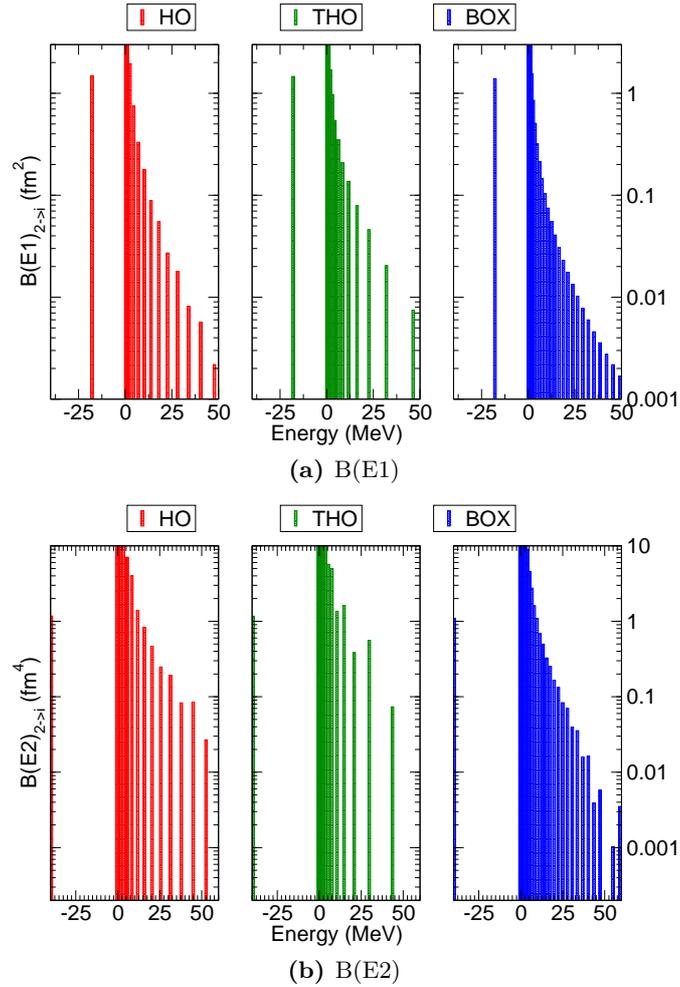

\begin{center}
\subfloat[B(E1)]{
\scalebox{0.32}{\includegraphics{1b_E1_from_2}}
}
\vspace*{0.2cm}
\subfloat[B(E2)]{
\scalebox{0.32}{\includegraphics{1b_E2_from_2}}
}
\caption{B(E1) and B(E2) results calculated starting from the third bound state $\Psi_{2}$  in HO, THO ($\gamma/b = 1.2 fm^{-1/2}$), and BOX cases.
\label{E1_E2_from_2}}
\end{center}
\end{figure}

We plot results for the electric dipole moment calculated from the ground state $\Psi_{gs}$ in Figs.\ \ref{E1_E2_HO_gs}a, \ref{E1_E2_THO_2c4_gs}a, \ref{E1_E2_THO_1c2_gs}a and \ref{E1_E2_BOX_gs}a for HO, THO ($\gamma/b = 2.4 fm^{-1/2}$ and $\gamma/b = 1.2 fm^{-1/2}$), and BOX cases respectively.
The electric quadrupole results calculated from the ground state $\Psi_{gs}$ are depicted in Figs.\ \ref{E1_E2_HO_gs}b, \ref{E1_E2_THO_2c4_gs}b, \ref{E1_E2_THO_1c2_gs}b and \ref{E1_E2_BOX_gs}b for the same cases as for the dipole moment.
The B(E1) and B(E2) results for HO, THO ($\gamma/b = 1.2 fm^{-1/2}$), and BOX cases are shown in Fig.\ \ref{E1_E2_from_1} for the calculations starting from the second bound state $\Psi_1$ and in Fig.\ \ref{E1_E2_from_2} for the calculations starting from the third and weakly--bound state $\Psi_2$.

The Total E1 value
\begin{equation}
\sum_{i} \mid \langle \Psi_{gs} \vert x \vert \Psi_{i} \rangle \mid ^{2}
\end{equation}
converges very fast to the value $0.978 fm^2$ with each method, in particular we obtain this result setting $N = 15$ for both the HO and THO methods and $x_b = 35 fm$ for BOX method. 
The convergence of the Total E2 transition intensities
\begin{equation}
\sum_{i} \mid \langle \Psi_{gs} \vert x^{2} \vert \Psi_{i} \rangle \mid ^{2} 
\end{equation}
as function of the basis dimension N for HO and THO and as a function of the box radius $x_b$ for the BOX method is shown in Tab.\ \ref{tab_E2}. 

\begin{table}[!htc]
\begin{center}\begin{tabular}{c|c|c|c||c|c}
\hline
\hline 
N & HO & THO ($r = 2.4 fm^{-1/2}$) & THO  ($r = 1.2 fm^{-1/2}$) & $x_b (fm)$ & BOX \\
15 & 2.96 & 2.95 & 3.18   & 35 & 2.95 \\
20 & 2.96 & - & 3.06	 & 45 & 2.96 \\
30 & 2.95 & - & 3.00	 & 55 & 2.95 \\
40 &    - &  - & 2.95	 & 65 & - \\
\hline
\hline
\end{tabular} 
\caption{Total E2 ($fm^4)$ convergence increasing the basis dimension N for HO and THO methods and increasing the box radius $x_b$ for the BOX method. \label{tab_E2}}
\end{center}
\end{table}

\begin{figure}[!htc]
\begin{center}
\subfloat[B(E1)]{
\scalebox{0.32}{\includegraphics{1b_E1_THO_1c2_gs}}
}
\hspace*{0.1cm}
\subfloat[B(E2)]{
\scalebox{0.32}{\includegraphics{1b_E2_THO_1c2_gs}}
}
\caption{B(E1) and B(E2) results calculated starting from the ground state $\Psi_{gs}$  as a function of the levels energies for different basis dimension N in the THO case with $\gamma/b = 1.2 fm ^{-1/2}$. 
\label{E1_E2_THO_1c2_gs}}
\end{center}
\end{figure}
\begin{figure}[!htc]
\begin{center}
\subfloat[B(E1)]{
\scalebox{0.32}{\includegraphics{1b_E1_from_1}}
}
\vspace*{0.2cm}
\subfloat[B(E2)]{
\scalebox{0.32}{\includegraphics{1b_E2_from_1}}
}
\caption{B(E1) and B(E2) results calculated starting from the second bound state $\Psi_{1}$ in HO, THO ($\gamma/b = 1.2 fm^{-1/2}$), and BOX cases.
\label{E1_E2_from_1}}
\end{center}
\end{figure}
\subsection{Resonances}

Some of the most striking effects in scattering theory are associated with resonances.
The appearance of a resonance in the continuum of a quantum system implies sudden changes of phase shifts and reflection/transmission coefficients at particular energy values, associated with the resonance.
This implies the occurrence of a "quasibound" state of the scattering potential, that is a long--lived state
which has sufficient energy to break--up into two  or more subsystems.

The usual approach to resonances implies the study of complex poles of the scattering matrix and complex eigenvalues of the system Hamiltonian $\epsilon = E - i \Gamma / 2 $ (Gamow--Siegert functions \cite{gamow}).
A standard approach to study resonances implies complex scaling, defining a complex scaled Hamiltonian and associating the resonance to a single square--integrable eigenfunction of this scaled Hamiltonian \cite{moise}.

Simpler alternatives to study resonances are quantum--mechanical stabilization calculations, using bound--state techniques \cite{Hol, Thompson, hazi_taylor}.
This approach is based on the fact that the resonance wave function in the potential region is akin to a bound state wave function.
When the energies of a system possessing resonances are plot as a function of the basis dimension (or other basis parameter) there are certain eigenvalues that vary very little compared to the rest: they mark the resonant state energies and thus the name stabilization plot \cite{lippman}.

Resonances are pervasive in 2D and 3D systems due to centrifugal barrier effects.
In order to study resonances using PS in a 1D system we consider the Hazi and Taylor Potential \cite{hazi_taylor} to reproduce their results. 
We also address an analytical case, as the Ginocchio Potential, which has been previously studied with THO states in 3D \cite{gin_res_manoli}, 
and as a third example we define a Woods--Saxon with barriers.

We show stabilization plots for each of the cases as a function of the relevant parameters and compute the resonance energy from the stabilization plots.
We also plot the resonant wave functions.

\begin{figure}[!htc]
\begin{center}
\scalebox{0.42}{\includegraphics{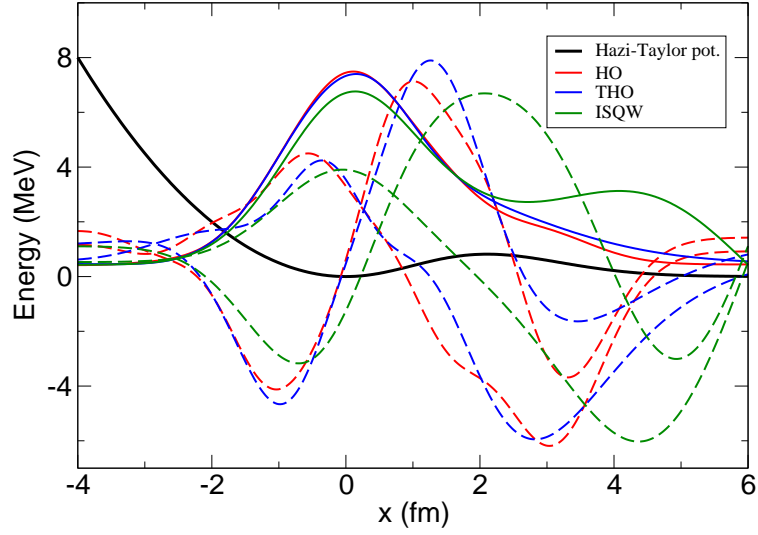}}
\caption{The Hazi--Taylor potentials with its first three wave functions calculated with each method, in solid line the resonant waves.\label{fig_ht_pot}}
\end{center}
\end{figure}

\begin{figure}[!htc]
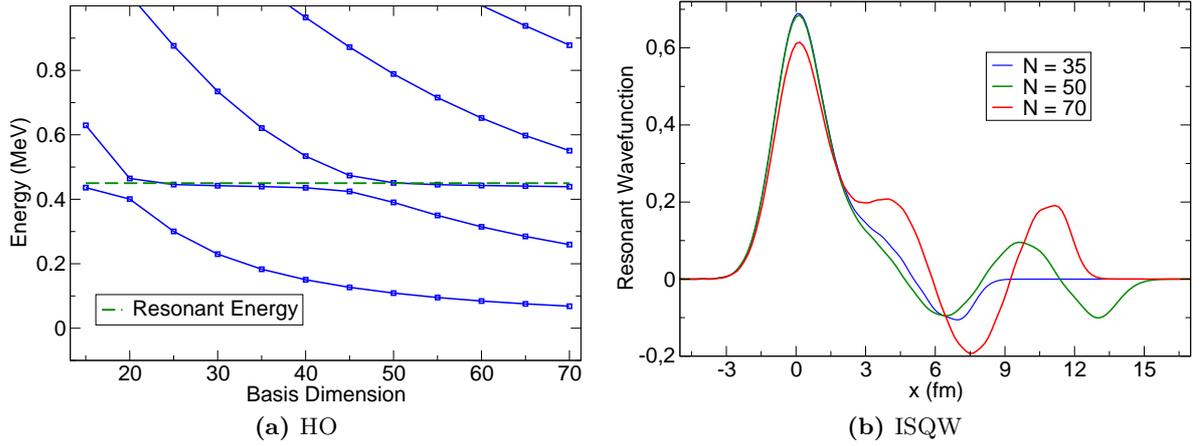

\begin{center}
\subfloat[HO]{
\scalebox{0.32}{\includegraphics{ht_ho_stabplot}}
}
\hspace*{0.1cm}
\subfloat[ISQW]{
\scalebox{0.32}{\includegraphics{ht_ho_resonant_wf}}
}
\caption{In the left panel: stabilization plot in the HO case as a function of N for the Hazi--Taylor potential. The green dashed line corresponds to the resonant energy. In the right panel: the resonance wave function for different basis dimension values. \label{fig_ht_ho_isqw_stabplot}}
\end{center}
\end{figure}

\begin{figure}[!htc]
\begin{center}
\scalebox{0.42}{\includegraphics{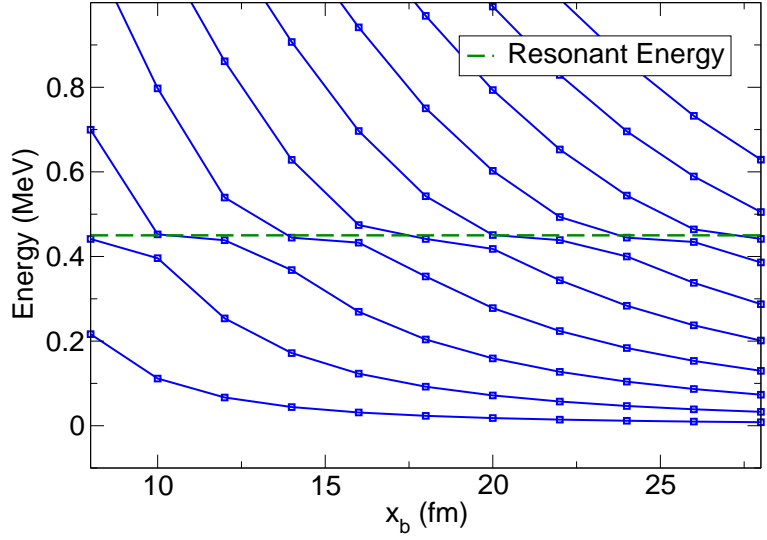}}
\caption{Stabilization plot in the BOX case as a function of the box radius $x_b$ for the Hazi--Taylor potential. The green dashed line corresponds to the resonant energy.\label{fig_ht_isqw}}
\end{center}
\end{figure}

\begin{figure}[!htc]
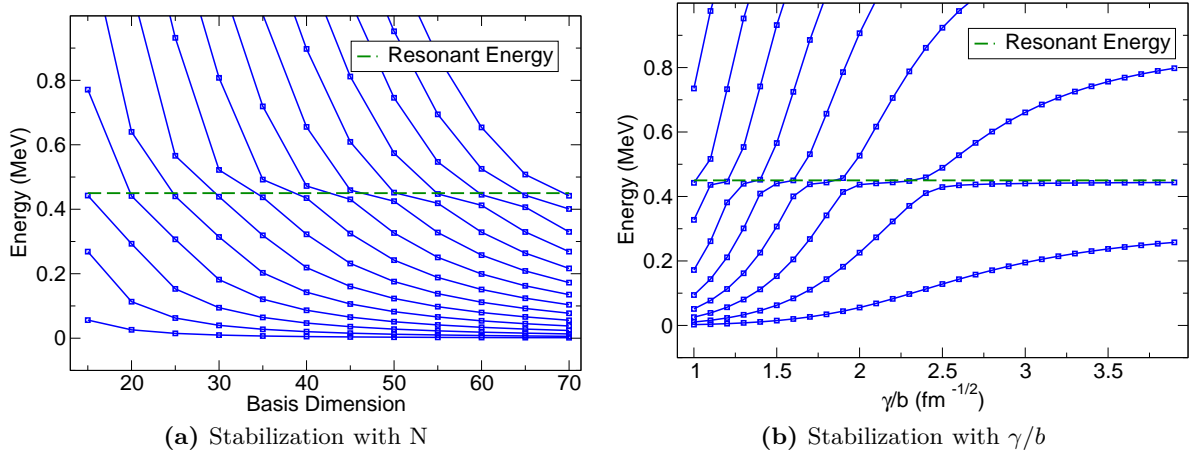

\begin{center}
\subfloat[Stabilization with N]{
\scalebox{0.32}{\includegraphics{ht_tho_stabplot_withN}}
}
\hspace*{0.1cm}
\subfloat[Stabilization with $\gamma /b$]{
\scalebox{0.32}{\includegraphics{ht_tho_stabplot_withRATIO}}
}
\caption{Stabilization plots in the THO case (in panel a as a function of N and in panel b as a function of $\gamma /b$) for the Hazi--Taylor potential. The green dashed line corresponds to the resonant energy.\label{fig_ht_tho_stabplot}}
\end{center}
\end{figure}

\begin{figure}[!htc]
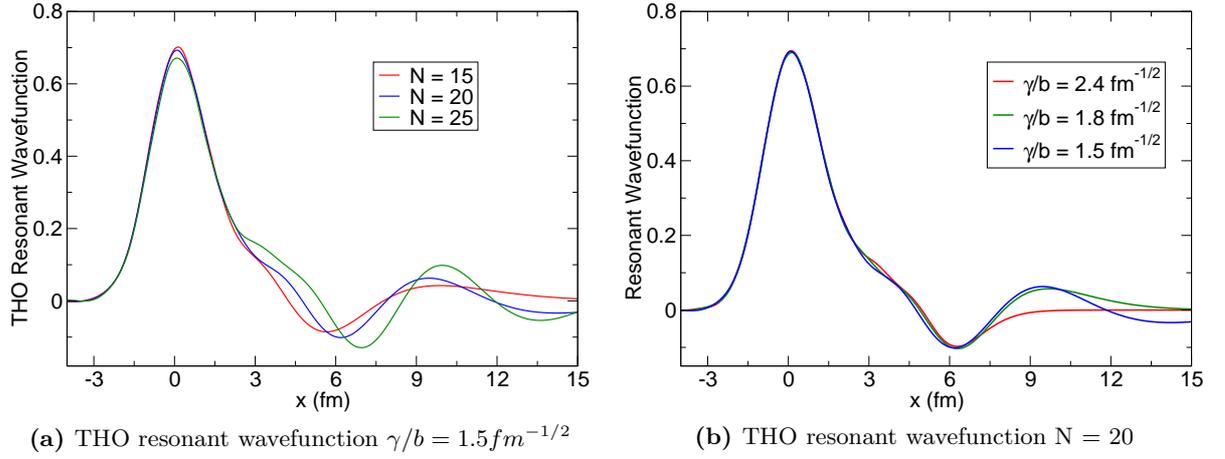

\begin{center}
\subfloat[THO resonant wavefunction $\gamma /b = 1.5 fm^{-1/2}$]{
\scalebox{0.32}{\includegraphics{ht_tho_resonant_wf_with_N}}
}
\hspace*{0.1cm}
\subfloat[THO resonant wavefunction N = 20]{
\scalebox{0.32}{\includegraphics{ht_tho_resonant_wf_with_ratio}}
}
\caption{Resonance wave function for the Hazi--Taylor potential in the THO case varying N (panel a) and varying $\gamma /b$ (panel b).\label{fig_ht_resonant_wf}}
\end{center}
\end{figure}

\paragraph{Hazi--Taylor potential resonances}
We have first reproduced  the Hazi--Taylor potential resonances \cite{hazi_taylor}.\\
The Hazi and Taylor potential is
\begin{equation}
V(x) =\bigg\{ \begin{array}{lr}
\frac{1}{2} x^{2}  & x  \leq  0\\
 \frac{1}{2} x^{2} e^{-\lambda x^{2}}  & x  \geq  0\\
\end{array}
\label{hazitaylor}
\end{equation}
where $ \lambda > 0$.
Using Ref.\ \cite{hazi_taylor} parameters and the HO method, energies and wave functions have been perfectly reproduced. 
In Figure \ref{fig_ht_pot} we plot the potential and its first wave functions.
The stabilization plot for the HO case as a function of the basis size is shown in Fig.\ \ref{fig_ht_ho_isqw_stabplot} (left panel),
while in the right panel the resonance wave function in shown for different N values.
The THO case results are shown in Fig.\ \ref{fig_ht_tho_stabplot} with stabilization plots as a function of the basis dimension (left panel) and $\gamma /b$ ratio (right panel).
The resonance wave function obtained with the THO basis is plot in Fig.\ \ref{fig_ht_resonant_wf} (in left panel varying N, in the right panel varying $\gamma /b$).
The stabilization plot with the box radius $x_b$ is shown in Fig.\ \ref{fig_ht_isqw} in the BOX case.
\vspace{0.3cm}
\vspace{0.3cm}
\vspace{0.3cm}
\vspace{2cm}

\begin{figure}[!htc]
\begin{center}
\scalebox{0.32}{\includegraphics{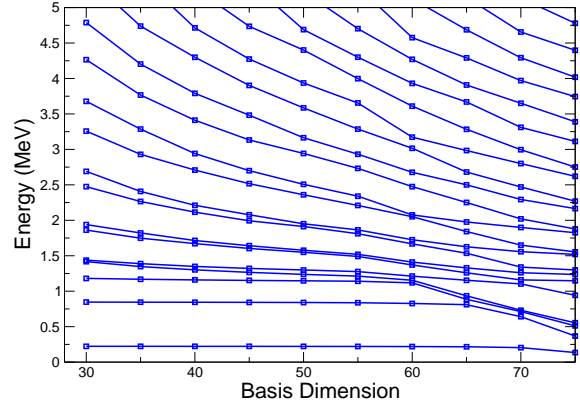}}
\caption{The stabilization plot in the HO case as a function of the basis dimension N for the Ginocchio potential.\label{fig_ho_gin}}
\end{center}
\end{figure}

\begin{figure}[!htc]
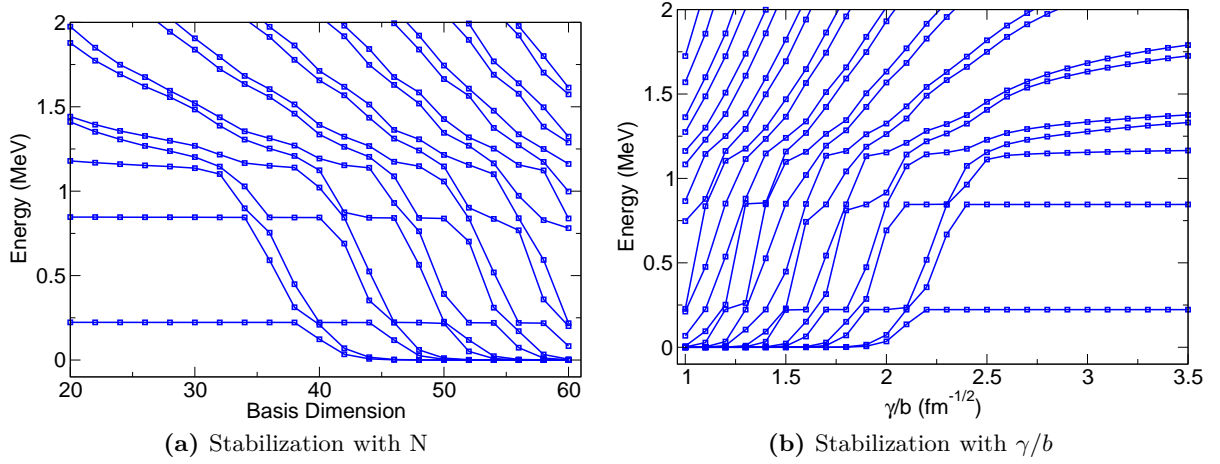

\begin{center}
\subfloat[Stabilization with N]{
\scalebox{0.32}{\includegraphics{gin_tho_stabplot_withN_zoom}}
}
\hspace*{0.1cm}
\subfloat[Stabilization with $\gamma /b$]{
\scalebox{0.32}{\includegraphics{gin_tho_stabplot_withRATIO}}
}
\caption{Stabilization plots in the THO case for the Ginocchio potential as a function of N (with the ratio fixed at 2.6 $ fm^{-1/2} $ ) in the right panel, and of the ratio (with N fixed at 30) in the left panel.\label{fig_tho_gin}}
\end{center}
\end{figure}

\newpage
\paragraph{Ginocchio potential resonances} 
Resonances for the Ginocchio potential \cite{ginocchio} have been reproduced according to \cite{gin_res_manoli} using HO and THO methods.\\
The Ginocchio potential is
\begin{equation}
\frac{V(x)}{V_0} = - \lambda ^2 \nu(\nu + 1) (1 - y^2) + \left(\frac{1-\lambda^2}{4}\right) (1 - y^2) [2 - (7-\lambda^2) y^2 + 5 (1-\lambda^2) y^4]
\label{ginocchio_pot_1}
\end{equation}
where
\begin{equation}
x(y) = \frac{1}{\lambda^2}{ arctanh(y) + [\lambda^2 -1]^{1/2} arctan( [\lambda^2 -1]^{1/2} y) }
\label{ginocchio_pot_2}
\end{equation}
the parameter $ \lambda $ regulates the wideness of the potential, $ \nu $ the the number of bound states.\\
We report the stabilization plots or both HO and THO cases in Figs.\ \ref{fig_ho_gin} and \ref{fig_tho_gin}. 
In particular in the THO case we plot the energy as a function of N (with  $\gamma /b$ fixed at 2.6 $ fm^{-1/2} $) or of the $\gamma /b$ (with N fixed at 30).\\


\newpage
\paragraph{Woods--Saxon potential with barriers resonances}
The Woods--Saxon with barriers potential is defined as
\begin{equation}
V(x) =  V_{WS}(x) + V_{1}e^{-(\mid x \mid - x_{0})^{2}}
\label{wsaxon_barriers}
\end{equation}
where $ V_{WS} $ is the standard potential of Eq.\ (\ref{wsaxon}). 
The parameters chosen are\\
\begin{subequations}
\begin{align}
V_1 &= 30.00\, {\mbox MeV}\\
x_{0} &= 4.00\,{\mbox fm}
\end{align}
\label{ws_bar_potpar}
\end{subequations}

\begin{figure}[!htb]
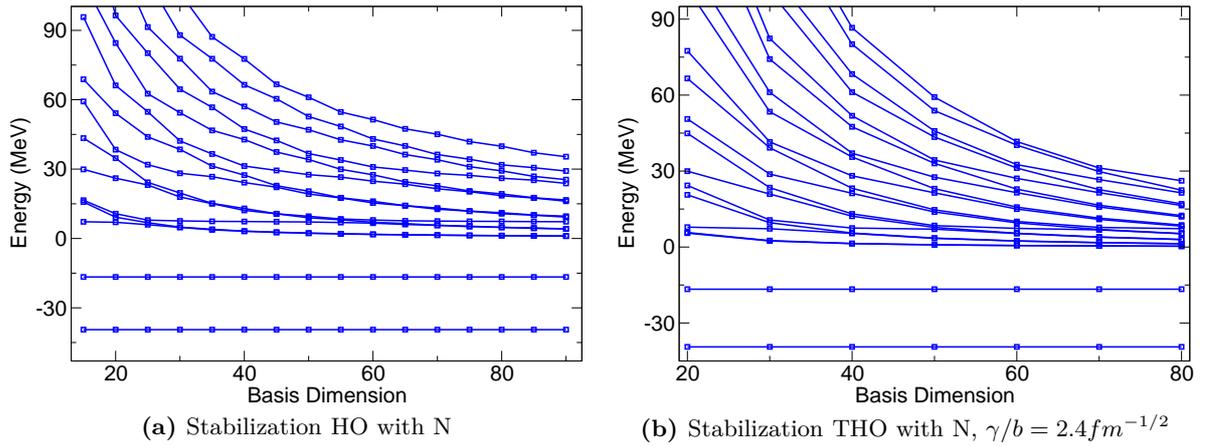

\begin{center}
\subfloat[Stabilization HO with N]{
\scalebox{0.32}{\includegraphics{ws_barriers_stab_plot}}
}
\hspace*{0.1cm}
\subfloat[Stabilization THO with N, $ \gamma /b = 2.4 fm^{-1/2} $]{
\includegraphics[scale=0.32]{ws_barriers_stab_plot_tho}
}
\caption{Left panel: stabilization plot in the HO case ad a function of the basis dimension for the Woods-Saxon potential with barriers. Right panel: stabilization plot in the THO case as a function of the basis dimension fixing $ \gamma /b = 2.4 fm^{-1/2} $ for the Woods-Saxon potential with barriers.\label{ws_barr_stab_plot}}
\end{center}
\end{figure}

\begin{figure}[!htb]
\centering
\includegraphics[scale=0.4]{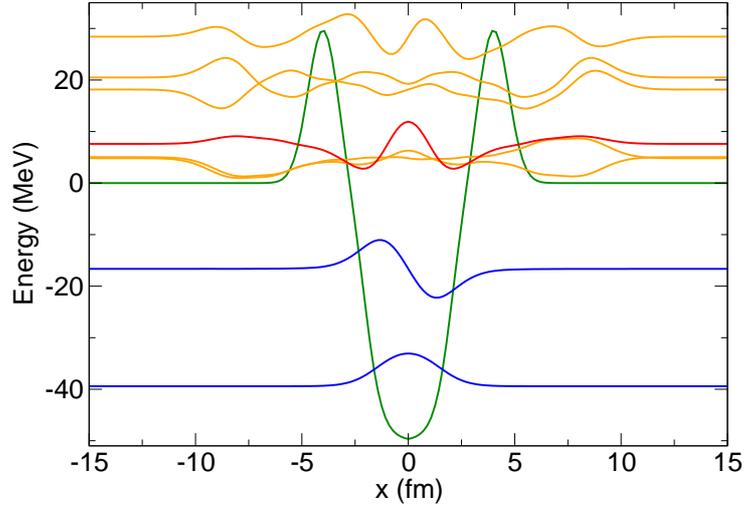}
\caption{The Woods--Saxon with barrier potential and the first wave functions calculated with the HO method. 
In red the resonant wave function; its probability density is higher into the potential.\label{ws_barr_wf}}
\end{figure}

This potential has resonances, we show them in the stabilization plots of Figure \ref{ws_barr_stab_plot} for the HO and THO ($ \gamma /b = 2.4 fm^{-1/2} $) cases.\\
In Figure \ref{ws_barr_wf} it can be appreciate the first resonant wave function (red), note that it is very narrow and concentrated into the potential range.

\vspace{1cm}

%
\subsubsection{Phase Shifts}

Resonances are characterized by sudden variation of the wave function phase shift, the slope of the variation being larger for narrow resonances.\\

The PSs obtained after the Hamiltonian matrix diagonalization, as will be shown in next section, are similar to the true continuum wave functions in the vicinities of the potential. The calculation of the phase shift $\eta$ from this normalized PSs, with zero asymptotic value, can be done using integral techniques. We make use of a formula derived by Schwinger \cite{Sch} and succerfully applied to the HO case in Ref.\ \cite{hazi_taylor}
\begin{equation}
\tan{\eta} = - \dfrac{\int_0 ^{\infty} \Psi^* _E [E - {\cal H}(x)]~\textit{f}(x)~\sin{kx}~dx }{\int_0 ^{\infty} \Psi^* _E [E - {\cal H}(x)]~\textit{f}(x)~\cos{kx}~dx}
\label{eq_13_ht}
\end{equation}
The function $ \textit{f} (x)  $ must satisfy the conditions
\begin{subequations}
\begin{align}
\lim_{x \rightarrow \infty} \textit{f} (x) = 1,\\
\textit{f} (0) = 0, \\
\left( \frac{d \textit{f}}{dx} \right)_{x=0} = 0,
\end{align}
\label{ws_bar_potpar}
\end{subequations}
but it is otherwise arbitrary. Following Ref.\ \cite{hazi_taylor} we used $ \textit{f} (x) = 1 - e^{- \lambda x ^2} $, chosing a $\lambda = 0.1 fm^{-2}$.
As in the previous section, we have first carried out calculation with the Hazi and Taylor potential in order to check the results with those in \cite{hazi_taylor}; the agreement is perfect for each choice of the initial parameters: the energies agree until the sixth decimal place and the phase shifts until the fourth decimal place.\\

\begin{figure}[!htc]
\begin{center}
\scalebox{0.4}{\includegraphics{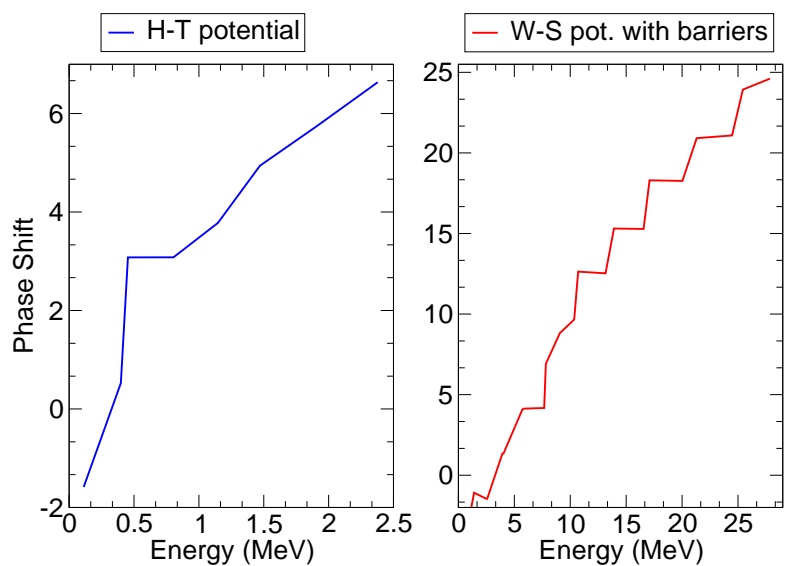}}
\caption{Phase shift as a function of the energy for the Hazy--Taylor and Woods--Saxon with barriers potentials. Calculations have been made with the HO basis using $N = 50$ and $N = 40$ basis states respectively.\label{phase_sh}}
\end{center}
\end{figure}

In Figure \ref{phase_sh} we present the phase shift as a function of the energy for the Hazi--Taylor and Woods--Saxon with barriers potentials. For the Hazi--Taylor potential a very steep increase can be noted around 0.4 MeV, which corresponds  to the resonance presented in Fig.\ \ref{fig_ht_resonant_wf}. In the case of Woods--Saxon potential with  barriers, one can note many evident steps, the first, around 7 MeV, corresponds to the resonant states shown in Fig.\ \ref{ws_barr_wf}.

%
%
%
\section{Comparison with other approaches}

\begin{figure}[!htc]
\begin{center}
\scalebox{0.32}{\includegraphics{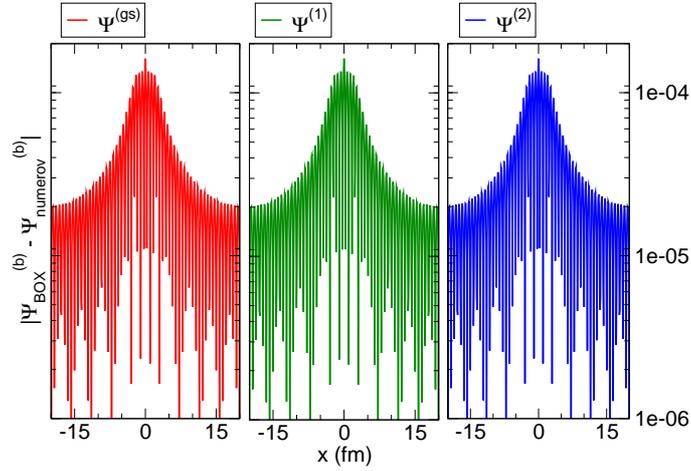}}
\caption{In these graphics are plotted the absolute value of the differences between the negative energy wave functions calculated with the Numerov and the ISQW methods.\label{fig_diff_numerical_method}}
\end{center}
\end{figure}

In the present section, for the sake of completeness, we compare some of the results presented in the one--body case with the results of alternative approaches.\\
We first compare the solution of the box case obtained diagonalizing the model Woods--Saxon potential in an ISQW basis with the results of a numerical integration of the 1D TISE
\begin{equation}
\left[ -\frac{\hbar^2}{2\mu} \frac{d^2}{dx^2} + V(x) \right] \psi_E(x) = E \psi_E(x)
\label{eq_avl}
\end{equation}
using a Numerov approach \cite{numerov}.
In Fig.\ \ref{fig_diff_numerical_method} we show the difference of the three bound states obtained with the two alterantive: Numerov and BOX cases. 
It is evident that the differences are comparable; the oscillations probably depend on the number of basis states in the ISQW, and so on the truncation of the wave functions.\\

We then proceed to compute, also using a Numerov approach, the true continuum solutions of the problem.
We obtain pairs of degenerate solutions with left-- and right--plane wave incoming conditions and build symmetric and antisymmetric combinations as detailed in App.\ \ref{appendix_binning}.\\

\begin{figure}[!htc]
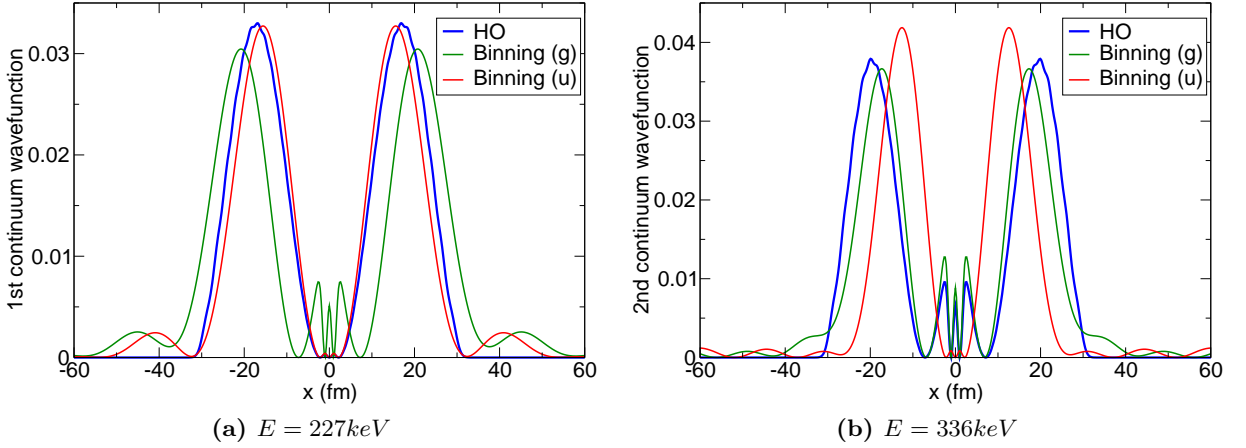

\begin{center}
\subfloat[$E = 227 keV$]{
\scalebox{0.32}{\includegraphics{1D_1body_cfr_binVSho_1st_cont}}
}
\hspace*{0.1cm}
\subfloat[$E = 336 keV$]{
\scalebox{0.32}{\includegraphics{1D_1body_cfr_binVSho_2nd_cont}}
}
\caption{Comparison between the probability density of the PSs calculated with the HO basis (N = 70) and the probability density  of the average functions obtained for the binning (both gerade and ungerade solutions).\label{fig_bin}}
\end{center}
\end{figure}

The standard procedure to discretize the continuum in continuum discretized coupled channel (CDCC) calculations is the average or binning method.
The method consists of averaging the true continuum wave functions in non--overlapping momentum intervals.
The average function does not oscillate for large coordinate values and it is square normalizable.
In case that the bin is defined for momenta $\{k_p \} _{p=1}^{N_b} $ the p-th function is 
\begin{equation}
\varphi _{bin, k_p} ^{\Gamma} (x) = \frac{1}{N_p} \int_{k_{p-1}} ^{k_p} \phi_k ^{\Gamma} (x) dk
\label{binning}
\end{equation}
where $ \Gamma = g,u $ and $N_p$ is a normalization constant such that the set $ \{ \varphi _{bin, k_p} ^{\Gamma} (x) \} _{p= 1} ^{N_b} $ is an orthonormal set.
We compute in Fig.\ \ref{fig_bin} the probability density $ \vert \Psi(x) \vert^2 $ with $E = 227 keV$ and $E = 336 keV$ obtained after diagonalization of the Woods--Saxon model potential with a HO basis having $N = 60$.
Thick blue lines are $ \vert \Psi(x) \vert^2 $ while red (green) lines are the probability density for the ungerade (gerade) symmetry average continuum function. The average functions $ \vert \varphi _{bin, k_p} ^{\Gamma} (x) \vert^2 $ in Fig.\ \ref{fig_bin} were obtained using a momentum interval centered at the momentum associated with the corresponding pseudostate energy. From Fig.\ \ref{fig_bin} it is clear the match between $ \vert \Psi(x) \vert^2 $ and the bin continuum function with the right symmetry.\\
\vspace{0.5cm}

Following Ref.\ \cite{tho_lst_3} we define the quantity
\begin{equation}
\rho ^{(BAS)} (k) = \sum_{n = 1} ^N  \rho_n ^{(BAS)} (k)  = \sum_{n = 1} ^N \left[ \langle \phi_k ^g \vert \psi_n ^{(BAS)} \rangle + \langle \phi_k ^u \vert \psi_n ^{(BAS)} \rangle \right]
\label{rho}
\end{equation}
where $BAS$ stands for HO, THO, BOX, and N is the total number of pseudostates included in the continuum discretizations.\\

\begin{figure}[!htc]
\begin{center}
\scalebox{0.32}{\includegraphics{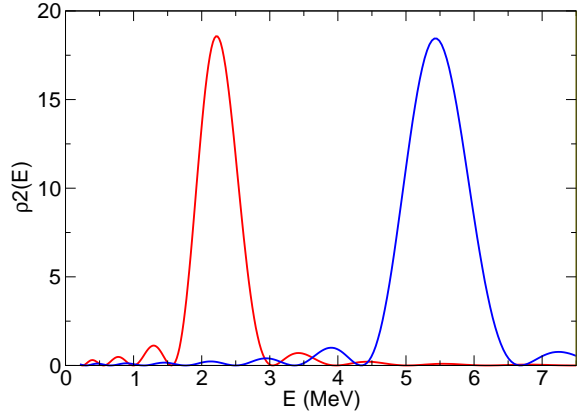}}
\caption{$ \rho 2 ^{(ISQW)} (E)  $ for $n = 11$ and $18$. It represents the probability to find the continuum wave function at a precise energy.\label{fig_rho}}
\end{center}
\end{figure}

\begin{figure}[!htc]
\begin{center}
\scalebox{0.32}{\includegraphics{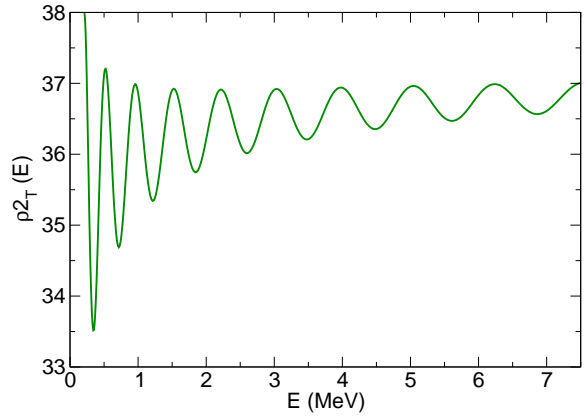}}
\caption{The total continuum contribution $ \rho2_T  ^{(ISQW)} (k) $.\label{fig_rho_tot}}
\end{center}
\end{figure}

In general $ \rho_n ^{(BAS)} (k) $ is the overlap between continuum wave functions and pseudostates and it is a complex quantity.
We can assess the contributions of the continuum wave functions to our pseudostates with the function
\begin{equation}
\rho 2 ^{(BAS)} (k) = \sum_{n = 1} ^N  \vert \rho_n ^{(BAS)} (k)  \vert ^2.
\label{rho2}
\end{equation}
We plot in Fig.\ \ref{fig_rho} $ \vert \rho_n ^{(BAS)} (k) \vert ^2$ for $BAS = $ ISQW and $n = 11$ and $18$ as a function of k.
It is shown how the maximum of these functions corresponds to energies equal or around the eigenenergy of the pseudostate.
The total continuum contribution $ \rho2_T  ^{(BAS)} (k)  $ is plotted in Fig.\ \ref{fig_rho_tot}.\\

\vspace{0.5cm}
\vspace{0.5cm}

In order to compare the electric dipole and quadrupole intensities, B(E1) and B(E2), obtained in Sec.\ \ref{1b_results} with values computed with the continuum eigenfunctions we need to perform some intermediate steps.

The E$\lambda$ transition between bound states and pseudostates is computed as 
\begin{equation}
B^{(N)} (E\lambda, i\rightarrow f) = \vert \langle \psi_f \vert {\cal M}(E\lambda) \vert \psi_i \rangle \vert ^2,
\label{B_E_lambda}
\end{equation}
while if the transition takes place between a bound state and the continuum the relevant quantity is
\begin{equation}
\dfrac{d B(E\lambda, , i\rightarrow k)}{d\varepsilon} = \frac{\mu k}{(2\pi)^3 \hbar^2} \left[ \vert \langle \phi_k ^g \vert {\cal M}(E\lambda) \vert \psi_i \rangle \vert ^2 + \vert \langle \phi_k ^u \vert {\cal M}(E\lambda) \vert \psi_i \rangle \vert ^2 \right].
\label{B_E_lambda_der}
\end{equation}
A simple way to compute the results of $ \dfrac{d B(E\lambda)}{d\varepsilon} $ from Eq.\ (\ref{B_E_lambda}) is as follows
\begin{equation}
\left( \dfrac{d B(E\lambda, , i\rightarrow k_n)}{d\varepsilon} \right)_{\varepsilon = E_n} = \frac{1}{\Delta n} B^{(N)} (E\lambda, i\rightarrow n) 
\label{B_E_lambda_der_bis}
\end{equation}
where $\Delta n = (E_{n+1} - E{n-1})/2$ and $k_n$ and $E_n$ are the momentum and energy of the n-th pseudostate.
We can also compute an approximation to (\ref{B_E_lambda_der}) folding our pseudostate results with the continuum wave functions.
Making use of the closure relation for the pseudostates and of Eq.\ (\ref{B_E_lambda_der_bis}) we obtain
\begin{equation}
\dfrac{d B ^{(N)} (E\lambda, , i\rightarrow k)}{d\varepsilon} = \frac{\mu k}{(2\pi)^3 \hbar^2}  \vert \sum_{j = 1} ^N \rho_j  ^{(BAS)} (k) \langle \psi_j \vert {\cal M}(E\lambda) \vert \psi_k \rangle \vert ^2.
\label{B_E_lambda_der_fold}
\end{equation}

\begin{figure}[!htc]
\begin{center}
\scalebox{0.32}{\includegraphics{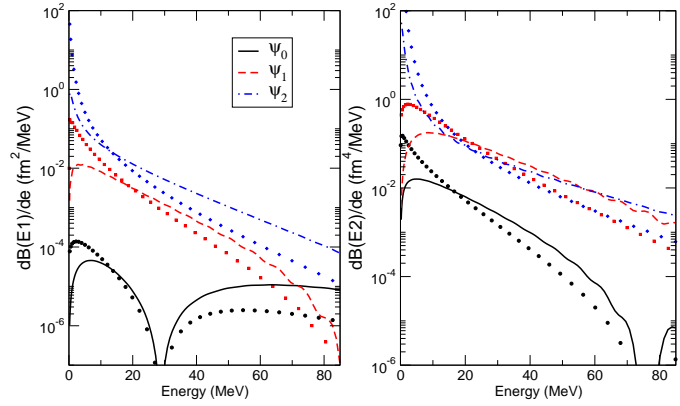}}
\caption{Folded transition intensities for E1 (panel a) and E2 (panel b) for the three bound states.\label{fig_bin_comp}}
\end{center}
\end{figure}

Preliminary results for this folded transition  intensities for E1 and E2 are depicted in Fig.\ \ref{fig_bin_comp}.
\vspace{0.5cm}
%

%
\chapter{Two--body problem\label{2b_probl}}
%
        \newpage

        \null 

        \thispagestyle{empty} 

        \newpage
Once that we have investigated different approaches to the continuum discretization in a one--body problem we proceed to a more complicate, and physically more enlightening, two--body system.
This problem consists of a fully occupied Woods--Saxon core plus two particles outside the core interacting via a matter density--dependent point contact residual interaction.\\

As explained in the introduction of this memory, our goal with this choice is to model a simplified (1D) Borromean nucleus, a halo nucleus with a two--particle halo which is a bound system notwithstanding the core plus one particle system is unbound (e.\ g.\ $^{11} Li$, $^6 He$).\\

This simplified model has already been presented and tested in some situations \cite{Vitturi, Hanigo, vitturi_1D_2b}.

In this chapter we present the two--body model in Sec.\ \ref{2b_model}, show results for the system eigenvalues and eigenvectors in Sec.\ \ref{2b_eigval_eigvec}, and compute some quantities of physical interest in Sec.\ \ref{2b_observables}.

\section{The two--body model\label{2b_model}}

The two--body problem consists of two valence
particles, moving in a one--dimensional Woods--Saxon potential core (\ref{wsaxon}) that is completely filled,
interacting via a density--dependent short--range attractive residual interaction 
\begin{equation}
V_{int}(x_1,x_2) = V_0 \delta(x_1-x_2)-V_{RI} \left[\frac{\rho[(x_1+x_2)/2]}{\rho_0}\right]^p \delta(x_1-x_2),
\label{rint}
\end{equation}
\noindent where $V_0$, $V_{RI}$, $p$, and $\rho_0$ are free parameters, and
$\rho(x)$ is the core density 
\begin{equation}
\rho(x) = \sum_{i=1}^{N_b}\psi^*_i(x)\psi_i(x),
\label{coreden}
\end{equation}
\noindent where $N_b$ is the number of bound states.
The formulation of this problem has been already presented in Ref.\ \cite{bertsch} 
for the 3D case. It is important to note that, by
definition, for a symmetric potential the core density (\ref{coreden})
is a symmetric function of $x$.
In the following, we assume that the volume term $V_0$ is zero and we
only deal with the matter density--weighted term.

Modeling a drip--line system, one can choose the Fermi
surface in such a way that there are no available bound states, and
the two unperturbed particles must be in the continuum. The final
state becomes bound due to the residual interaction between the two
particles, akin to a 1D ``Borromean'' system. Thus, the
two--body Hamiltonian $H_{2b}$ is built combining the one body Hamiltonian
(\ref{wsaxon}) with the residual interaction (\ref{rint})
\begin{equation}
H_{2b}(x_1,x_2) = H_{1b}(x_1) + H_{1b}(x_2) + V_{int}(x_1,x_2).
\label{2bham}
\end{equation}

We diagonalize the two--body Hamiltonian (\ref{2bham}) in a two--body
basis built with states that are above the Fermi energy surface.
We proceed to detail the basis construction.\\
The full 1D one--body wave function has two components, a spatial part and
a spinor part
\begin{equation}
\Psi^{(1b)}_{n,m_s}(x) = \psi_n(x) \chi_{m_s}^{(s)} ,
\end{equation}
\noindent where the one-body spatial component has been previously
obtained using any of the methods presented in Sec.\ \ref{discr_methods}.

A two-body wave function involves the combination of one--body wave
functions to obtain
\begin{equation}
\Psi^{(2b)}_{n_1,n_2,S,m_S}(x_1,x_2) = \psi_{n_1}(x_1)\psi_{n_2}(x_2)
\sum_{m_{s_1},m_{s_2}} \langle s_1 m_{s_1}s_2 m_{s_2}| s_1 s_2 S m_S\rangle \chi_{m_{s_1}}^{(s_1)} \chi_{m_{s_2}}^{(s_2)} .
\label{2bwf}
\end{equation}
Assuming we are dealing with fermions, the full wave function
(\ref{2bwf}) should be antisymmetric under the interchange of the
labels $1$ and $2$. Thus, if we consider the singlet $S=0$ wave
function, the spin degrees of freedom are antisymmetric\footnote{The
  triplet contribution of the spin can be considered when the spatial
  part can be antisymmetrized, thus the interaction should not be of a
  contact type and $x_1\neq x_2$.}
\begin{equation}
\Psi^{(2b)}_{n_1,n_2,0,0}(x_1,x_2) = \psi_{n_1}(x_1)\psi_{n_2}(x_2)
\left[\frac{1}{\sqrt{2}}\left( 
\chi_{1/2}^{(1/2)} \chi_{-1/2}^{(1/2)}  - \chi_{-1/2}^{(1/2)} \chi_{1/2}^{(1/2)} \right)\right].
\end{equation}
The spatial part should be symmetrized, and the original dimension of
the problem for $N$ one-body spatial wave functions goes down from
$N^2$ to $N(N+1)/2$ for the symmetric two-body spatial wave functions $\psi^{(2b)}_{n_1,n_2}(x_1,x_2)$
\begin{equation}
\psi^{(2b)}_{n_1,n_2}(x_1,x_2) =
\frac{\sqrt{2-\delta_{n_1,n_2}}}{2}\left[\psi_{n_1}(x_1)\psi_{n_2}(x_2)
+ \psi_{n_2}(x_1)\psi_{n_1}(x_2) \right].
\label{2bodybasis}
\end{equation}
And using the ket notation
\begin{equation}
\psi^{(2b)}_{n_1,n_2}(x_1,x_2)  \rightarrow |(s) n_1 n_2\rangle = \frac{\sqrt{2-\delta_{n_1,n_2}}}{2}\left( | n_1 n_2\rangle +  |n_2 n_1\rangle \right).
\label{ket_notation}
\end{equation}

The matrix elements of the Hamiltonian (\ref{2bham}) in the
symmetrised basis are
\begin{align}
\langle (s) n'_1 n'_2 | H_{2b}|(s) n_1 n_2 \rangle & = 
\frac{\sqrt{(2-\delta_{n_1,n_2})(2-\delta_{n'_1,n'_2})}}{2} \nonumber\\
&\times
(\epsilon_{n_1}+\epsilon_{n_2})(\delta_{n_1,n'_1}\delta_{n_2,n'_2} +
\delta_{n_1,n'_2}\delta_{n_2,n'_1}) \label{2bmatel}\\  
&  + 2 \langle (s) n'_1 n'_2 | V_{int} |(s) n_1 n_2 \rangle,\nonumber
\end{align}
\noindent where the matrix element of the residual interaction is 
\begin{equation}
\langle (s) n'_1 n'_2 | V_{int} |(s) n_1 n_2\rangle =  -V_{RI} \int_{-\infty}^{+\infty}dx\,
\psi^*_{n'_1}(x)\psi^*_{n'_2}(x)\left[\frac{\rho(x)}{\rho_0}\right]^p\psi_{n_1}(x)\psi_{n_2}(x) .
\end{equation}
As the core density (\ref{coreden}) is symmetric, the integrand has to
be a symmetric function too, which implies the selection rule
\begin{equation}
n_1+n_2+n'_1+n'_2 = 2n ; ~~~ n = 0,1,2, \ldots
\end{equation}
As we are dealing with a contact interaction, it is important to
define an energy threshold, $E_{th}$, beyond which the two-body basis components
are not taken into account. Thus, only basis states $|(s) n_1
n_2\rangle$ such that $\epsilon_{n_1}+\epsilon_{n_2} \le E_{th}$ enter
into the calculation. This is due to the special characteristics of the contact
interaction that forbids convergence when the full space is considered \cite{bertsch}.\\

In each case we first use one of the considered bases (HO, THO, BOX) to solve the one--body
problem (\ref{1Dham}) to obtain a set of bound states and a set of pseudostates
representing the continuum. Then, depending on the existance or not of
free bound states (states above the Fermi energy surface), the two--body
basis is built and the two--body Hamiltonian (\ref{2bham}) is diagonalized,
computing the matrix elements (\ref{2bmatel}). This second part is
common to all the methods.

The residual interaction parameter values selected in the present work are as follows
\begin{align}
V_0 &= 0  \,\mbox{MeV} &V_{RI} &= -38 \,\mbox{MeV} \nonumber\\
\rho_0 &= 0.15  & p &= 1\label{ripar}\\
E_{th}&=50 \,\mbox{MeV} & \mu  &= 0.975\, \mbox{amu}\nonumber
\end{align}

\section{Two--body model Energies and Wavefunctions\label{2b_eigval_eigvec}}
As in the one--body case, we check the bound state energy convergence and the wave functions tails.
We should emphasize that this is a more complicate task, due to the two--body nature of the system, which implies larger bases and more cumbersome calculations.

A proper behavior in the wave functions tail region is essential: it is involved in the description of the pairing field and of the two--particle transfer processes in heavy--ion induced reactions.

In the first place we check the dependence of the two--body Hamiltonian
eigenvalues with the truncated 1D HO basis
dimension. We then display the change with the basis size N of the
ground state wave function spatial dependence.
We then repeat the calculations for the THO and BOX bases.

\begin{figure}[hc]
\begin{center}
\includegraphics[scale = 0.32]{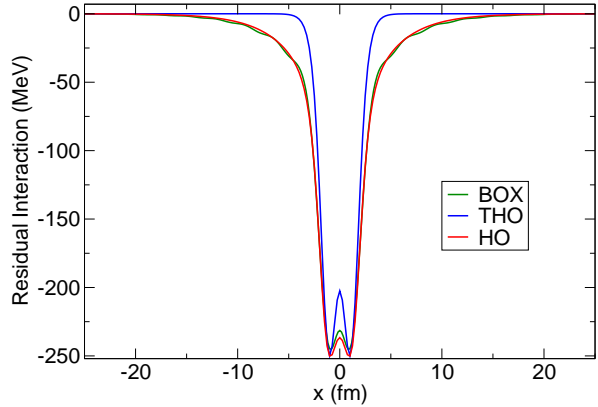}
\caption{\label{fig_res_int} Spatial dependence of
  the residual interaction for parameters in Eq.\ \ref{ripar} for the HO ($N = 70$),  
  THO($\gamma/b = 1.2 fm^{-1/2}$ and $N = 50$), and BOX ($x_b = 50 fm$) cases.}
\end{center}
\end{figure}

The description of the one--body bound states obtained in Chap.\ \ref{2b_probl} is quite accurate even for a
small basis dimension. Therefore the residual interaction is not
expected to vary with N. This can be checked in
Fig.\  \ref{fig_res_int} where the resulting residual interaction
$ -V_{RI}[\rho(x)/\rho_0]^p$ is plotted for the HO, THO and BOX cases there are some
differences between the three cases, which for large values of $x$ can be significant.

\subsection{Eigenenvalues}

\begin{figure}[!htc]
\begin{center}
\scalebox{0.32}{\includegraphics{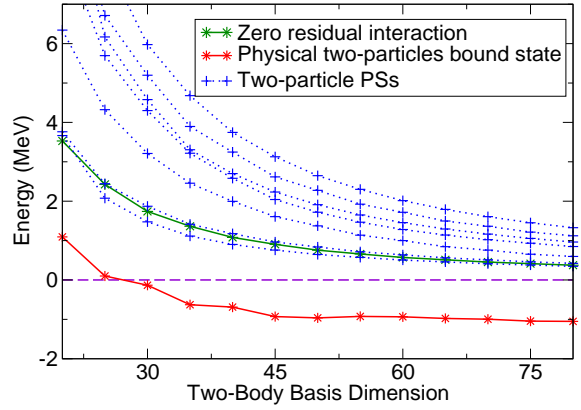}}
\caption{Two-body Hamiltonian
  energies for the first eigenstates as a function of the dimension of
  the truncated 1D HO basis.\label{fig_2b_ho_energy}}
\end{center}
\end{figure}

\begin{figure}[!htc]
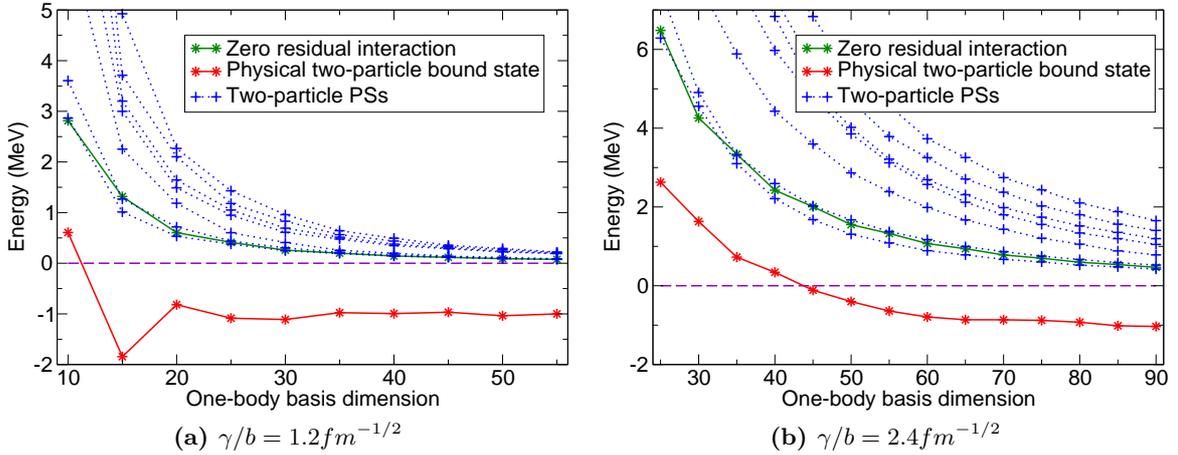

\begin{center}
\subfloat[$\gamma/b = 1.2 fm^{-1/2} $]{
\scalebox{0.32}{\includegraphics{1D_2b_energyVS1bN_THO_1c2}}
}
\hspace*{0.1cm}
\subfloat[$\gamma/b = 2.4 fm^{-1/2} $]{
\scalebox{0.32}{\includegraphics{1D_2b_energyVS1bN_THO_2c4}}
}
\caption{Two-body Hamiltonian
  energies for the first eigenstates as a function of the dimension of
  the truncated 1D THO basis. 
  The results with two different values of the ratio parameter are reported.\label{fig_tho_energy_conv}}
\end{center}
\end{figure}

\begin{figure}[!htc]
\begin{center}
\scalebox{0.32}{\includegraphics{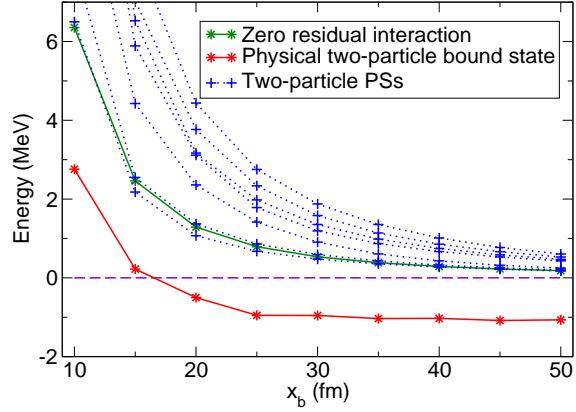}}
\caption{Two-body Hamiltonian
  energies for the first eigenstates as a function of the radius $x_b$ 
  of the box method; the one--body basis dimension has been set as large as possible.\label{fig_2b_isqw_energy}}
\end{center}
\end{figure}

The dependence of the two-body Hamiltonian eigenvalues with the
truncated 1D basis dimension is depicted in the Figs.\
\ref{fig_2b_ho_energy} and \ref{fig_tho_energy_conv} for the HO and THO methods, 
while for the BOX method the dependence on the box radius is reported in Fig.\ \ref{fig_2b_isqw_energy}.
In the case of the BOX method the one--body basis dimension has been set large enough to obtain convergence in the one body eigenstates below the energy threshold $E_{th}$.

As can be seen from those figures, the energy is
converging, more or less rapidly  to a limiting value $E_b = -1.01 MeV$. 
Note that without the residual interaction (green lines) the two--body system would be unbound, and it is the attractive residual interaction between the two valence neutron that makes the system can be bound (red lines).

\begin{figure}[!htc]
\begin{center}
\scalebox{0.32}{\includegraphics{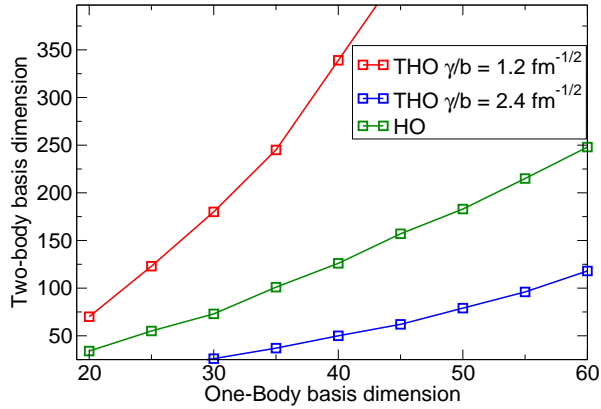}}
\caption{Two--body basis dimension as a function of the one--body basis dimension for the HO and THO methods.\label{fig_2b1bdim}}
\end{center}
\end{figure}

In order to give an idea of the increased complexity of
the problem we plot in Fig.\ \ref{fig_2b1bdim} the dependence of the two--body symmetrized basis dimension as a function of the one--body basis dimension.
As discussed in Chapt.\ \ref{1b_probl}, the $\gamma/b = 2.4 fm^{-1/2}$ THO case behaves in a similar way to the HO, and variation of $\gamma/b$ gives us an extra degree of freedom in the investigation of the system.

\subsection{Eigenstates}
It is important to check also the behavior of the resulting new bound
eigenstate. In this case, we expect a sensitive dependence on the
parameters, specially with regard to the tails of the wave
function.\\
In the next figures the two--body bound state wave function for $x_1= x_2 = x$ 
is depicted with respect to the parameters of each method.

\begin{figure}[!h]
\begin{center}
\includegraphics[width=0.75\textwidth]{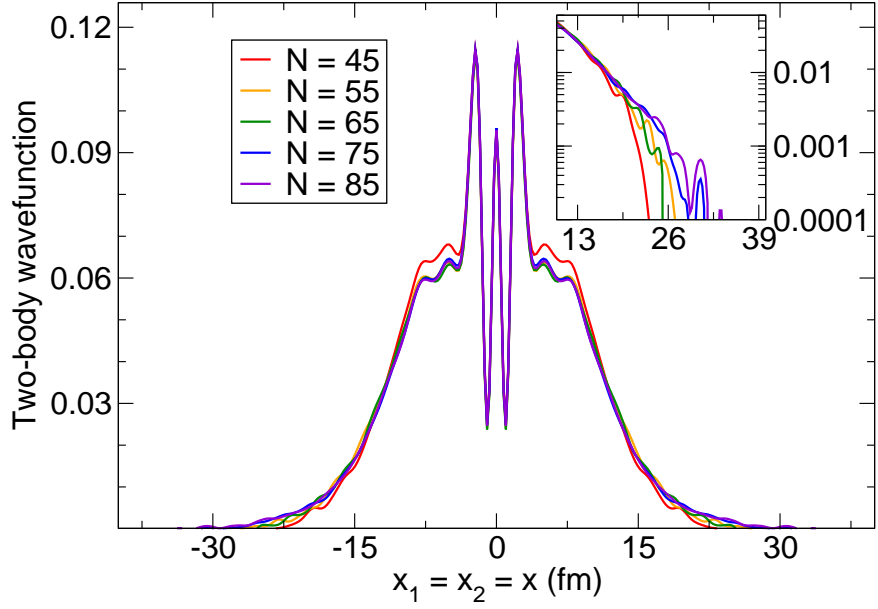}
\caption{\label{fig_2b_ho_bound_wf} Two-body bound state
  for $x_1=x_2$ obtained after Hamiltonian diagonalization in
  a truncated one-body 1D HO basis with different dimensions.}
  \end{center}
\end{figure}

\begin{figure}[!h]
\begin{center}
\includegraphics[width=0.75\textwidth]{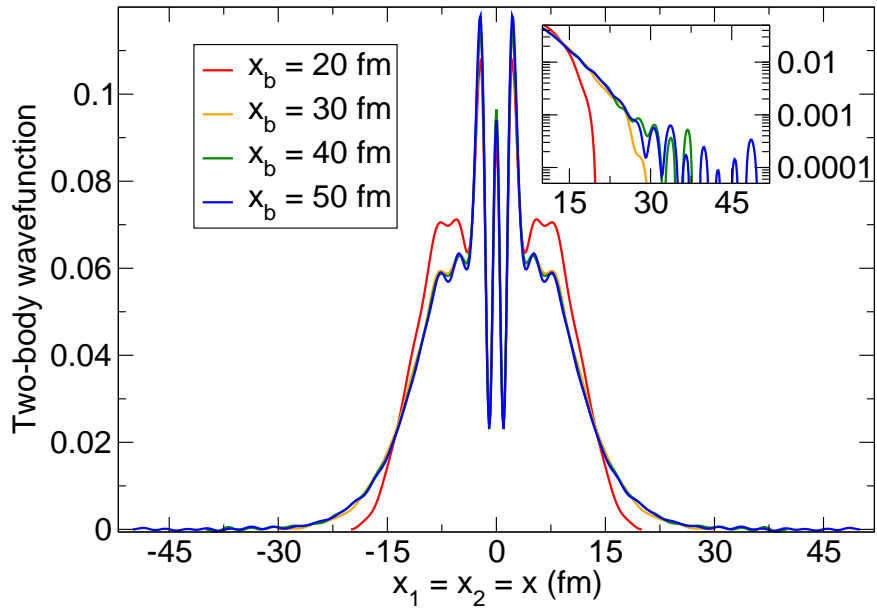}
\caption{\label{fig_2b_isqw_bound_wf} Two-body bound state
  for $x_1=x_2$ obtained after Hamiltonian diagonalization in 
  a Box with different radius (ISQW method). }
  \end{center}
\end{figure}

\begin{figure}[!htc]
\begin{center}
\subfloat[$\gamma/b = 1.2 fm^{-1/2} $]{
\scalebox{0.32}{\includegraphics{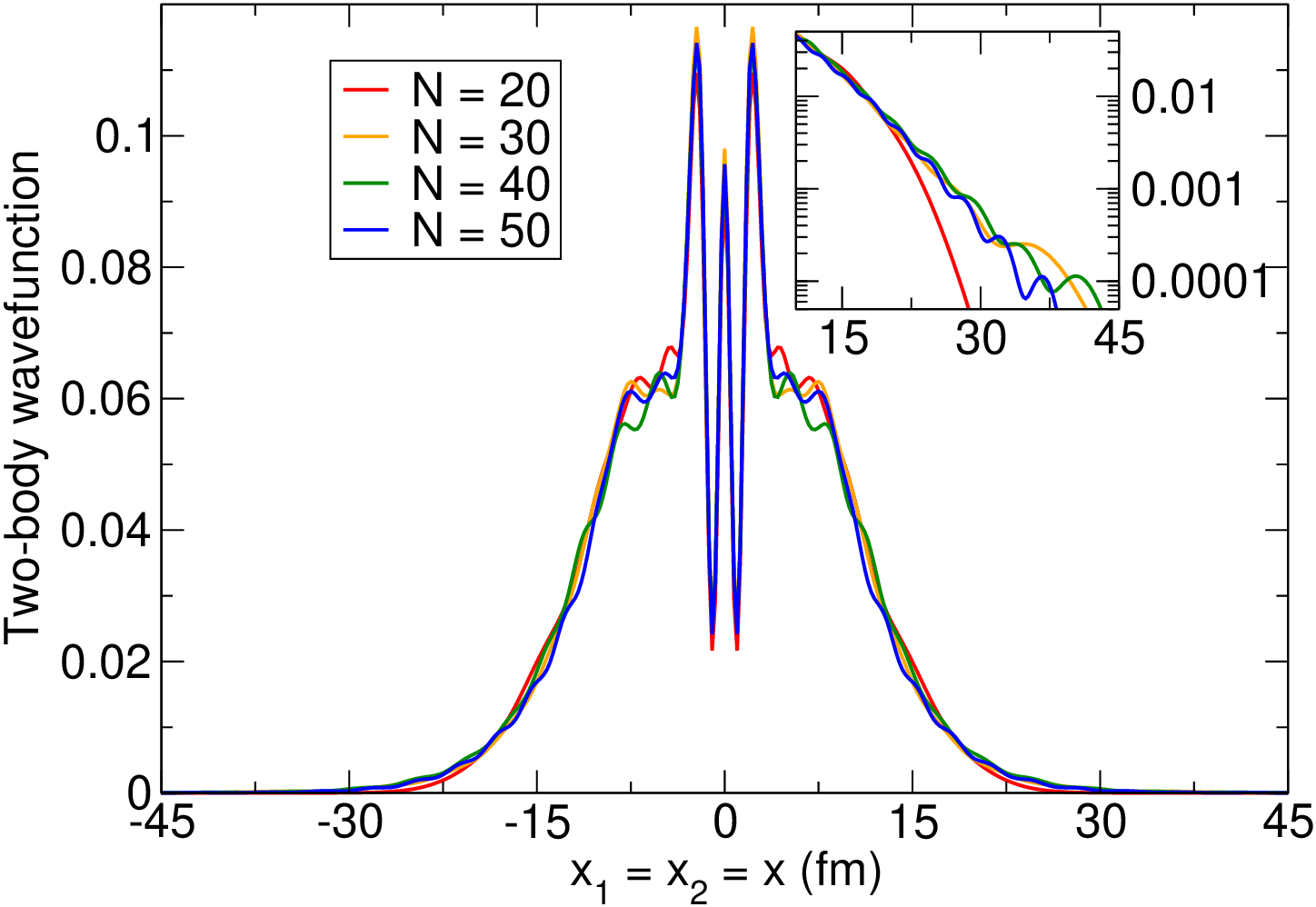}}
}
\vspace*{0.2cm}
\subfloat[$\gamma/b = 2.4 fm^{-1/2} $]{
\scalebox{0.32}{\includegraphics{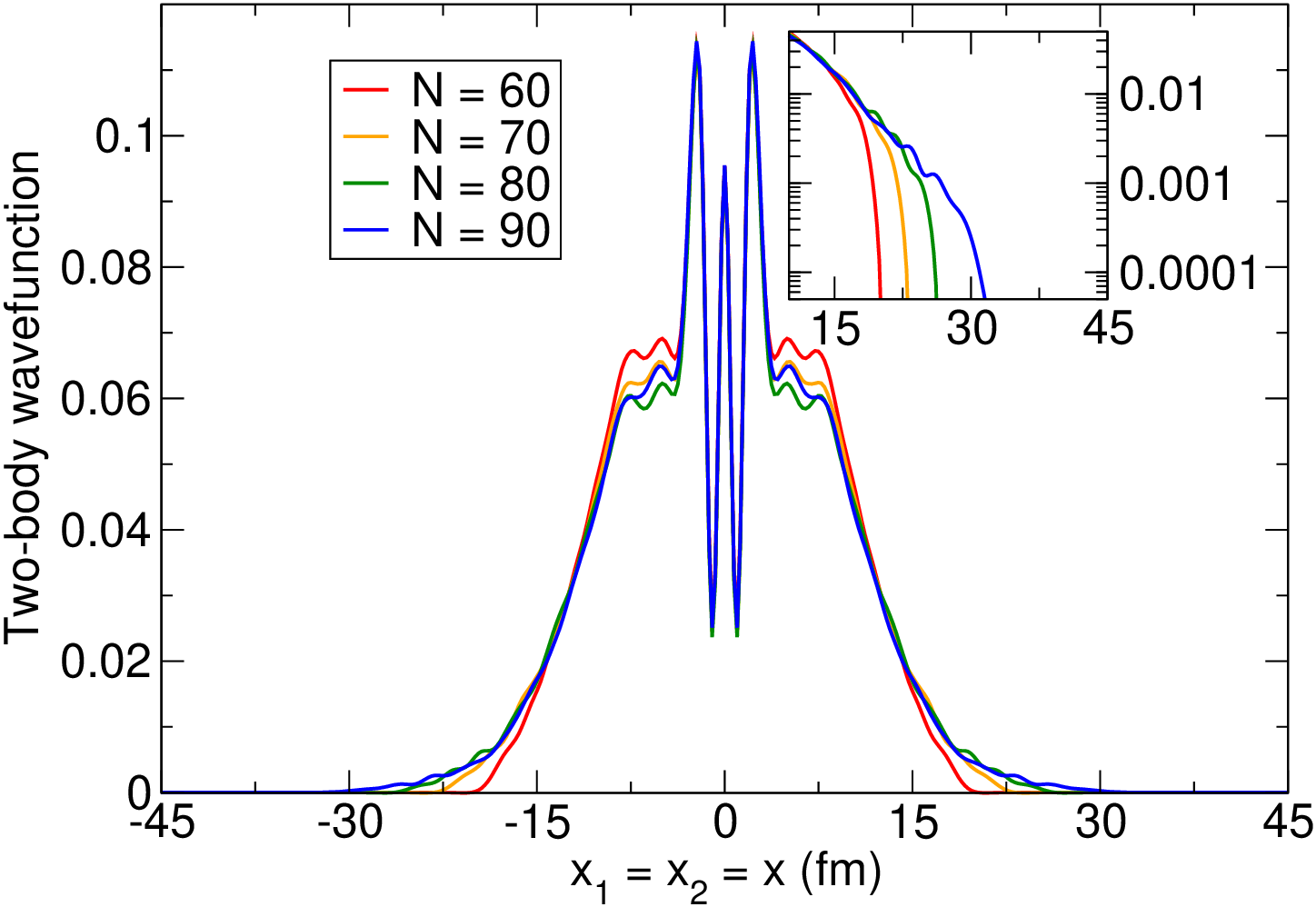}}
}
\caption{Two-body bound state
  for $x_1=x_2$ obtained after Hamiltonian diagonalization in
  a truncated one-body 1D THO basis with different dimensions.\label{fig_tho_bound_wf}}
\end{center}
\end{figure}

We plot in Figs.\ \ref{fig_2b_ho_bound_wf} and \ref{fig_tho_bound_wf} the bound wave function as a function of $x$ for several basis dimension values for the HO and THO methods, respectively. In Fig.\ \ref{fig_2b_isqw_bound_wf} we plot the bound eigenfunction in the BOX case for different $x_b$ values.

The difference between the eigenfunctions is
more apparent in the inset panel with a zoom for the large $x$ region depicted in logaritmic scale.

\begin{figure}[!htc]
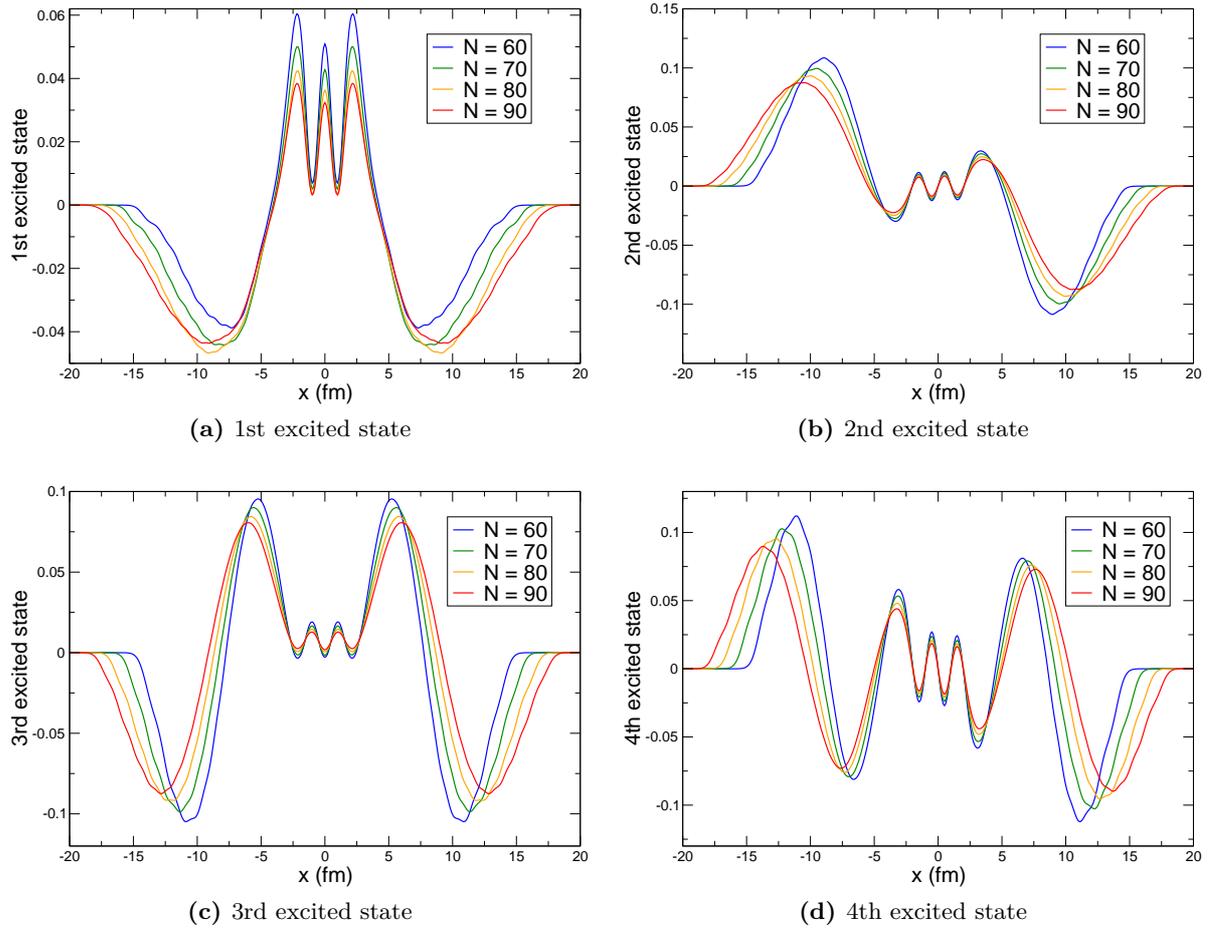

\begin{center}
\subfloat[1st excited state]{
\scalebox{0.32}{\includegraphics{1D_2b_1st_excit_WF_withN1b_HO}}
}
\hspace*{0.1cm}
\subfloat[2nd excited state]{
\scalebox{0.32}{\includegraphics{1D_2b_2nd_excit_WF_withN1b_HO}}
}
\vspace{0.5cm}
\subfloat[3rd excited state]{
\scalebox{0.32}{\includegraphics{1D_2b_3rd_excit_WF_withN1b_HO}}
}
\hspace*{0.1cm}
\subfloat[4th excited state]{
\scalebox{0.32}{\includegraphics{1D_2b_4th_excit_WF_withN1b_HO}}
}
\caption{First four continuum two-body states
  for $x_1=x_2$ obtained after Hamiltonian diagonalization for
  a truncated one-body 1D HO basis with different dimensions.\label{fig_continuum_1D_2body_ho}}
\end{center}
\end{figure}

Diagonalizing the Hamiltonian matrix for the two--body problem we also obtain a set of discretized wave of the continuum.
We depict continuum two--body PS, as an example with the HO method, in Fig.\ \ref{fig_continuum_1D_2body_ho}
for $x_1= x_2 = x$.

\subsection{\label{2bPair}Pairing Effect}
\begin{figure}[!htc]
\begin{center}
\subfloat[HO correlated case]{
\scalebox{0.32}{\includegraphics{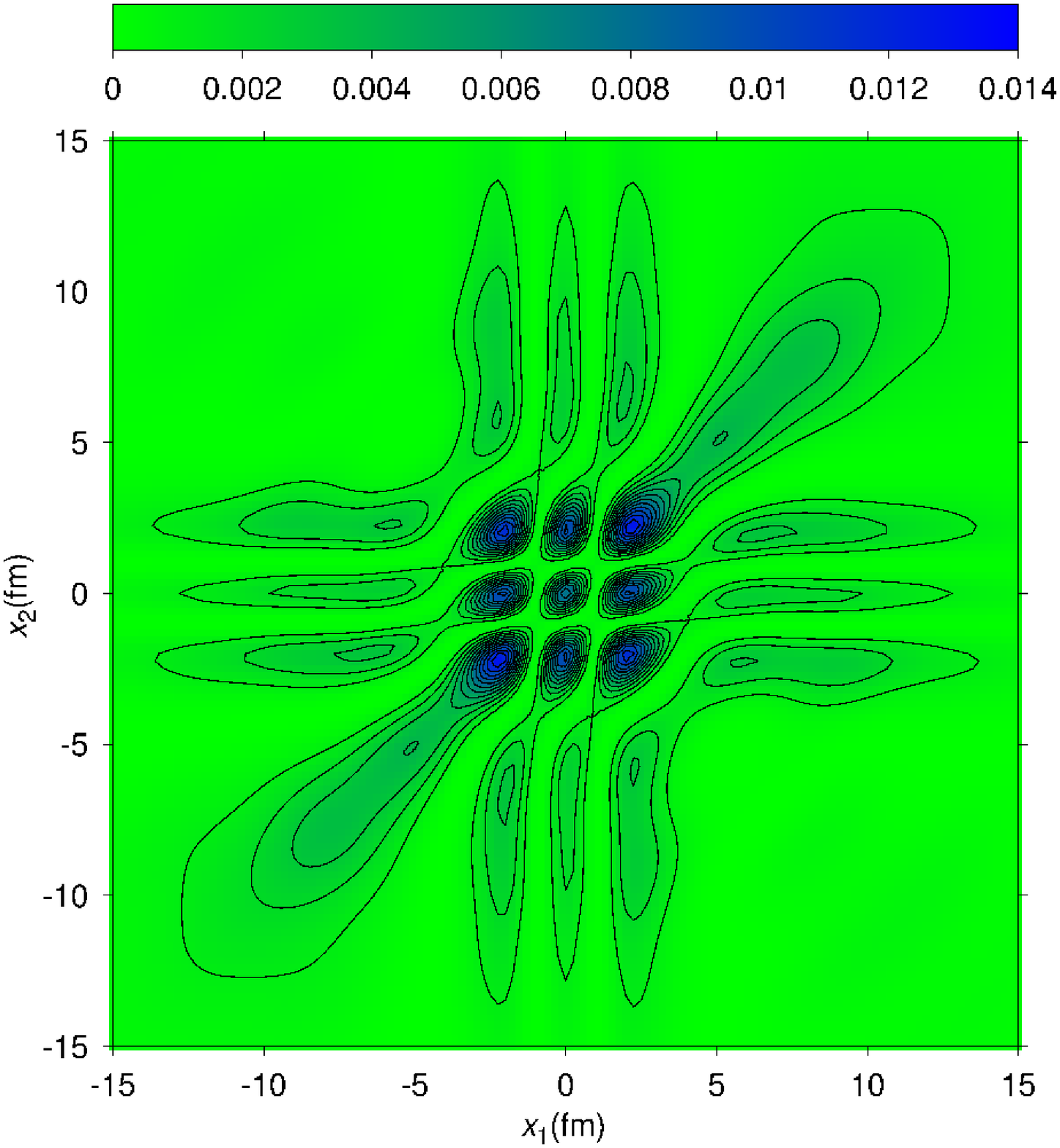}}
}
\hspace*{0.1cm}
\subfloat[THO correlated case]{
\scalebox{0.32}{\includegraphics{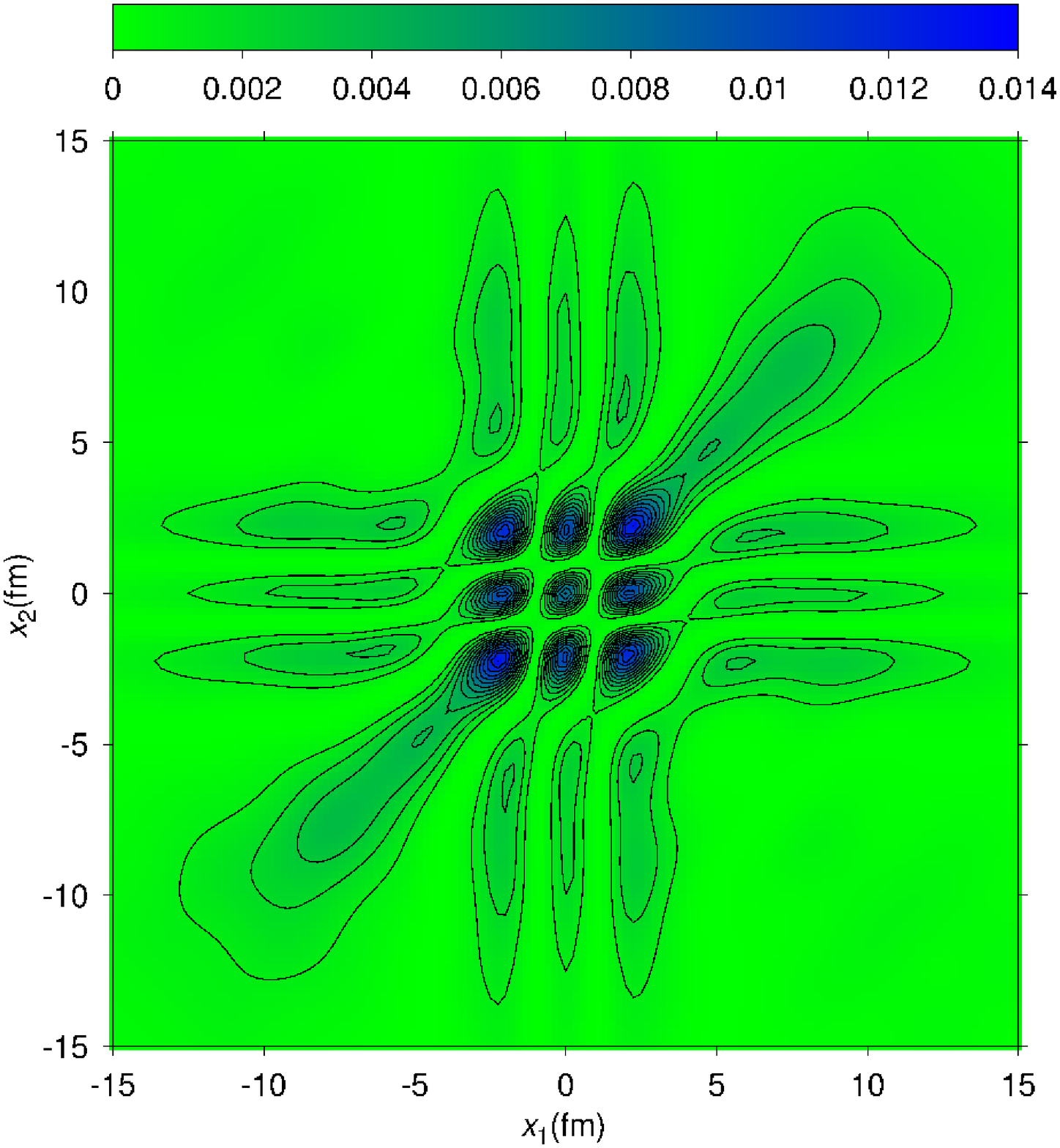}}
}
\vspace*{0.2cm}
\subfloat[BOX correlated case]{
\scalebox{0.32}{\includegraphics{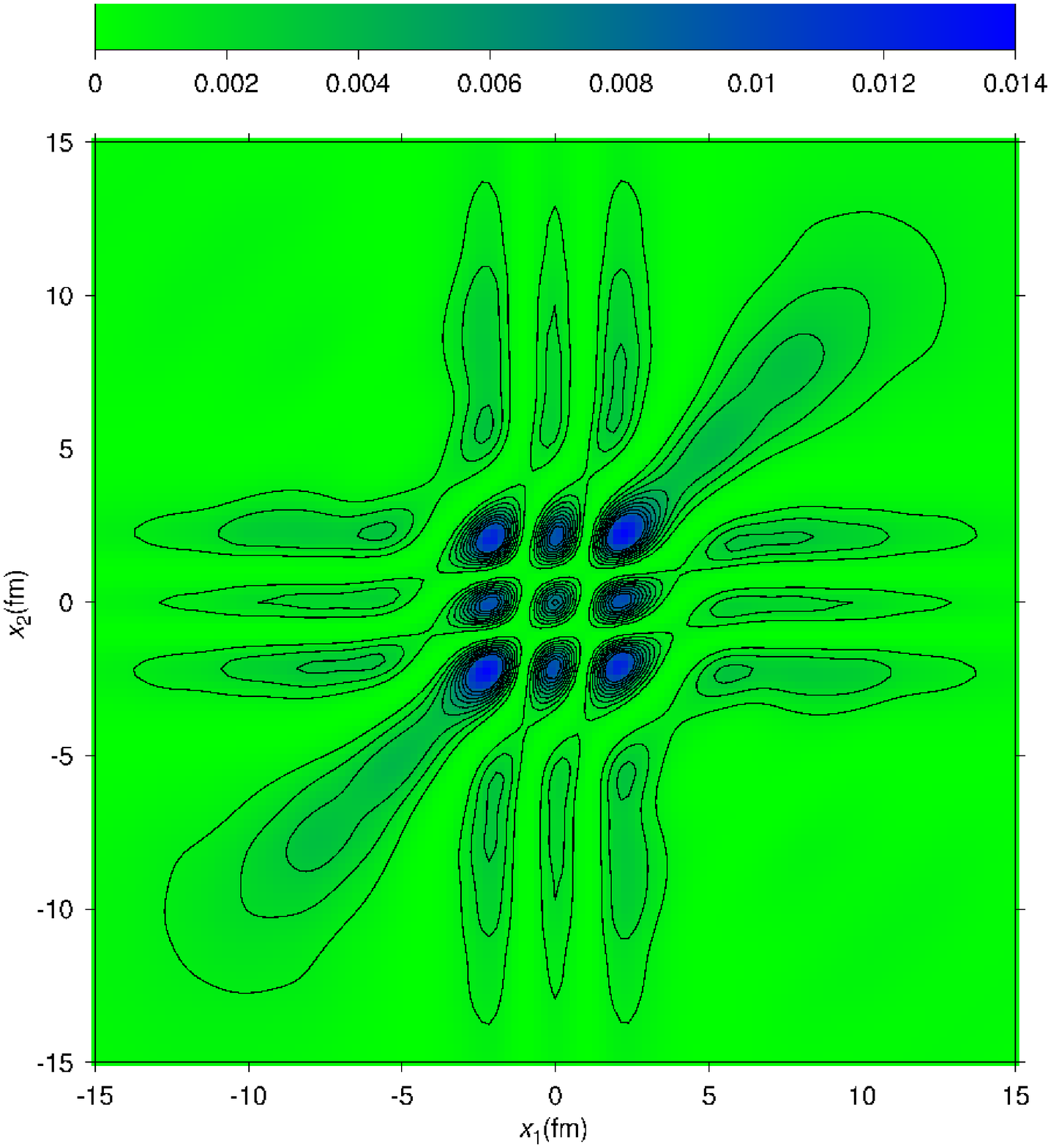}}
}
\hspace*{0.1cm}
\subfloat[Uncorrelated case]{
\scalebox{0.32}{\includegraphics{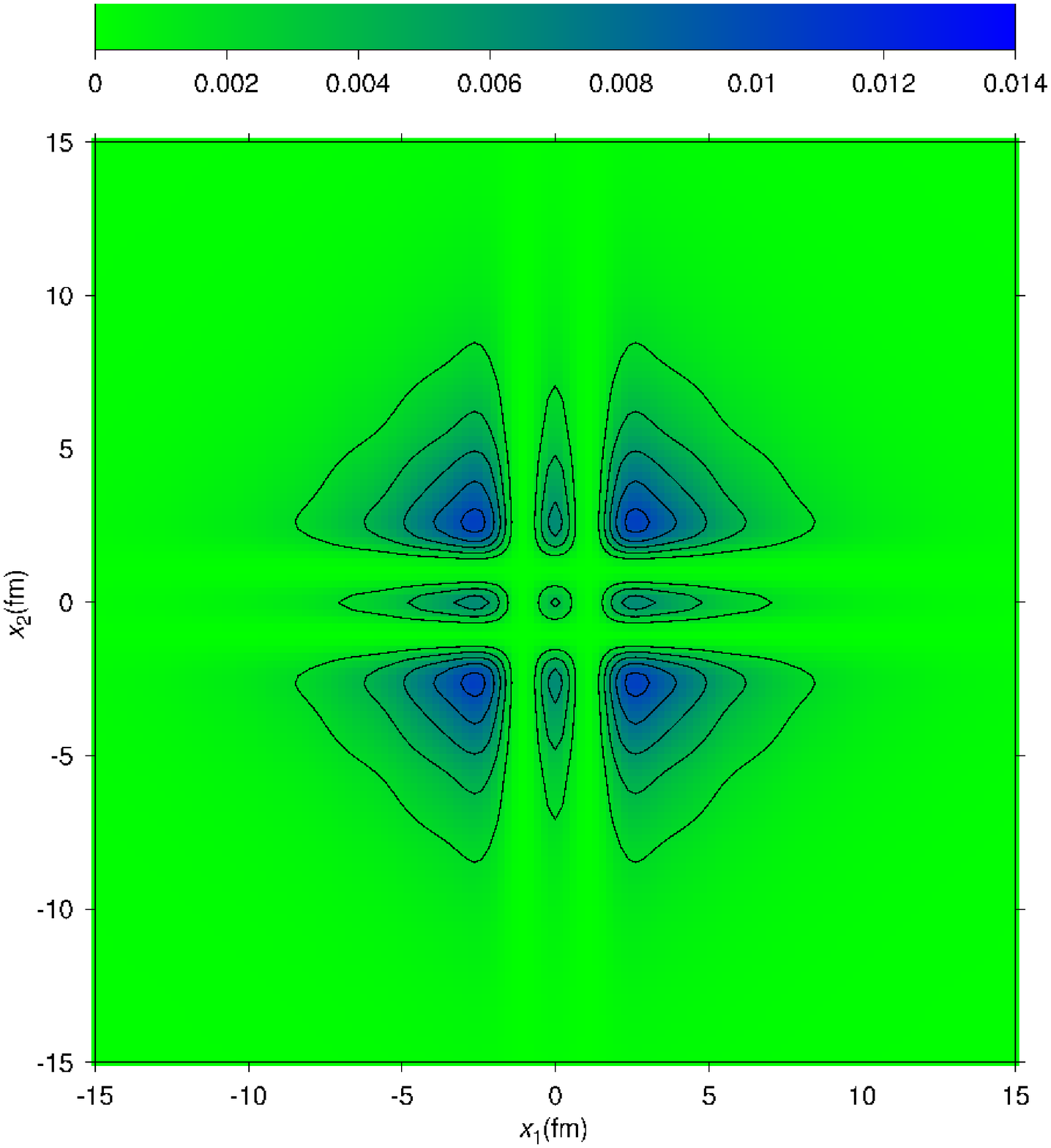}}
}
\caption{The contour
  plots of the probability density for the two-body bound state with
  binding energy $E = -1.01 $ MeV using the correlated Hamiltonian
  (\ref{2bham}) constructed with the different bases are reported:
  HO basis with $N = 70$ in panel (a), THO ($\gamma/b = 1.2 fm^{-1/2}$) with $N = 50$ in panel (b),
  and a ISQW basis with $N = 70$ and $x_b = 60 fm$ in panel (c).
  In panel (d) is reported the 
  uncorrelated Hamiltonian (zero residual interaction) with a single
  particle potential depth modified to obtain the same binding energy
  for the system ground state diagonalized, e.\ g.\ in the BOX case with $x_b = 60 fm$.
  \label{correlation}}
\end{center}
\end{figure}

An insight on the effect of the pairing correlations can be obtained by looking at the wave function (or its modulus squared) as a function of coordinates $ x_1 $ and $ x_2 $. The results obtained for the correlated two--particle ground--state displayed in Figs.\ \ref{correlation}a,b,c can be compared with the uncorrelated case displayed in Fig.\ \ref{correlation}d. 

In order to better pinpoint the effect of correlations by separating possible features coming from the weak binding of the two--particle ground state, we define an uncorrelated case with zero residual interaction and a mean field such to obtain a two--particle uncorrelated wave function with the same binding energy ($-1.01 MeV$) as the correlated one.

It is clear that the residual interaction has created a spatial correlation between the two particles 
which is proved by an increased probability along the bisector line $x_1 = x_2$ for small relative distances.
On the contrary, in the uncorrelated case the wave functions symmetry implies equal probabilities independent of the particles relative distances.

\vspace{0.5cm}

One should keep in mind that the resulting spatial correlation in the paired case has an important role beyond the pure structure properties, being responsible in enhancing reaction processes such as two--particle transfer, two--particle break--up or two--particle knock--out reactions and in characterizing the angular correlation between the two emitted particles in the two latter cases.

%
\section{Results for Anomalous Density and Electric Transitions\label{2b_observables}}
As in the one--body case, several quantities of interest can be investigated.
In this section we obtain the Anomalous Density for the ground two--body state and first
continuum states and the dipole and quadrupole
electric transition operators expectation values for a transition
between the system two-body ground state and the continuum.

\subsection{Anomalous Density}
The Anomalous Density $T_0$ is defined as follows \cite{Ring}

\begin{equation}
T_0 =  \left|\int_{-\infty}^{+\infty} dx\, \delta\rho_{pair}(x) \right|^2~,
\label{T0}
\end{equation}
\noindent where $\int dx\, \delta\rho_{pair}(x)=\langle A+2|a^\dagger a^\dagger| A\rangle $ and thus
\begin{equation}
 \delta\rho_{pair}(x) =  \sum_{i_1,i_2} X_{i_1,i_2} \phi_{i_1}(x) \phi_{i_2}(x)~.
\label{deltarho}
\end{equation}
\noindent The coefficients $X_{i_1,i_2}$ are obtained by diagonalization of the Hamiltonian (\ref{2bham}).
Large values of the anomalous density imply a collective character of the state under study, and its  evaluation implies the integration of the two-body eigenfunction in points $x_1=x_2=x$, depicted in Fig.\ \ref{fig_2b_ho_bound_wf}.

It is interesting to take into account that, due to the orthonormality
of the one--body basis functions $\phi_{i}(x)$, the anomalous density
can be written as 
\begin{equation}
T_0 =  \left|\sum_{i} X_{i,i}\right|^2~,
\label{T0bis}
\end{equation}
\noindent which implied that only the eigenvector components
associated to the $|(s) n n\rangle$ basis states should be taken into
account. \\

Anomalous density values can also be computed for the excited
two--body pseudostates. We will use the notation $T_0^{(i)}$ for the
anomalous density for the $i$-th two--body excited state where
Eq.\ (\ref{deltarho}) is redefined as
\begin{equation}
 \delta\rho^{(i)}_{pair}(x) =  \sum_{i_1,i_2} X^{(i)}_{i_1,i_2} \phi_{i_1}(x) \phi_{i_2}(x)~,
\label{deltarhoi}
\end{equation}
\noindent where $X^{(i)}_{i_1,i_2}$ are the coefficients for the
$i$-th excited state obtained by diagonalization of the
Hamiltonian. As in the ground state case, taking into account the
basis orthonormality in the integral evaluation the anomalous density
can be rewritten as\\
\begin{equation}
T^{(i)}_0 =  \left|\sum_{k} X^{(i)}_{k,k}\right|^2~.
\label{Texc}
\end{equation}

\begin{figure}[!htc]
\begin{center}
\scalebox{0.32}{\includegraphics{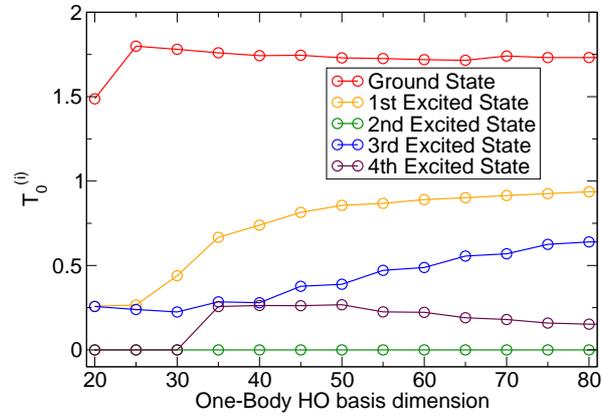}}
\caption{The Anomalous Density result for the HO method as a function of the one--body basis dinension N.\label{fig_2b_ho_anomdens}}
\end{center}
\end{figure}

\begin{figure}[!htc]
\begin{center}
\scalebox{0.32}{\includegraphics{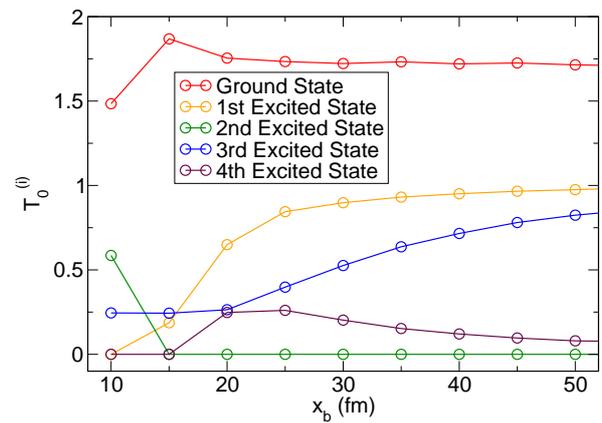}}
\caption{The Anomalous Density result for the BOX method as a function of the box radius $x_b$.\label{fig_2b_isqw_anomdens}}
\end{center}
\end{figure}

\begin{figure}[!htc]
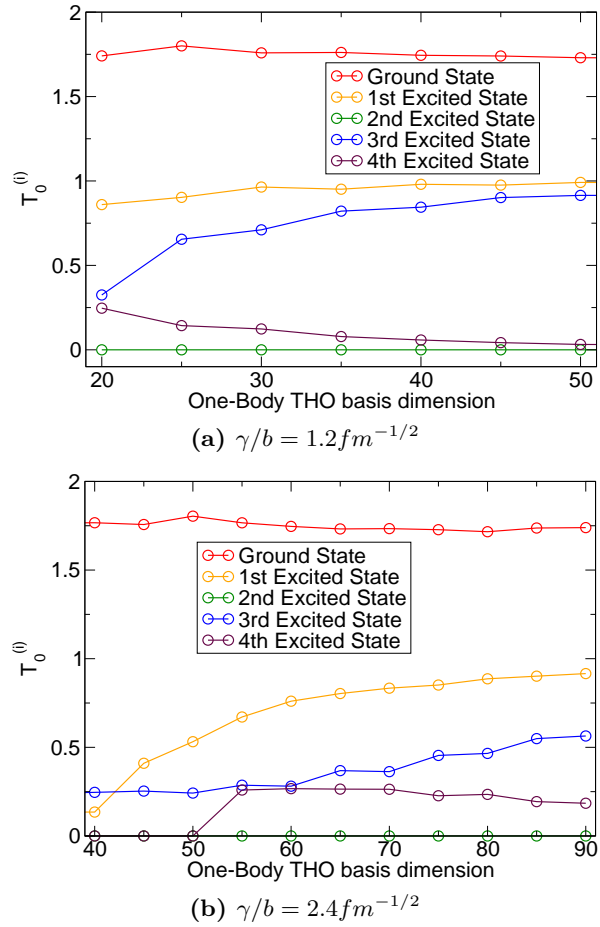

\begin{center}
\subfloat[$\gamma/b = 1.2 fm^{-1/2} $]{
\scalebox{0.32}{\includegraphics{1D_2b_anomdensVS1bN_THO_1c2}}
}
\vspace*{0.2cm}
\subfloat[$\gamma/b = 2.4 fm^{-1/2} $]{
\scalebox{0.32}{\includegraphics{1D_2b_anomdensVS1bN_THO_2c4}}
}
\caption{The Anomalous Density result for the THO method as a function of the one--body basis dinension N for $\gamma/b = 1.2 fm^{-1/2} $ (left panel) and $\gamma/b = 2.4 fm^{-1/2} $ (right panel).\label{fig_tho_anomdens}}
\end{center}
\end{figure}

Anomalous Density results for HO (Fig.\ \ref{fig_2b_ho_anomdens}), THO (Fig.\ \ref{fig_tho_anomdens}), and 
BOX (Fig.\ \ref{fig_2b_isqw_anomdens}) agree satisfactory and indicate, as expected, the collective character of the two--body system ground state.


%



\subsection{\label{ssec:2bTop}Electric Transitions}

We proceed to extend the results obtained for the electric dipole and quadrupole transition intensities to the two--body problem. In the case of halo nuclei these quantities are of major importance for an adequate description of radiative capture and break--up processes.

Following Eq.\ (\ref{MElambda}) we define
\begin{align}
{\cal M}(E1) = x_1 + x_2\\
{\cal M}(E2) = x_1 ^2 + x_2 ^ 2
\end{align}
and we compute the transition integrals
\begin{equation}
\langle \Psi_b \vert {\cal O} \vert \Psi_i \rangle = \int _{-\infty} ^{+\infty} dx_1 dx_2 ~ \Psi_b ^{\ast} (x_1,x_2) {\cal O} \Psi_i (x_1,x_2)
\end{equation}
where ${\cal O} = {\cal M}(E1), {\cal M}(E2)$, $ \Psi_b(x_1,x_2) $ is the weakly--bound ground state of the two--body system.\\

\begin{figure}[!htc]
\begin{center}
\scalebox{0.32}{\includegraphics{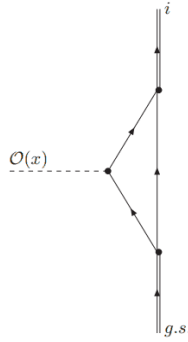}}
\caption{Interaction of the transition operator ${\cal O}(x)$ with the two--particle correlated system. Following the Nuclear Field Theory conventions, single arrowed lines indicate single--particle states and arrowed double lines indicate correlated two--particle states. The dashed line indicates the action of the one---body "dipole" or "quadrupole" fields.\label{feyn_diag}}
\end{center}
\end{figure}

In this section we assume that the core does not play any role in the electromagnetic excitation process and is kept frozen. Thus the excitation process involves only the two valence nucleons as schematically depicted in Fig.\ \ref{feyn_diag}.
\vspace{0.5cm}

The two--body system ground state and pseudostates can be written as
\begin{equation}
\Psi_k(x_1,x_2) = \sum_{i=1}^{N}\beta_{ik}\psi^{(2b)}_{(n_1,n_2)_i}(x_1,x_2)~~.
\end{equation}

The ${\cal M}(E1)$ and ${\cal M}(E2)$ observable values for any pair of
states can then be computed as 
\begin{align}
\langle \Psi_k|D_{12}|\Psi_{gs}\rangle &= \langle \Psi_k|x_1 + x_2|\Psi_{gs}\rangle~, \\
\langle \Psi_k|Q_{12}|\Psi_{gs}\rangle &= \langle \Psi_k|x^2_1 + x^2_2|\Psi_{gs}\rangle~.
\end{align}

Taking into account Eq.\ (\ref{ket_notation}).
\begin{align}
\langle \Psi_k|D_{12}|\Psi_{gs}\rangle &= \langle \Psi_k|x_1 +
x_2|\Psi_{gs}\rangle~, \nonumber\\
&=  \sum_{i,j=1}^{N}\beta_{ik}\beta_{igs} {_i\langle} (s) n_1 n_2|x_1 + x_2|(s) n_1 n_2\rangle_j,\\
\langle \Psi_k|Q_{12}|\Psi_{gs}\rangle &= \langle \Psi_k|x^2_1 + x^2_2|\Psi_{gs}\rangle \nonumber\\
&=  \sum_{i,j=1}^{N}\beta_{ik}\beta_{igs} {_i\langle} (s) n_1 n_2|x^2_1 + x^2_2|(s) n_1 n_2\rangle_j~.
\end{align}
Using Eqs.\ (\ref{1bodywfHO}), (\ref{1bodywfBOX}), and (\ref{2bodybasis})
 the matrix elements can be computed using the
one-- and two--body eigenvector matrices and the matrix elements of
$x$ and $x^2$ given in subsection \ref{1b_E1_E2}.

\begin{figure}[!htc]
\begin{center}
\scalebox{0.32}{\includegraphics{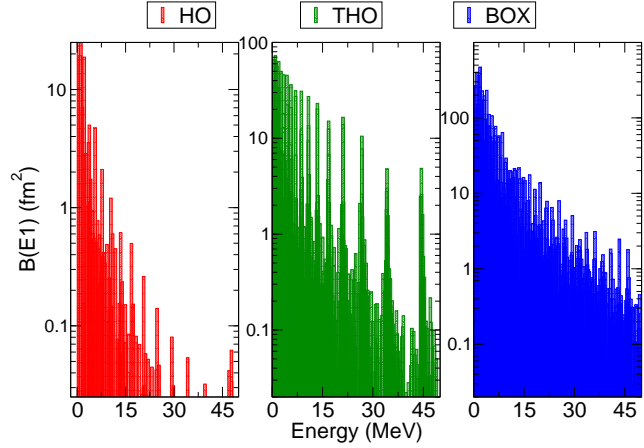}}
\caption{Electric dipole transition calculated for the HO ($N = 60$), THO ($\gamma/b = 1.2 fm^{-1/2}$ and $N = 50$), and BOX ($x_b = 50 fm$) cases.\label{fig_E1_1D_2body}}
\end{center}
\end{figure}

\begin{figure}[!htc]
\begin{center}
\scalebox{0.32}{\includegraphics{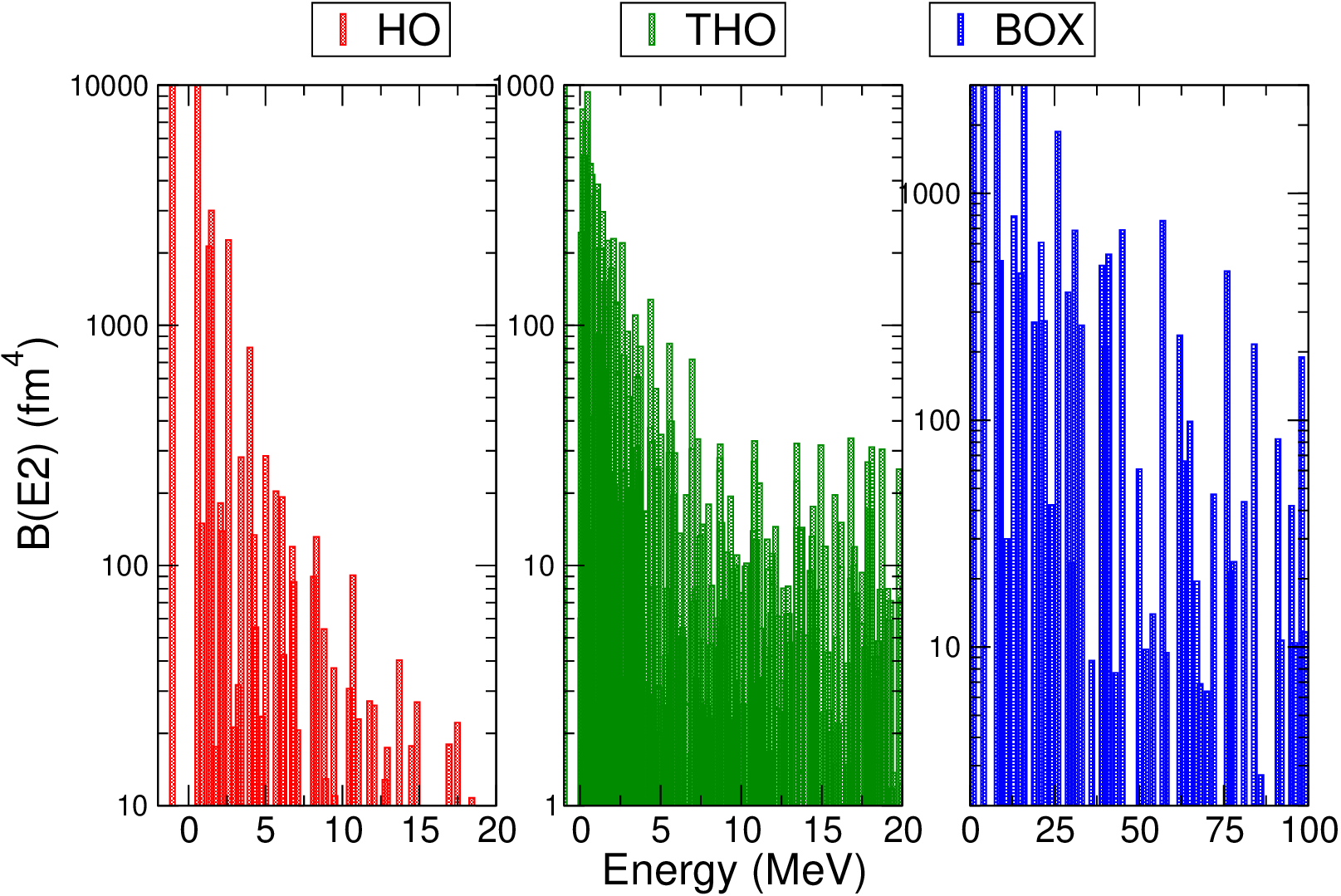}}
\caption{Electric quadrupole transition calculated for the HO ($N = 60$), THO ($\gamma/b = 1.2 fm^{-1/2}$ and $N = 50$), and BOX ($x_b = 50 fm$) cases.\label{fig_E2_1D_2body}}
\end{center}
\end{figure}

\begin{figure}[!htc]
\begin{center}
\scalebox{0.32}{\includegraphics{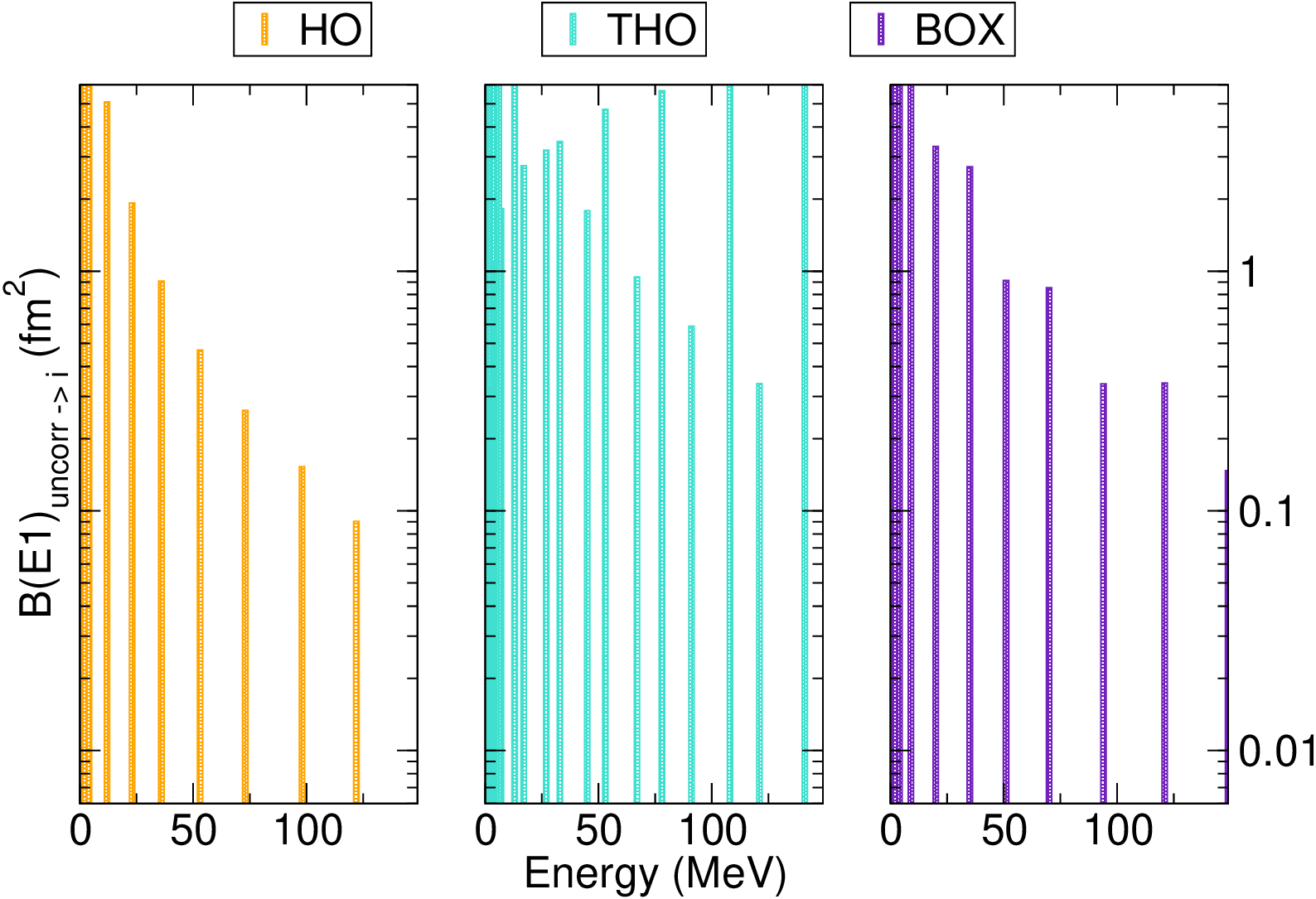}}
\caption{Electric dipole transition calculated for the HO ($N = 70$), THO ($\gamma/b = 1.2 fm^{-1/2}$ and $N = 50$), and BOX ($x_b = 45 fm$) cases in the uncorrelated situation.\label{fig_E1_uncorr}}
\end{center}
\end{figure}

\begin{figure}[!htc]
\begin{center}
\scalebox{0.32}{\includegraphics{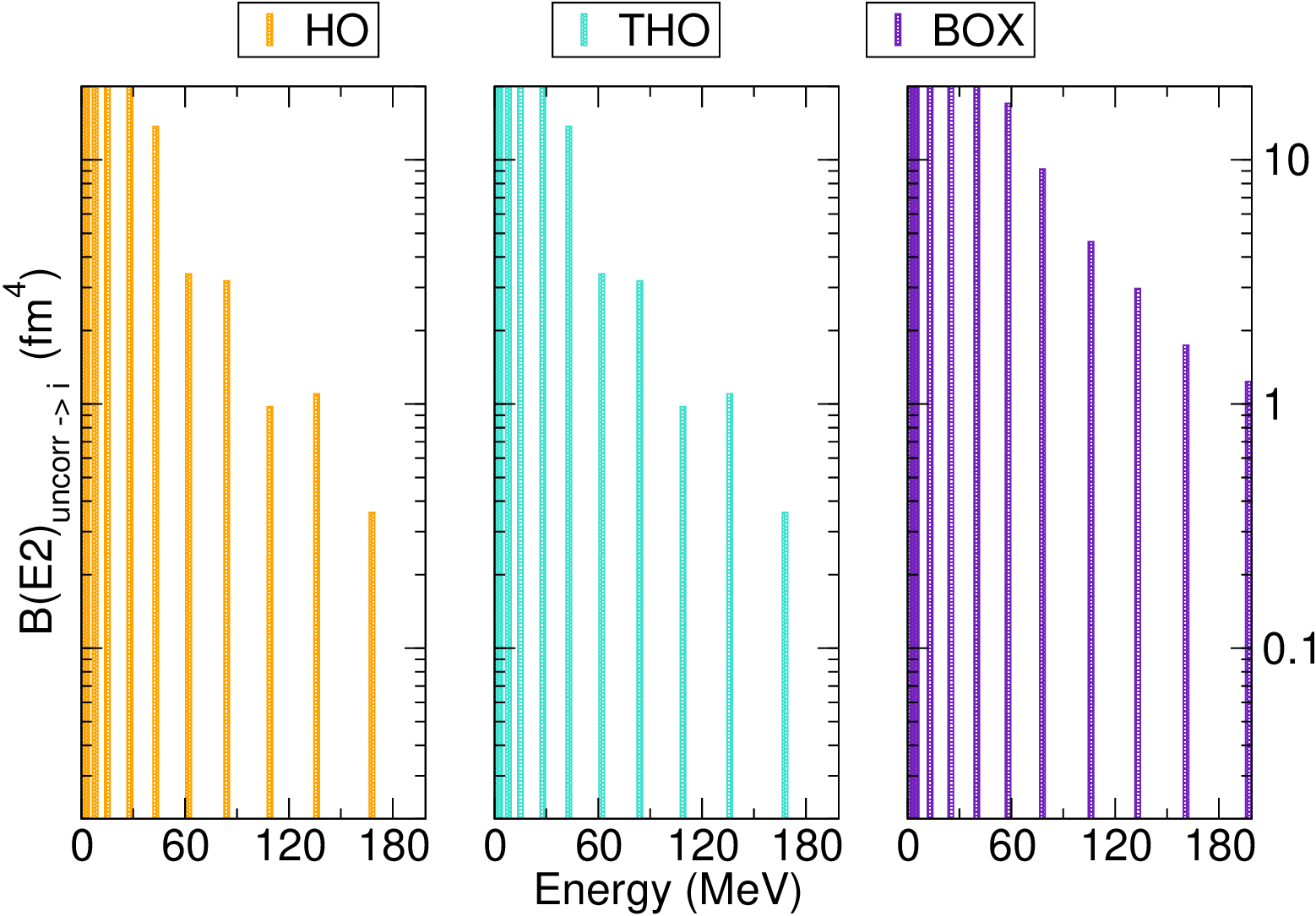}}
\caption{Electric quadrupole transition calculated for the HO ($N = 70$), THO ($\gamma/b = 1.2 fm^{-1/2}$ and $N = 50$), and BOX ($x_b = 45 fm$) cases in the uncorrelated situation.\label{fig_E2_uncorr}}
\end{center}
\end{figure}

We plot results for the electric dipole transition in Fig.\ \ref{fig_E1_1D_2body} for the HO, THO ($\gamma/b = 1.2 fm^{-1/2}$), and BOX cases.
The electric quadrupole results are depicted in Fig.\ \ref{fig_E2_1D_2body} for the HO, THO ($\gamma/b = 1.2 fm^{-1/2}$), and BOX cases.
The B(E1) and B(E2) for the uncorrelated case are shown in Figs.\ \ref{fig_E1_uncorr} and \ref{fig_E2_uncorr} for the HO, THO ($\gamma/b = 1.2 fm^{-1/2}$), and BOX cases; as we define in section \ref{2bPair}, the uncorrelated case has zero residual interaction and a mean field such to obtain a two--particle uncorrelated wave function with the same energy binding ($-1.01 MeV$) as the correlated one.

\vspace{0.5cm}
\vspace{0.5cm}

\vspace{0.5cm}
\vspace{0.5cm}
%
\chapter{Comparison between continuum discretization methods\label{HOvsTHO_cfr}}
        \newpage

        \null 

        \thispagestyle{empty} 

        \newpage
We have performed the one--body and two--body continuum discretization using pseudostates obtained with three different methods
\begin{itemize}
\item Truncated Harmonic Oscillator basis (HO),
\item Truncated Harmonic Oscillator basis + LST (THO),
\item Potential in a Box (BOX).
\end{itemize}
Each of the three approaches considered has its pros and cons.
The HO method is computationally very efficient and simple. The calculation convergence needs to be explored only as a function of the basis size, N, what simplifies the task. This method has been applied in the 70's to the calculation of resonances with stabilization plots \cite{hazi_taylor} with satisfactory results. 
The main drawback of the HO method is the Gaussian asymptotic dependence of the basis functions.
This makes basis states tend to zero much faster than the exponentially  decaying bound states.
A sensible election of the inverse oscillator length parameter $a$ can help to tackle with weakly--bound states, as explained in App.\ \ref{inv_osc_len}.
However, the HO approach requires larger N values to obtain the bound system eigenvalues with a certain degree of accuracy compared to the other two methods.

The main advantage of the THO method is that it corrects the asymptotic Gaussian dependence of basis functions of the HO case including the useful possibility of tuning via the $\gamma/b$ parameter value the distribution of pseudostates in the positive energy region.
As in the HO case, the only parameter that needs to be varied looking for convergence is N, the basis size.
The LST makes calculations more involved, but the degree of complexity is not too high and computing time increments are almost negligible.
This is certainly so with the analytical LST from \cite{tho_lst_4} compared to the original formulation of the THO method \cite{sum_rules}.
As we have already stated, another difference with the HO method is that we need to assign values to the $\gamma/b$ ratio parameter.
The best option is to fix it considering the continuum region which is of major interest in each particular case and give $\gamma/b$ values according to this \cite{tho_lst_3}.

The BOX case differs from the previous two in the sense that the convergence should not only be proved with respect to N, the basis dimension, but also as a function of the box radius $x_b$.\\

We concentrate on the comparison of the HO versus THO approaches to ascertain if there is an advantage in the inclusion of the LST.
As a help for the comparison of the approaches considered in the one-- and two--body cases we present in Tab.\ \ref{tab1} the basis dimension in the HO and THO cases such that for a $ \Delta N = 10 $ the variation in the bound eigenvalues of the Woods--Saxon model potential is $ \Delta E_{b} < 5 keV $. 
The HO$^{(1)}$ and HO$^{(2)}$ notation stands for truncated Harmonic Oscillator basis in both cases: in the HO$^{(1)}$ case the inverse oscillator length $a$ is assessed with $N = 1$ calculation, though in the HO$^{(2)}$ case the value of $a$ is fixed using a recipe described in App.\ \ref{inv_osc_len}, aiming to improve the description of weakly--bound states.

The THO$^{(1)}$ and THO$^{(2)}$ notation stands for THO with $\gamma/b = 2.4 fm^{-1/2}$ and $\gamma/b = 1.2 fm^{-1/2}$, respectively.

The value of $x_{max}$ is the integration interval in both cases; integrals stretch from $-x_{max}$ to $+x_{max}$ and we check that this value is large enough in each case computing the basis states normalization.\\

\begin{table}[!htc]
\begin{center}
\begin{tabular}{lcc}
\hline 
\hline
             &  N   &   $ x_{max} (fm) $  \\
HO$^{(1)}$   &  90  &    55.0         \\
HO$^{(2)}$   &  50  &    60.0         \\
THO$^{(1)}$  &  50  &    75.0         \\
THO$^{(2)}$  &  20  &    75.0         \\
\hline 
\hline
\end{tabular} 
\caption{One body dimension N and integration length $ x_{max} $ required to achieve converged energies for the different methods. In the HO$^{(1)}$ case the inverse oscillator length $a$ is assessed with $N = 1$ calculation, though in the HO$^{(2)}$ case the value of $a$ is fixed using a recipe described in App.\ \ref{inv_osc_len}. The THO$^{(1)}$ and THO$^{(2)}$ notation stands for THO with $\gamma/b = 2.4 fm^{-1/2}$ and $\gamma/b = 1.2 fm^{-1/2}$, respectively. \label{tab1}}
\end{center}
\end{table}

\begin{figure}[!htc]
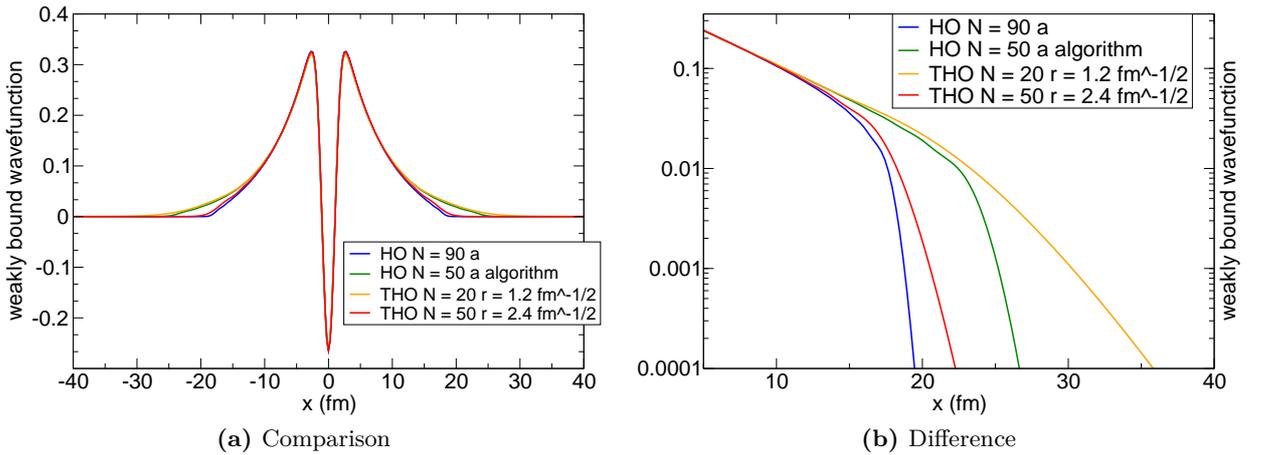

\begin{center}
\subfloat[Comparison]{
\scalebox{0.32}{\includegraphics{1D_1b_wb_wf_hovstho}} 
}
\hspace*{0.1cm}
\subfloat[Difference]{
\scalebox{0.32}{\includegraphics{1D_1b_wb_wf_hovstho_zoom}}
}
\caption{In the left part are reported the weakly--bound wave functions for the 1D one--body problem calculated with HO and THO basis using different parameters. In the right part the logaritmic scale of the wave functions tails; the main differences are located on the tails region and in particular it can be noticed that with THO$^{(2)}$ one can reproduce the proper tail using a smaller number of basis states.\label{fig_ho_tho_1b_wf}}
\end{center}
\end{figure}

In  Fig.\ \ref{fig_ho_tho_1b_wf} the least--bound state of the model Woods--Saxon potential $\Psi_2 (x)$ is depicted for HO$^{(1,2)}$ and THO$^{(1,2)}$, the left panel shows the full spatial dependence of $\Psi_2 (x)$ and the right panel shows a zoom of the wave function tails, where differences are more noticeable.\\

\vspace{0.5cm}

\begin{figure}[!htc]
\begin{center}
\scalebox{0.32}{\includegraphics{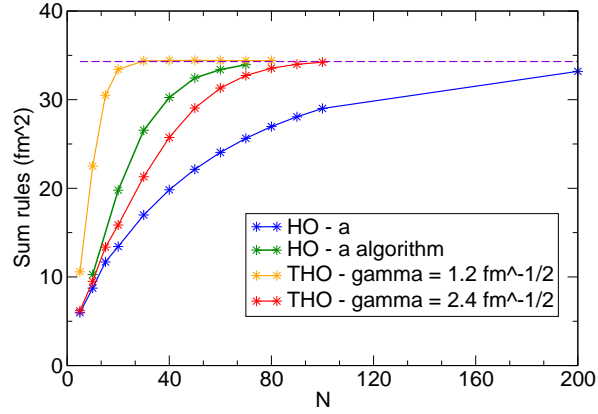}}
\caption{The total strength sum rule convergence calculated with the weakly bound state for the 1D one--body problem in a Woods--Saxon potential.\label{sum_rules_cfr}}
\end{center}
\end{figure}

We also include in the one--body case the results obtained for the Total Strength sum rule of $\Psi_2 (x)$ with the four approaches considered in Fig.\ \ref{sum_rules_cfr}.\\

\begin{figure}[!htc]
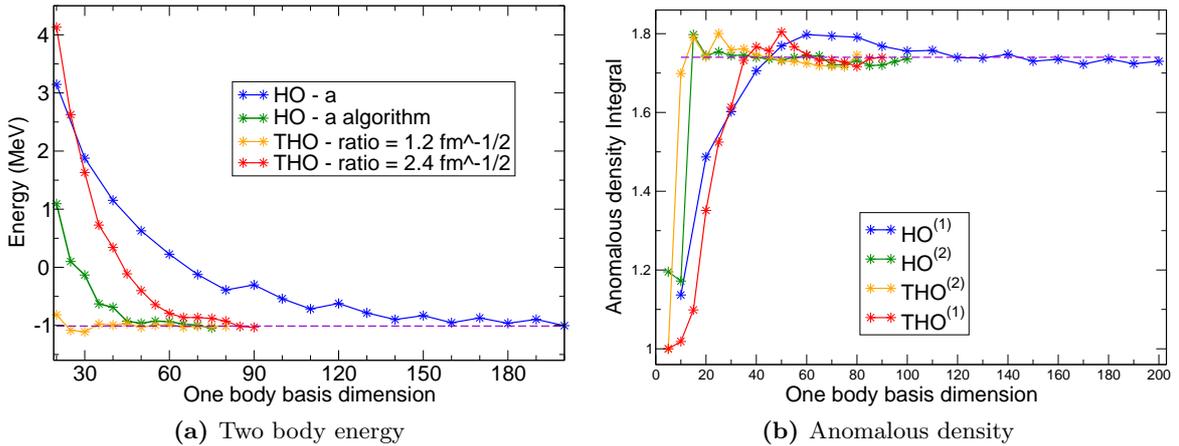

\begin{center}
\subfloat[Two body energy]{
\scalebox{0.32}{\includegraphics{cfr_tho_ho_2b_en_conv}}
}
\hspace*{0.1cm}
\subfloat[Anomalous density]{
\scalebox{0.32}{\includegraphics{cfr_tho_ho_2b_anomdens_conv}}
}
\caption{In the left part the energy convergence for the 1D two--body problem in a Woods--Saxon potential. In the right part the anomalous density for the ground state convergence.\label{1D_2b_energy_anomdens}}
\end{center}
\end{figure}

Finally in the two--body case we include the weakly--bound ground state energy dependence with N in Tab.\ \ref{tab2} as well as a plot of the ground state energy (left panel) and ground state anomalous density value (right panel) in Fig.\ \ref{1D_2b_energy_anomdens}.

\begin{table}[!htc]
\begin{center}
\begin{tabular}{lcc}
\hline 
\hline
                     &  $ N_{1b} $   &   $ x_{max} (fm) $  \\
HO$^{(1)}$  &  200   &    30.0         \\
HO$^{(2)}$  &  45    &    75.0         \\
THO$^{(2)}$ &  50    &    75.0         \\
THO$^{(1)}$ &  85    &    75.0         \\
\hline 
\hline
\end{tabular} 
\caption{One body dimension N and integration length $ x_{max} $ required to achieve converged energy for the different methods.
 In the HO$^{(1)}$ case the inverse oscillator length $a$ is assessed with $N = 1$ calculation, though in the HO$^{(2)}$ case the value of $a$ is fixed using a recipe described in App.\ \ref{inv_osc_len}. The THO$^{(1)}$ and THO$^{(2)}$ notation stands for THO with $\gamma/b = 2.4 fm^{-1/2}$ and $\gamma/b = 1.2 fm^{-1/2}$, respectively.\label{tab2}}
\end{center}
\end{table}

As can be seen in Tabs.\ \ref{tab1} and \ref{tab2} and in Figs.\ \ref{fig_ho_tho_1b_wf}, \ref{sum_rules_cfr}, \ref{1D_2b_energy_anomdens} results from the three approaches are in good agreement, as expected.
The THO$^{(2)}$ seems the most flexible method, as it manages to obtain converged energies and a fairly good description of asymptotic wave functions. This is the case with the fastest convergence and a best asymptotic wave function description in the one--body case.\\

The difference in basis sizes in Tabs.\ \ref{tab1} and \ref{tab2} seems not too dramatic in this case, but we should take into account that these continuum discretization methods are of interest for three--body studies and beyond \cite{Manoli,Matsu_1,Matsu_2,Matsu_3} and in this case having a small one--body basis confers a very important advantage in the full problem.\\

The THO method also, as seen in Fig.\ \ref{fig_ht_tho_stabplot}, is more adequate to find resonances in stabilization plots.\\

The convenience of THO$^{(2)}$ is further confirmed in the one--body case by fast convergence of the Total Strength sum rule for the $\Psi_2 (x)$ weakly--bound state, and in the two--body case for the fast energy and anomalous density convergence.\\

It is important to emphasize that all methods agree on the large N limit for the different quantities computed.
%
\chapter{Summary and Conclusions}
\clearpage
        \newpage

        \null 

        \thispagestyle{empty} 

        \newpage
The goal of the present memory is to investigate a weakly--bound 1D quantum two--body system.
This system is composed by a core plus two valence nucleons bound by a matter density--dependent delta residual interaction.
The inclusion of the continuum in the description of the chosen  system is absolutely mandatory and we have explored three different bases to discretize the one--body system (mean field potential modeled by a Woods--Saxon) and obtain both bound states (states with negative energies) and pseudostates (positive energy states). The pseudostate are considered a finite and discrete representation of the continuum spectrum.\\

In the one--body case, in addition to the calculation of the system eigenstates we have computed several magnitudes of interest to assess the goodness of the continuum description that we have achieved with the different methods.
In particular, we have computed the Total Strength for the ${\cal O} = x^2$ operator, the Energy Weighted sum rule, the Electric Dipole and Quadrupole transition  intensities, and we have shown the possibility of dealing with resonances and phase shifts in the pseudostate approach.\\

Still in the one--body problem, we have compared the results obtained with the pseudostate approach with calculations performed using other methods.
We have compared the results obtained in the BOX case using an Infinite Square Well basis to build the model Woods--Saxon potential plus a Box Hamiltonian with the solution of the problem using a Numerov algorithm to integrate the 1D time independent Schroedinger Equation. We have also computed, using the the Numerov algorithm, the true continuum wave functions (momentum normalized and symmetrized).
We have checked the result of several of our calculations with the result obtained with these continuum non--normalizable states.
We have defined a continuum density operator for each pseudostate that gives a pictorial description of how pseudostates discretize the continuum computing the overlap between continuum wave functions and pseudostates.
Using the computed overlap between continuum wave functions and pseudostates we have obtained preliminary results for folded E1 and E2 electric transition probabilities that qualitatively agree with the pseudostate method results.
We have also compared the pseudostates with the square integrable wave functions obtained using the average method, the standard method employed in CDCC calculations to discretize the continuum. \\

In the two--body case, we have built the basis and the system Hamiltonian matrix obtained in the one--body system. We have computed the system eigenvalues and eigenvectors, studied the bound states energy convergence and the nature of its wave functions as well as its degree of collectivity (via anomalous density).
We have also computed E1 and E2 transition intensities for the two--body system.\\

As already mentioned in the introduction, we have developed the computer codes required to perform the calculations included in the memory.\\

The results obtained in the considered (HO, THO, and BOX) one-- and two--body cases agree satisfactorily.
The pseudostates method is a computationally efficient method to deal with weakly--bound systems.
The 1D simplified model allows a very clear description of the relevant physical parameters without such mathematical complexities.\\

In the one--body case the system is so simple that large (several hundreds elements) bases can be used with a very small computing time required. However, for the two--body system analysis, despite the symmetrization and energy threshold, the dimension of the involved matrices is much larger and the problem is computationally much heavier.
In fact, this is the main reason to support the convenience of the case of a THO basis.
Despite the necessity of calculating an optimal $\gamma/b$ ratio for the problem under study, the THO basis offers two important advantages.
The first is the possibility of tuning the density of continuum states, making possible to enlarge the density pseudostates at energies relevant for the process under study. The second advantage offered by the THO approach is the exponentially asymptotic behavior of its basis elements. Due to this the THO approach achieves convergence for weakly--bound states faster than the other two methods. 
Though in the present case this advantage is not decisive, in more involved multichannel many--body  calculations it can be of major relevance.\\

The present results open many possible lines to proceed the investigation on weakly--bound systems.
We enumerate some promising research lines for the future:

\begin{itemize}
\item study the role of resonances when moving from a one--body problem with resonant states to a two--body system as we have done in the memory;
\item in connection to the previous point, the present model is a convenient way of reckoning the importance of the pairing interaction in the continuum and the effect of resonant and non--resonant states \cite{pairing};
\item the present approach simplicity allows its use to model transfer reactions and break--up  processes in a simplified and schematic way;
\item the inclusion of core excitations in the model opens also a very enticing line of research. These core excitations have been proved to play an important role in the determination of the structure of some halo nuclei \cite{dinam_core_1,dinam_core_2} and this model could offer a convenient (and simple) test ground for their study;
\item the extension of the model to 2D and 3D systems, with the inclusion of angular momentum in the picture, will allow to treat more realistic systems.
\end{itemize}
%
\addcontentsline{toc}{chapter}{Acknowledgments}
\chapter*{Acknowledgments}
        \newpage

        \null 

        \thispagestyle{empty} 

        \newpage

Ringrazio il professor Andrea Vitturi per i suoi consigli e per l'entusiasmo che mi trasmette sempre con i suoi incoraggiamenti.\\
Agradezco tambi\'en al profesor Francisco P\'erez Bernal por brindarme todo su apoyo en esta tarea adem\'as de su amistad y confianza; 
sin olvidar la excelente acogida del Departamento de F\'isica Aplicada de la Universidad de Huelva donde recib\'i particularmente el apoyo del profesor Jos\'e Enrique Garc\'ia Ramos a quien tambi\'en me gustar\'ia agradecer.\\
Por otro lado me gustar\'ia dar las gracias al Departamento de F\'isica At\'omica, Molecular y Nuclear de la Universidad de Sevilla, en donde me he encontrado conversaciones fruct\'iferas con los profesores Jos\'e Arias y Antonio Moro, y en donde tambi\'en me he encontrado la ayuda de Manoli.\\

%
\begin{appendices}
%
\chapter{The inverse oscillator length\label{inv_osc_len}}
	
	The inverse oscillator length parameter $a = \left( \dfrac{k \mu}{\hbar^2} \right) ^{1/4}$ determines the curvature of the HO potential at the origin and thus, how wide is the potential. 
	
	In accordance to other cases \cite{tho_lst_3} the value of this parameter is fixed to minimize the
	ground state energy with a small HO basis. In fact, we use a $N = 1$ basis, that is, a basis with the HO ground state as its only
	component to obtain an approximation to the system's ground
	state energy. The value of $a$ is varied to minimize this
	energy. 
	
	In cases with several bound states that include weakly--bound states, as the chosen model Woods--Saxon potential, the $a$ value obtained using the ground state is too large (Harmonic potential too narrow) and a very large number of HO states in the basis is request to sample the large $x$ value where the weakly--bound state wave function tails are still significant.
	To overcome this problem one should diminish the value of the $a$ value of the oscillator in order to include the whole range of the significant states.
		
	This can be done fixing the value by hand though we define the following recipe or algorithm.
	We first make the minimization explained above; 
	then with the value of $a$ obtained we build the basis and the system Hamiltonian is diagonalized.
	At this point it is possible to evaluate the expectation value of the $ x^2 $ operator for the weakly bound wave function e.\ g.\ $ \langle \Psi_2 			
	\vert x^2 \vert \Psi_2 \rangle $. We then compare this result with the same matrix element calculated for an Harmonic Oscillator basis.
	We can set a new inverse oscillator length $ a $ calculated analytically following Eq.\ (\ref{x2homatel}), $ \langle \Phi_2 ^{(HO)}			
	\vert x^2 \vert \Phi_2 ^{(HO)} \rangle $ with $ i = j = 2 $.
	Once obtained the new parameter we reconstruct the basis and diagonalize again before proceeding with the observables calculation.
	
\begin{figure}[!htb]
	\centering
	\includegraphics[scale=0.4]{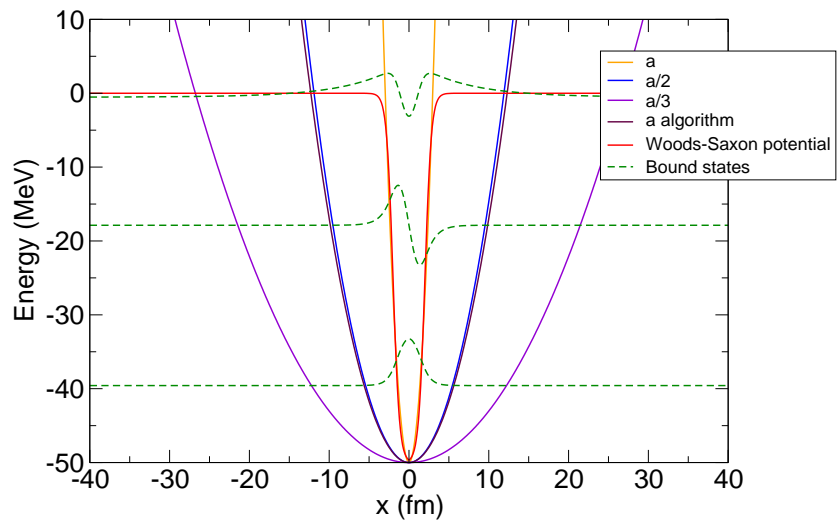}
	\caption{The Woods--Saxon potential with the wave functions and the different tries for the basis oscillator.\label{fig_ho_bas}}
	\end{figure}

	In Figure \ref{fig_ho_bas} is shown the comparison between the model Woods--Saxon potential and the Harmonic Oscillator potentials obtained by minimization (label \textit{a}), by fixing the value by hand after minimization (labels \textit{a/2} and \textit{a/3}) and by following the procedure described.\\

	It is evident that the variation of the inverse oscillator length allows to construct an Harmonic Oscillator basis which stretches to the full bound states range limiting the number of necessary HO functions in the basis.
\chapter{Energy--Weighted Sum Rule demonstration\label{app_ewsr}}

We take into account that for large N values the basis is complete
\begin{equation}
\sum _i \vert \Psi_i \rangle \langle \Psi_i \vert = 1
\end{equation}
and the TISE 
\begin{equation}
{\cal H} \vert \Psi_b \rangle = E_b \vert \Psi_b \rangle
\end{equation}
where $ {\cal H} = - \frac{1}{2} \frac{\hbar ^2}{\mu}\dfrac{d^2}{dx^2} + V(x) $.

\begin{align}
{\cal E}^{(b)}_W({\cal O}(x)) &= \lim _{N \rightarrow\infty} {\cal E}^{(b)}_W({\cal O}(x), N)\\
 &= \sum_i^N (E_i - E_b) \mid  \langle \Psi_{i} \vert {\cal O}(x) \vert \Psi_{b} \rangle \mid ^2 ,\notag\\
	&=  \sum_{j}^N (E_i - E_b) \langle \Psi_{b} \vert {\cal O}(x)\vert \Psi_{j} \rangle  \langle \Psi_{j}  \vert {\cal O}(x) \vert \Psi_{b} \rangle ,\notag \\
	& = \langle \Psi_{b} \vert {\cal O}(x) \sum_j^N \vert \Psi_{j} \rangle \langle \Psi_{j} \vert ({\cal H} - E_b) \sum_{j'}^N \vert \Psi_{j'} \rangle \langle \Psi_{j'} \vert {\cal O}(x) \vert \Psi_{b} \rangle, \notag\\
	& = \frac{1}{2}  \langle \Psi_{b} \vert 2 {\cal O}(x) ({\cal H} - E_b) {\cal O}(x) \vert \Psi_{b} \rangle, \notag\\
	& = \frac{1}{2} \langle \Psi_{b} \vert 2 {\cal O}(x) ({\cal H} - E_b) {\cal O}(x) - ({\cal H} - E_b){\cal O}(x)^2 - {\cal O}(x)^2 ({\cal H} - E_b) \vert \Psi_{b} \rangle, \notag\\
	& = \frac{1}{2} \langle \Psi_{b} \vert \left[ {\cal O}(x), ({\cal H} - E_b){\cal O}(x) - {\cal O}(x)({\cal H} - E_b) \right] \vert \Psi_{b} \rangle, \notag\\
	& = \frac{1}{2} \langle \Psi_{b} \vert \left[ {\cal O}(x), \left[{\cal H} - E_b, {\cal O}(x) \right] \right] \vert \Psi_{b} \rangle, \notag\\
\end{align}
And acting with the double commutator on a function $ \varphi (x) $ one finally obtains
\begin{align}
\left[ {\cal O}(x), \left[{\cal H} - E_b, {\cal O}(x) \right] \right] \varphi (x)
	&= - \frac{1}{2} \frac{\hbar ^2}{\mu} \left[ {\cal O}(x), \left[ \dfrac{d^2}{dx^2} , {\cal O}(x) \right] \right] \varphi (x) ,\notag\\
	&= - \frac{1}{2} \frac{\hbar ^2}{\mu} \left[ 2 {\cal O}(x) \dfrac{d^2}{dx^2} ({\cal O}(x) \varphi (x)) - {\cal O}(x)^2 \dfrac{d^2}{dx^2} \varphi (x) - \dfrac{d^2}{dx^2} {\cal O}(x)^2  \varphi (x) \right] ,\notag\\
	& = \frac{\hbar ^2}{\mu} \left( \dfrac{d{\cal O}(x)}{dx} \right) ^2 \varphi (x) ,\notag\\
\end{align}
In this case the operator is $ {\cal O}(x) = x $ so in the $ N \rightarrow \infty $ limit the result expected is $ {\cal E}^{(b)}_W({\cal O}(x)) =\frac{1}{2} \frac{\hbar ^2}{\mu} $.\\

\chapter{Continuum one-body eigenstates\label{appendix_binning}}
\begin{figure}[!htb]
\centering
\includegraphics[scale=0.4]{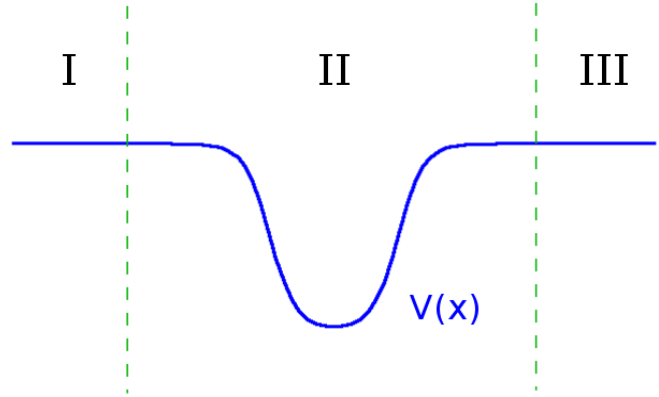}
\caption{A potential V(x) divided in three regions, the one above the potential (II) and the asymptotic regions (I and III).\label{app_pot}}
\end{figure}

In order to compare the pseudostate approach with the true continuum wavefunctions we consider the problem of the scattering in a general symmetric  1D potential $V(x)$ \cite{Tannor}.
We divide the real axes in two asymptotic regions labeled (I) and (III) and the potential region (II), as shown schematically in Fig.\ \ref{app_pot}.

The continuum wave functions are those solutions of the Schroedinger equation (\ref{eq_avl}) with positive energy ($E > 0$). In this case all energy is allowed and levels are doubly degenerate. For each energy one should consider a wave incoming from the left and another from the right. 
In the asymptotic region solutions are 
\begin{equation}
\phi _{\pm k}(x) = \dfrac{e^{\pm ikx}}{\sqrt{2\pi}}.
\label{app_en_norm}
\end{equation}
These continuum waves $ \phi _{k}(x) $ are momentum--normalized, thus
\begin{equation}
\langle \phi _{k} \vert \phi _{k'} \rangle  = \delta(k - k').
\label{app_en_norm}
\end{equation}
As $p = \hbar k$ and $E = \dfrac{\hbar^2 k^2}{2 \mu}$ there can be defined also energy normalized states 
\begin{equation}
\langle \phi _{k} (E) \vert \phi _{k'} (E') \rangle  = \delta(E - E')
\label{app_en_norm}
\end{equation}
and 
\begin{equation}
\phi _{\pm k} (E) = \left( \frac{m}{2 \pi \hbar ^2 |k|} \right) ^{1/2} e^{\pm ikx}.
\label{app_en_norm}
\end{equation}
The left incoming wave will have the asymptotic behaviors
\begin{subequations}
\begin{align}
\phi _{k} ^{I} (x) = \frac{1}{\sqrt(2 \pi)} \left( e^{ikx} + r e^{-ikx} \right) ~~~~~~\rightarrow ~incident ~ + ~ reflected\\
\phi _{k} ^{III} (x) = \frac{1}{\sqrt(2 \pi)} t e^{ikx} ~~~~~~\rightarrow ~transmitted
\end{align}
\label{app_eq_I_III}
\end{subequations}
\noindent where, in general, $ t, r \in  \mathbb{C} $ and $ \vert t \vert ^2 + \vert r \vert ^2 = 1 $.\\
We define $ \alpha _k $ as the logaritmic derivative of the numerical solution $ \psi_k (x) $ obtained applying Numerov algorithm with initial conditions \ref{app_eq_I_III}
\begin{equation}
\alpha = \dfrac{\psi'(x_m)}{\psi(x_m)}
\label{app_alfa_def}
\end{equation}
and $A$ as a complex normalization constant
\begin{equation}
A \psi_k(x_m) = \frac{1}{\sqrt(2 \pi)} \left( e^{ikx_m} + r e^{-ikx_m} \right)~~~and~~~A \psi(x_M) = \frac{1}{\sqrt(2 \pi)} t e^{ikx_M} 
\label{app_A_def}
\end{equation}
where $x_m$ and $x_M$ are the minimum and maximum $x$ values.
Using Eqs.\ \ref{app_alfa_def} and \ref{app_A_def} we obtain
\begin{subequations}
\begin{align}
r & = \dfrac{1+i\frac{\alpha}{k}}{1-i\frac{\alpha}{k}} e^{2ikx_m}, \\
A & = \frac{1}{\sqrt(2 \pi)} \frac{1}{\psi(x_m)}  \left( e^{ikx_m} + r e^{-ikx_m} \right),\\
t & = sqrt(2 \pi) A \psi(x_M) e^{-ikx_M}.
\end{align}
\label{app_par}
\end{subequations}

The degenerate partner is obtained forcing the asymptotic behavior  
\begin{subequations}
\begin{align}
\phi _{-k} ^{I} (x) = \frac{1}{\sqrt(2 \pi)} t e^{-ikx},\\
\phi _{-k} ^{III} (x) = \frac{1}{\sqrt(2 \pi)} \left( e^{-ikx} + r e^{ikx} \right),
\end{align}
\label{app_eq_I_III_partner}
\end{subequations}
and in this case 
\begin{subequations}
\begin{align}
r & = \dfrac{1-i\frac{\alpha}{k}}{1+i\frac{\alpha}{k}} e^{-2ikx_m}, \\
A & = \frac{1}{\sqrt(2 \pi)} \frac{1}{\psi(x_M)}  \left( e^{-ikx_M} + r e^{ikx_M} \right),\\
t & = \sqrt(2 \pi) A \psi(x_m) e^{ikx_m}.
\end{align}
\label{app_par}
\end{subequations}

One can calculate the symmetric and antisymmetric combinations
\begin{subequations}
\begin{align}
\psi _{\pm k} ^g (x) = \frac{1}{\sqrt(2 \pi)} \left[ \psi _{+ k}(x) + \psi _{- k}(x) \right] \\
\psi _{\pm k} ^u (x) = \frac{1}{\sqrt(2 \pi)} \left[ \psi _{+ k}(x) - \psi _{- k}(x) \right]
\end{align}
\label{app_sym_asym}
\end{subequations}
such that 
\begin{equation}
\langle \psi _{k} ^{\Gamma} \vert \phi _{k'} ^{\Gamma '} \rangle  = \delta_{\Gamma \Gamma '} \delta(k - k')
\label{app_scal_prod}
\end{equation}
where $ \Gamma = g, u$, and the solutions are still momentum-normalized.

\begin{equation}
\varphi _n (x) = \frac{1}{C_{norm}} \int_{k_n - \Delta_k} ^{k_n + \Delta_k} \phi ^{\pm k} _{S,A} (x) dk,
\label{binning}
\end{equation}
which results in a set of normalizable waves 
\begin{equation}
\langle \varphi _n \vert \varphi _m \rangle  = \delta_{nm}
\label{n_norm}
\end{equation}
decaying to zero. The size of $ \Delta_k $ regulates the distance at which the waves decay to zero.

\end{appendices}
%
\addcontentsline{toc}{chapter}{\bibname}
\bibliographystyle{unsrt} 
\bibliography{bib_tesi}
\end{document}